\PassOptionsToPackage{unicode}{hyperref}
\PassOptionsToPackage{hyphens}{url}
\PassOptionsToPackage{dvipsnames,svgnames,x11names}{xcolor}
\documentclass[12pt, a4paper]{article}
\usepackage[top=1.54in, bottom=1.55in, left=1.35in, right=1.35in]{geometry}

\usepackage{xr-hyper}
\usepackage{amsmath,amssymb,titlesec}
\usepackage{iftex}
\ifPDFTeX
\usepackage[T1]{fontenc}
\usepackage[utf8]{inputenc}
\usepackage{textcomp, orcidlink} 
\else
\usepackage{unicode-math}
\defaultfontfeatures{Scale=MatchLowercase}
\defaultfontfeatures[\rmfamily]{Ligatures=TeX,Scale=1}
\fi
\usepackage{lmodern}
\ifPDFTeX\else  
\fi
\IfFileExists{upquote.sty}{\usepackage{upquote}}{}
\IfFileExists{microtype.sty}{
\usepackage[]{microtype}
\UseMicrotypeSet[protrusion]{basicmath}
}{}
\makeatletter
\@ifundefined{KOMAClassName}{
\IfFileExists{parskip.sty}{
\usepackage{parskip}
}{\setlength{\parindent}{0pt}
\setlength{\parskip}{6pt plus 2pt minus 1pt}}
}{
\KOMAoptions{parskip=half}}
\makeatother
\usepackage{xcolor}
\makeatletter
\ifx\paragraph\undefined\else
  \let\oldparagraph\paragraph
  \renewcommand{\paragraph}{
    \@ifstar
      \xxxParagraphStar
      \xxxParagraphNoStar
  }
  \newcommand{\xxxParagraphStar}[1]{\oldparagraph*{#1}\mbox{}}
  \newcommand{\xxxParagraphNoStar}[1]{\oldparagraph{#1}\mbox{}}
\fi
\ifx\subparagraph\undefined\else
  \let\oldsubparagraph\subparagraph
  \renewcommand{\subparagraph}{
    \@ifstar
      \xxxSubParagraphStar
      \xxxSubParagraphNoStar
  }
  \newcommand{\xxxSubParagraphStar}[1]{\oldsubparagraph*{#1}\mbox{}}
  \newcommand{\xxxSubParagraphNoStar}[1]{\oldsubparagraph{#1}\mbox{}}
\fi
\makeatother

\usepackage{longtable,booktabs,array}
\usepackage{calc} 
\usepackage{etoolbox}
\makeatletter
\patchcmd\longtable{\par}{\if@noskipsec\mbox{}\fi\par}{}{}
\makeatother
\IfFileExists{footnotehyper.sty}{\usepackage{footnotehyper}}{\usepackage{footnote}}
\makesavenoteenv{longtable}
\usepackage{graphicx}
\makeatletter
\def\maxwidth{\ifdim\Gin@nat@width>\linewidth\linewidth\else\Gin@nat@width\fi}
\def\maxheight{\ifdim\Gin@nat@height>\textheight\textheight\else\Gin@nat@height\fi}
\makeatother
\setkeys{Gin}{width=\maxwidth,height=\maxheight,keepaspectratio}
\makeatletter
\def\fps@figure{htbp}
\makeatother

\makeatletter
\@ifpackageloaded{caption}{}{\usepackage{caption}}
\AtBeginDocument{
\ifdefined\contentsname
\renewcommand*\contentsname{Table of contents}
\else
\newcommand\contentsname{Table of contents}
\fi
\ifdefined\listfigurename
\renewcommand*\listfigurename{List of Figures}
\else
\newcommand\listfigurename{List of Figures}
\fi
\ifdefined\listtablename
\renewcommand*\listtablename{List of Tables}
\else
\newcommand\listtablename{List of Tables}
\fi
\ifdefined\figurename
\renewcommand*\figurename{Figure}
\else
\newcommand\figurename{Figure}
\fi
\ifdefined\tablename
\renewcommand*\tablename{Table}
\else
\newcommand\tablename{Table}
\fi
}
\@ifpackageloaded{float}{}{\usepackage{float}}
\floatstyle{ruled}
\@ifundefined{c@chapter}{\newfloat{codelisting}{h}{lop}}{\newfloat{codelisting}{h}{lop}[chapter]}
\floatname{codelisting}{Listing}

\makeatother
\makeatletter
\@ifpackageloaded{caption}{}{\usepackage{caption}}
\@ifpackageloaded{subcaption}{}{\usepackage{subcaption}}
\usepackage[font=small]{caption}
\makeatother

\ifLuaTeX
\usepackage{selnolig}
\fi
\usepackage[]{natbib}
\usepackage{bookmark}

\IfFileExists{xurl.sty}{\usepackage{xurl}}{} 
\hypersetup{
  pdftitle={Title},
  pdfauthor={Author 1; Author 2},
  pdfkeywords={3 to 6 keywords, that do not appear in the title},
  colorlinks=true,
  linkcolor={blue},
  filecolor={Maroon},
  citecolor={Blue},
  urlcolor={Blue},
  pdfcreator={LaTeX via pandoc}}

\newcommand{\anon}{1}

\usepackage{xcolor}
\definecolor{darkblue}{rgb}{0.1, 0.3, 0.75}
\usepackage{fontenc,fontaxes,amsmath,amsfonts,mathtools,arydshln,setspace,bbm,tikz,multirow,longtable,subcaption,graphicx,booktabs,bm,dsfont,microtype,booktabs,longtable,subcaption,ragged2e,paralist,slashbox}
\usepackage{orcidlink}
\usepackage[inline,shortlabels]{enumitem}

\graphicspath{{plots/}}
\mathtoolsset{mathic=true}
\DeclareMathOperator{\ran}{ran}

\DeclareMathOperator{\rank}{rank}

\DeclareMathOperator{\sgn}{sgn}

\newcommand{\hdd}{\bar{h}_T} 
\newcommand{\hddd}{\widetilde{h}_T} 
\newcommand{\htd}{\widetilde{h}} 
\newcommand{\hd}{\bar{h}}
\newcommand{\cd}{{c}_{D}}

\newcommand{\asd}[1]{\hspace{-0.05em}(\text{{$\hspace{-0.05em}#1\hspace{-0.05em}$}})}

\DeclareMathOperator{\seq}{seq}
\def\thickhline{\noalign{\hrule height.3pt}}

\newenvironment{figures}[1][htbp]
  {\begin{figure}[#1] % Start a figure-like environment
   \centering} % Center the content inside the environment
  {\end{figure}} % End the figure environment

\newenvironment{proofs}[1][Proof]
  {\begin{trivlist}
   \item[\hskip \labelsep {\bfseries #1.}] % Custom title for proof (default "Proof")
   \itshape} % Use italic font for the body of the proof
  {\hfill $\square$ \end{trivlist}} % End the proof with a box (for completion)

\newtheorem{lemmas}{Lemma}[section]
\newtheorem{corollarys}{Corollary}[section]

\newtheorem{definitions}{Definition}
\newtheorem{propositions}{Proposition}[section]
\newtheorem{remarks}{Remark}[section]

\graphicspath{{plots/}}

\newtheorem{assumpM}{Assumption}

\newtheorem{assumptionV}{Assumption}

\newtheorem{assumpS}{Assumption}

\newcommand{\commRV}[1]{{\leavevmode\color{black}#1}} 
\newcommand{\revlab}[1]{\phantomsection\label{#1}} 
\newcommand{\commRVV}[1]{{\leavevmode\color{black}#1}} 
\newcommand{\commRVS}[1]{{\leavevmode\color{black}#1}} %% for small correction of typos, rephrase, etc. 
\begin{document}

\def\spacingset#1{\renewcommand{\baselinestretch}%
{#1}\small\normalsize} \spacingset{1}

\if1\anon
{
  \title{\bf Testing for integer integration in functional time series}
  \author{\normalsize Won-Ki Seo \orcidlink{0000-0001-6629-3027} \thanks{
\normalsize    This article benefited from the helpful suggestions for improvement made by the Associate Editor and anonymous reviewers, to whom we express our thanks.}\\
\normalsize     School of Economics, University of Sydney\\
\\
\normalsize     Han Lin Shang \orcidlink{0000-0003-1769-6430} \thanks{
\normalsize     Shang gratefully acknowledges the support of the Australian Research Council through a Future Fellowship (grant no. FT240100338) for this research.} \\
\normalsize     Department of Actuarial Studies and Business Analytics, Macquarie University}
\date{}
  \maketitle
} \fi

\if0\anon
{
  \bigskip
  \bigskip
  \bigskip
  \begin{center}
    {\LARGE\bf Testing for integer integration in functional time series}
\end{center}
  \medskip
} \fi

\bigskip
\begin{abstract}
	We develop a statistical testing procedure to examine whether the curve-valued time series of interest is integrated of order $d$ for a nonnegative integer $d$.  The proposed procedure can distinguish between integer-integrated time series and fractionally-integrated ones, and it has broad applicability in practice. Monte Carlo simulation experiments show that the proposed testing procedure performs reasonably well. We apply our methodology to Canadian yield curve data and French sub-national age-specific mortality data. We find evidence that these time series are mostly integrated of order one, while some have fractional orders exceeding or falling below one.
\end{abstract}

\vfill

\newpage
\spacingset{1.8} % DON'T change the spacing!

\section{Introduction} \label{secintro}
	
	Recent advances in computing and storage technologies have facilitated the presence of functional data with \commRVS{graphical representations} of curves \citep{RS05}, images \citep{SGS13}, and shapes \citep{SK16}. Unlike conventional multivariate analysis, each datum in functional data analysis consists of random functions, which are realizations of a stochastic process commonly residing in a Hilbert space. In many scientific fields ranging from genomics to finance, analyzing functional data has had a significant impact on statistical methods and thinking, permanently changing how we display, model and forecast high-frequency data.  For an overview of functional data analysis, \cite{RS05} presented several state-of-the-art statistical techniques while \cite{FV06} listed a range of nonparametric techniques. \cite{HE15} provided theoretical foundations with an introduction to linear operators, while \cite{MG22} studied spatially-dependent functional data.
	
	\cite{KS23} presents a survey of second-generation functional data, including the analysis of temporally dependent functional data. For an overview of functional time series, \cite{KR17} introduced several state-of-the-art statistical techniques. Functional time series analysis balances functional data and time series analyses. Similar to its univariate or multivariate counterparts, a temporal dependence structure exists in functional observations. It arises in many fields, including demography \citep{CM09}, environmental science \citep{DG02}, finance \citep{KZ12} and transportation \citep{KKW17}. Depending on whether the continuum is also a time variable, functional time series can be classified into two groups. 
	
	%biology \citep{BNF+23}, 
	
	For analyzing functional time series, a growing body of literature has developed statistical methods and theory\commRVS{, the bulk of which} assumes stationarity over the temporal domain, i.e., short-range dependence. These include the functional autoregressive model \citep{Bosq00,KR13}, linear regression with dependent functional variables \citep{seong2021functional},   nonparametric functional regression \citep{FV06}, functional factor model \citep{BYZ10}, functional moving average model \citep{KK17}, and functional autoregressive moving average model \citep{KKW17}. 
	
	On the other hand, there have been many notable developments in long memory \citep[see, e.g.,][]{LRS20, SK23, BKM24} and cointegration \citep[see, e.g.,][]{CKP16, CHP17} of functional time series.  The long-memory curve time series describes processes with greater persistence than short-range dependent ones. In the stationary case, autocovariance decays very slowly, and the spectral density is unbounded, typically at zero frequency. \cite{CKP16} considered cointegration for functional time series and developed statistical methods based on functional principal component analysis. \cite{BSS17} and \cite{SEO2019} extended the Granger-Johansen representation theorem for nonstationary functional time series taking values in a Hilbert space and a Bayes Hilbert space, respectively; more recent extensions in this area include \cite{BS20}, \cite{FP20}, and \cite{seo_2022}.
	% While \cite{BSS17} provided a representation of I$(1)$ autoregressive Hilbertian process, \cite{BS20} presented a representation of I$(1)$ and I$(2)$ autoregressive Hilbertian processes; see also \cite{FP20} and \cite{seo_2022} for similar representation results in a more general setting.
	
	{A research gap in the literature concerning stationary curve time series is how to determine if the time series of interest is I$(0)$ (i.e., short-range dependent) or I$(d)$ for $d \in (-0.5, 0.5) \setminus \{0\}$ (i.e., fractionally-integrated). To fill this gap, we offer our solution by presenting a functional version of the KPSS-type test, which has been widely used for examining the stationarity of scalar- or vector-valued time series. A similar gap exists in the literature on nonstationary curve-valued time series, where it is often assumed that such a time series is I$(1)$ or integrated of any integer order, and hence, appropriately differenced sequences are assumed to be stationary I$(0)$ sequences. We thus extend our proposed test to assess the widespread assumption of integer integration for nonstationary time series. A useful testing procedure based on sequential applications of our proposed tests is also provided; it not only assesses the statistical plausibility of integer integration but also enables practitioners to identify an admissible range for the integration order of functional time series. \revlab{r1_major1}\commRV{In the literature, the assumption of integer integration is commonly employed, which may be because it enables the direct use of existing methods for stationary series after taking integer-order differences. The fractional integration framework, which allows $d$ to take real values within a restricted range, challenges this assumption; however, it remains at an early stage of development, and many applied studies continue to rely on integer integration. In this context, our methods provide a practical way to test the integer-integration hypothesis against fractional alternatives, addressing a question that may often be of interest to practitioners. Furthermore, prior approaches to estimating the integration order $d$ require prior knowledge of an interval containing $d$, which can also affect the asymptotic properties of the estimators (see, e.g., \citealp{seoshang22}). Our procedure serves as a complementary tool to systematically infer the plausible range of the memory parameter $d$.}

The remainder of the paper is outlined as follows. In Section~\ref{sec:2},  we define notation and introduce fractionally-integrated functional time series. Section~\ref{sec:3} outlines assumptions and develops the proposed tests for examining integer integration; in Section~\ref{sec_frac01}, we test the presence of short-range dependence against dependence implied by fractional integration in a stationary functional time series. Its extension to the nonstationary case is documented in Section~\ref{sec_frac02}. Some extensions of the proposed tests, including a sequential testing procedure, are considered in Section~\ref{sec:4.3}. In Section~\ref{sec_sim} we present a series of Monte Carlo simulation studies. In Section~\ref{sec:6}, we apply our hypothesis tests to Canadian yield curve data and French sub-national age-specific mortality rates. Concluding remarks are given in Section~\ref{sec:7}.
	
	\section{Preliminary}\label{sec:2}
	
	\subsection{Notation}\label{sec:2.1}
	
	We let $\mathcal H$ be a Hilbert space of square-integrable functions on a compact interval, equipped with inner product $\langle\cdot,\cdot\rangle$ and norm $\|\cdot\|$. Let $\mathcal L_{\mathcal H}$ denote the space of continuous linear operators equipped with the usual operator norm $\|\cdot\|_{\mathcal L_{\mathcal H}}$. For any $A \in \mathcal L_{\mathcal H}$, \revlab{r2major5}\commRV{$\ran A$  denotes the range of $A$. The dimension of $\ran A$ is called the rank of $A$ and is denoted by $\rank A$ (i.e., $\dim(\ran A)=\rank A$).} The adjoint $A^\ast$ of $A$ is the unique element of $\mathcal L_{\mathcal H}$ satisfying that $\langle A h_1 ,h_2 \rangle=\langle h_1 ,  A^\ast h_2 \rangle$ for all $h_1,h_2 \in \mathcal H$. If $A=A^\ast$, it is said to be self-adjoint. We say that $A$ is nonnegative (resp.\ positive) if $\langle A h,h \rangle \geq 0$ (resp.\ $\langle A h,h \rangle > 0$) for all nonzero $h\in \mathcal H$.  An element $A \in \mathcal L_{\mathcal H}$ is a compact operator if $A = \sum_{j=1}^\infty a_jh_{1j} \otimes h_{2j} $ for some orthonormal bases $\{h_{1j}\}_{j \geq 1}$ and $\{h_{2j}\}_{j \geq 1}$ and some sequence of real numbers $\{a_j\}_{j \geq 1}$ tending to zero, where $\otimes$ denotes the tensor product on $\mathcal H$. If $A$ is nonnegative, self-adjoint and compact, we may assume that $h_{1j}=h_{2j}$ and $a_j\geq 0$, and in this case $\{a_j\}_{j \geq 1}$ \commRVS{becomes} the eigenvalues of $A$. %; for such an operator, we let $\lambda_{\max}(A)$ denote the maximal eigenvalue.
	
	A $\mathcal H$-valued random element, say $U$, is a measurable map from the underlying probability space to $\mathcal H$. We say that $U$ is square-integrable if $E[\|U\|^2]<\infty$, and for such a random element, we may define its expectation $E[U]$ (satisfying $E[\langle U,h \rangle]=\langle E[U],h\rangle$ for any $h\in \mathcal H$) and covariance operator $C_U \coloneqq E[(U-E[U])\otimes (U-E[U])]$.
	
	\subsection{Fractionally-integrated functional time series}\label{sec:2.2}
	We consider a $\mathcal H$-valued, possibly nonstationary, time series \revlab{r1_minor2a}\commRV{$Y_t$ (for $t \in \mathbb{Z}$) whose $d$-th (fractional) difference after initialization, $\Delta^{d} (Y_t-Y_0)$ for $t \geq 1$, becomes stationary I($0$) time series}, \revlab{r1_minor1}\commRV{where $\Delta^{d}$ is understood as a power series of the lag operator $L$ defined by  $
			\Delta^{d} = \sum_{j=0}^{\infty} \frac{\Gamma(j-d)}{\Gamma(-d)\Gamma(j+1)} L^j$ and $\Gamma(\cdot)$ is the so-called Gamma function \citep[see, e.g., Section 5.2 of ][]{hassler2018time}. } % and $\Delta^0$ is understood as the identity, and hence $\Delta^{0} (Y_t-Y_0)=Y_t-Y_0$;} 
Hereafter, such a time series is conveniently called a Fractionally-Integrated (FI) Hilbert-valued Time Series (HTS). The formal definition of the I(0) property in this paper is given as follows:
	\begin{definitions}\label{def1}
		%We say that a sequence $\mathbf{Y}=\{Y_t\}_{t\geq t_0}$ is I($0$) (and write $\mathbf{Y} \sim$ I($0$)) if it can be made a stationary sequence with nonzero long-run covariance under a suitable initial condition. $\mathbf{Z}$ is said to be I($d$) if $\Delta^{d} \mathbf{Y} = \{\Delta^{d} Y_t\}_{t\geq t_0}$ is I(0). 
		The sequence $Z_t$ is I(0) if the following are satisfied: 
		\begin{inparaenum}
			\item[(i)] for each $t \geq 1$, $Z_t$ allows the representation $Z_t = \sum_{j=0}^\infty A_j \eta_{t-j}$ for a mean-zero independently and identically distributed (iid) sequence $\eta_t$ with covariance $C_{\eta}$ and a sequence of bounded linear operator $\{A_j\}_{j\geq 1} \subset \mathcal L_{\mathcal H}$ with $\sum_{j=0}^\infty \|A_j\|_{\mathcal L_{\mathcal H}} < \infty$; \item[(ii)] its long-run covariance $(\sum_{j=0}^\infty A_j) C_\eta (\sum_{j=0}^\infty A_j)^\ast$ is a nonzero operator. 
		\end{inparaenum}
	\end{definitions}
	The I($0$) property is adopted from recent articles (e.g., \citealp{BSS17, BS20, seo_2022}) concerning integer-integrated functional time series. As will be discussed in detail later, however, the meaning of $\Delta^{d} (Y_t - Y_0)$ \revlab{r1_minor2b}\commRV{for $t\geq 1$} being I($0$) is more general and differs in the present paper. This difference arises not only because $d$ is allowed to be a fraction, but also, more importantly, because we require only that the highest memory of the time series $\Delta^{d} (Y_t - Y_0)$ be zero, while allowing it to contain anti-persistent components, provided that stationarity is preserved. Specifically, there may be a projection map $P$ such that $\Delta^{\widetilde{d}} P({Y}_t - Y_0)$ is I($0$) for some $\widetilde{d} \in (-1/2, 0]$ in our setup. This treatment is not only more realistic but also necessary for our subsequent analysis in developing a novel test for integer-order integration of functional time series.
	
	We are interested in examining integer-order integration hypotheses, i.e.,  $Y_t\sim I(d)$, where 
	\begin{equation*}
		H_0: \text{$d$ is an integer} \quad \text{ against } \quad H_1: \text{$d$ is a fraction}.
	\end{equation*}
Testing the above hypotheses is important for practitioners because it helps determine an important characteristic of functional time series and identify a suitable statistical model.
	
	The cases with $d=0$ and \commRVS{$d=1$} may be the most empirically relevant, and thus, we \commRVS{focus} on these cases subsequently.  Moreover, \commRVS{for} technical reasons, we restrict the range of admissible values of $d$ \commRVS{to} $d>-1/2$. To sum up, we will mainly examine if~$d$ is an integer contained in the interval $(-1/2,2)$; however, our test to be developed can be extended to allow the case with $d\geq 2$ in an obvious way, which will be discussed in Section~\ref{sec:3} (see Remark~\ref{rem2}). 
	%\begin{remark} \normalfont
	%Note that The reason why we allow an I($0$) time series to potentially contain anti-persistent component  originates from the latter test examining I($1$)-ness against I($d$) with $d \in (0.5, 1.5)\setminus\{1\}$. Note that if $Y_t\sim I(d)$ for some $1/2<d < 1$, the first different time series will be I(d) 
	%\end{remark}%
	
	\section{Testing integer integration}\label{sec:3}
	\subsection{Assumptions}\label{sec:3a}

	We will consider the time series of functional observations $\mathbf{Y}_0= \{{Y}_{t}-Y_0\}_{t=1}^T$. For some nonnegative integer $\delta \geq 0$,  we hereafter let $\mathbf Y_{\delta}=\{	Y_{\delta,t}\}_{t= \delta + 1}^T$ be the $\delta$-th difference of $\mathbf{Y}_0$, i.e.,  
	\begin{equation} \label{eqinitial}
		Y_{\delta,t} = \Delta^{\delta} ( {Y}_t-  {Y}_0), \quad t \geq \delta+1, 
	\end{equation}
	where $\Delta^{0}$ is understood as the identity operator.
	%Note that we particularly consider the initialized variables ${Y}_t- {Y}_0$ in the case $\delta=0$ while such an initialization is redundant when $\delta \geq 1$ (since $\Delta^{\delta} (Y_t- Y_0) = \Delta^{\delta} {Y}_t$). %\bt{WK:Our test statistics will be constructed from $\mathbf{Y}_{\delta,T}$, which makes our tests to be developed not dependent on potential existence of unknown mean or intercept.} 
	For our asymptotic analysis, we will assume that the time series $\mathbf{Y}_0$ satisfies some regularity conditions, introduced below:
	
	\begin{assumpM}\label{assum1} 
		\commRVS{The sequence $\mathbf{Y}_0 = \{Y_{0,t}\}_{t=1}^T$} is a FIHTS satisfying the following properties:
		\begin{enumerate}[(i)]
			\item \label{assum1a} For some $K>0$, there exist a $\mathcal H$-valued sequence $\{X_{t}\}_{t\geq 1}$ and orthogonal projections $P_1,\ldots,P_K$ such that $\sum_{j=1}^K P_j = I_{\mathcal H}$ (the identity map on $\mathcal H$) and $Y_{0,t}=  \sum_{j=1}^K P_j X_{t}$ for $t\geq 1$. Moreover, for each $j=1\,\ldots,K$ and $$d_1>d_2>\ldots> d_K > -1/2,$$ there exists a stationary I($0$) sequence $\{v_{j,t}\}_{t\in \mathbb{Z}}$ satisfying the following:  
			\begin{align}
				&v_{1,t} = \Delta^{d_1} P_1X_{t} =  \sum_{k=0}^\infty \psi_{1,k} \varepsilon_{t-k}, \label{eqfi01}\\
				%&v_{2,t} \coloneqq \Delta^{d_2} P_2X_{t} =  \sum_{j=0}^\infty \psi_{2,j} \varepsilon_{t-j}, \label{eqfi02}\\
				&\quad  \vdots \notag \\
				&v_{K,t} = \Delta^{d_K} P_KX_{t} =  \sum_{k=0}^\infty \psi_{K,k} \varepsilon_{t-k}, \label{eqfi03}
			\end{align}
			where $\{\psi_{j,k}\}_{j\geq 1}$  is a sequence in $\mathcal L_{\mathcal H}$ with $\sum_{k=0}^\infty k \|\psi_{j,k}\|_{\mathcal L_{\mathcal H}} < \infty$ for all $j$, and $\varepsilon_t$  \revlab{r1_minor2c}\commRV{(for $t\in \mathbb{Z}$)} is an iid copies of $\varepsilon$ with mean zero and positive definite covariance $C_{\varepsilon}$. 
			\item \label{assum1b} $E[\|\varepsilon_t\|^p]<\infty$ for some $p > \max\{4,2/(2d_K+1)\}$.
			\item  \label{assum1c}  $\mathcal H_1 \coloneqq \ran P_1$ is finite dimensional, i.e., $\dim(\mathcal H_1)=\rank P_1 < \infty$.
			\item \label{assum1d}  $d_1 \in (-1/2, 2)$.
		\end{enumerate}
		%\item \label{assum1b} $\gamma_{j,k}(s) \sim C s^{2d-1}\sigma_{j,k}^{1/2}$ if $\phi_j, \phi_k \in \ran P_D$ and $\gamma_{j,k}(s) \leq C s^{2d-1}\sigma_{j,k}^{1/2}$ . Moreover, $\sum_{r,s,u=-\infty}^\infty |\kappa_{j,k,\ell,m} (r,s,u)|\leq C \sigma_{j,k,\ell,m}^{1/2}$ and the spectral density $f$ of $x_{t,j}$ satisfies $f(\lambda)=C|\lambda|^{-2d}$ if $d<0$, and $\sup_{r} \sum_{s,u=-N}^N |\kappa_{j,k,\ell,m} (r,s,u)| \leq C N^{2d}\sigma_{j,k,\ell,m}^{1/2}$ if $d\geq 0$. 
		%\item \label{assum1b} Let $\mathcal H_D$ be the collection of vectors $h$ such that $\langle \Delta^d X_{d,t},h \rangle \sim$ I($d$). For any arbitrary $h \in \ran \mathcal H_D$ and $\ell \geq  0$,
		%\begin{align}
		%\frac{1}{T} \sum_{t=\ell+1}^T \langle X_{d,t},h \rangle \langle X_{d,t-\ell},h\rangle  \to_p \langle C_{\ell}h,h\rangle 
		%\frac{1}{T} \sum_{t=\ell+1}^T  X_{d,t} \otimes X_{d,t-\ell}  \to_p  C_{\ell}.
		%\end{align} 
	\end{assumpM}
	%Some comments on Condition \ref{assum1} are in order. In order to accommodate a more general and realistic scenario, we let $Z_t$, which will be set to $Y_{\delta,t}$ for some $\delta \geq 0$, have a possibly nonzero intercept $\mu$. We also note that  $\mathbf{Z}=\{Z_t\}_{t\geq 1}$ given in Condition \ref{assum1} is a FIHTS and $\mathcal H_1 \coloneqq \ran P_1$ is its dominant subspace (\citealp{LRS20,LRS,seoshang22}). %Under (i), it can be shown that $\mathcal H_D$ is given by the closure of $\ran \psi(1)$ (Proposition 3.2 of \citealp{BSS17}), and thus (ii) guarantees that the existence of orthogonal projection $P_D$ onto $\mathcal H_D$. 
	Some comments on Assumption~\ref{assum1} are in order. We first note that $\mathbf{Y}_0$ is a FIHTS with the dominant subspace $\mathcal H_1 = \ran P_1$ (\commRVS{see, e.g.,} \citealp{LRS20,LRS,seoshang22}).
	The case where $\mathcal H_1$ is a finite dimensional subspace has been considered in the literature on FIHTS, and thus Assumption~\ref{assum1}\ref{assum1c} does not \commRVS{seem restrictive}. In our theoretical development, finite dimensionality of $\mathcal H_1$ is only used to easily obtain a stochastic bound of a certain operator (\commRVS{see, e.g.,} \eqref{eqpfadd01} of the Supplementary Material). It should also be noted that we do not require finite dimensionality of the other subspaces (i.e., $\ran P_2,\ldots,\ran P_K$), and thus the time series described by Assumptions~\ref{assum1}\ref{assum1a}-\ref{assum1c}  is still quite general than those considered in the recent literature. If $d_1=0$, the time series $X_t$ defined in Assumption~\ref{assum1} is I($0$) in the sense of Definition~\ref{def1}. That is, the time series basically allows potential memory reduction on $\mathcal H$ and thus our statistical inference needs to be based on elements in $\mathcal H_1$; if we use $(I_{\mathcal H}-P_1)Z_t$, then we may mistakenly conclude that its memory is negative even if $\mathbf{Y}_0$ is I($0$) and, due to this nature, it is important to base our statistical inference on the dominant subspace $\mathcal H_1$. %In (iii), we require the sample autocovariances of $\langle X_{d,t},h\rangle$ converge to their true counterparts. Even if this may seem a slightly higher-level condition, this is not only a weak requirement given that $\{\langle X_{d,t},h\rangle\}_{t\geq 1}$ is stationary, but also a convenient assumption to facilitate our theoretical analysis without specifying the detailed behavior of $X_{d,t}$ under stationary long-memory alternative.
	Assumption~\ref{assum1}\ref{assum1d} is introduced to simplify the subsequent discussion. Under this assumption, the order of integer integration is either $0$ or $1$. This assumption does not place any practical restriction given that examples of I($d$) processes with $d\geq 2$ appear to be scarce and thus they have been seldom discussed in the literature on function time series analysis. Moreover, even if we mainly develop our theoretical results under this assumption, it is straightforward to extend the subsequent results to the case with $d_1 \in (-1/2, m)$ for some other integer $m>2$ (see Remark~\ref{rem2}).
	
	It will be convenient to introduce some additional notation. For any time series $\mathbf{Z} = \{Z_{t}\}_{t=t_0}^T$ (which will be set to $\mathbf{Y}_0$, $\mathbf{Y}_1$ or their projections in the subsequent discussion), where the realizations are given for $t=t_0,\ldots,T$ (e.g., $t_0=1$ if $\mathbf{Z}=\mathbf{Y}_0$ while $t_0=2$ if $\mathbf{Z}=\mathbf{Y}_1$), we define 
	\begin{equation*}
		{\mathcal K}_T\asd{\mathbf{Z}} \coloneqq  \sum_{t=t_0}^T \left(\sum_{s=t_0}^t Z_s  \otimes  \sum_{s=t_0}^t  Z_s\right).
		%&\widehat{\Lambda}_v = \frac{1}{T} \sum_{t=1}^T \nu_t  \otimes  \nu_t +  \frac{1}{T} \sum_{t=1}^T \sum_{s=0}^{h} \left(1 - \frac{s}{q}\right) \left(\nu_t  \otimes  \nu_{t-s} + \nu_{t-s}  \otimes  \nu_{t}\right). 
	\end{equation*}
	We also let
	\begin{equation}\label{eqlrv}
		%{\Lambda}_T\asd{\mathbf{Z}} &\coloneqq  T^{-1}\sum_{t=t_0}^T (Z_t-\bar{Z}_T)  \otimes  (Z_t-\bar{Z}_T) \\ & +  T^{-1}\sum_{t=s+t_0}^T \sum_{s=1}^{q} w(s,q) \left((Z_t-\bar{Z}_T) \otimes (Z_{t-s}-\bar{Z}_T)+ (Z_{t-s}-\bar{Z}_T) \otimes (Z_t-\bar{Z}_T)\right),
		{\Lambda}_T\asd{\mathbf{Z}} \coloneqq  T^{-1} \sum_{s=-q}^q w(s,q) \sum_{t_0 \leq t, t-s \leq T} Z_t  \otimes  Z_{t-s},
	\end{equation}
	where $w(s,q)=1 - {|s|}/(q+1)$, which corresponds to the Bartlett kernel weight. ${\mathcal K}_T\asd{\mathbf{Z}}$ (resp.\ ${\Lambda}_T\asd{\mathbf{Z}}$) is the unnormalized sample covariance (\commRVS{resp.}\ long-run covariance) operator of $\sum_{s=t_0}^tZ_s$ (resp.\ $Z_{t}$).  % We first establish a preliminary result on ${\mathcal K}_T\asd{\mathbf{Z}}$:

	\subsection{Testing I($0$)-ness against FI alternatives} \label{sec_frac01}
	We first consider testing I($0$)-ness (i.e., $d_1=0$) against stationary FI alternatives (\commRVS{i.e., $d_1 \in (-1/2,1/2)\setminus\{0\}$}), and \commRVS{extend this} to the case accommodating nonstationary FI alternatives (\commRVS{i.e., $d_1\geq 1/2$}). The test is constructed from the time series $\mathbf{Y}_0$.

	% $\langle {\mathcal K}_T\asd{\mathbf{Y}_{0}} \bar{h}, \bar{h}\rangle$  is nonnegative self-adjoint, we find that the operator norm of ${\mathcal K}_T\asd{\mathbf{Y}_{0}}$ satisfies the following:
	%\begin{align}
	%\|{\mathcal K}_T\asd{\mathbf{Y}_{0}}\|_{\mathcal L_{\mathcal H}} = \sup_{\|h\|\leq 1}  \langle  {\mathcal K}_T\asd{\mathbf{Y}_{0}} h, h \rangle = \lambda_{\max}({\mathcal K}_T\asd{\mathbf{Y}_{0}}).
	%\end{align}
	%We let $\hdd\asd{\mathbf{Y}_{0}}$ be the dominant eigenvector (the eigenvector associated with the maximal eigenvalue $\lambda_{\max}({\mathcal K}_T\asd{\mathbf{Y}_{0}}$)) of ${\mathcal K}_T\asd{\mathbf{Y}_{0}}$, i.e., 
	%\begin{align}
	% \lambda_{\max}({\mathcal K}_T\asd{\mathbf{Y}_{0}}) = \langle{\mathcal K}_T\asd{\mathbf{Y}_{0}}\hdd\asd{\mathbf{Y}_{0}}, \hdd\asd{\mathbf{Y}_{0}} \rangle.
	%\end{align}
	%We let $\hdd(\mathbf{Y})$ be called the dominant eigenvector of $\widehat{\mathcal K}(\mathbf{Y})$ hereafter.  %as follows:
	%\begin{align}
	%\widehat{\mathcal K}_{0} \widehat{h}_j = \widehat{\lambda}_j \widehat{h}_j, \quad  %\widehat{\lambda}_1\geq\widehat{\lambda}_2\geq\ldots.
	%\end{align}
	Under Assumption~\ref{assum1}, we know that $v_{1,t} = \Delta^{d_1} P_1X_{t}$ is a stationary I($0$) sequence. We let  $\Lambda_{v_1}$ denote the population long-run covariance operator of $v_{1,t}$, i.e., 
	\begin{equation*}
		\Lambda_{v_1} = \sum_{s=-\infty}^\infty E\left[v_{1,t} \otimes v_{1,t-s} \right].
	\end{equation*}
	%\item \bt{If $d_2$ is the memory of $(I-P_D)X_t$, $T^{-2-2d}\|\widehat{\mathcal K}_0 -P_D\widehat{\mathcal K}_0P_D\|= O_p(T^{d_2-d})$ and $\widehat{h}= h + O_p(T^{d_2-d})$ for some $h \in \ran P_D$.}
	%\item \label{lem1c} Let $\gamma_h(j) = E[\langle X_{d,t},h \rangle \langle X_{d,t-s},h \rangle]$. Under Assumption \ref{assum1}, 
	Let $\hd$ be any unit-norm element of the dominant subspace $\mathcal H_1$. We first establish the following preliminary result, which helps in understanding the practical challenges involved in developing a test for examining integer integration: below and hereafter, we let $W$ denote the standard Brownian motion on $[0,1]$ and let $\int$ represent $\int_{0}^1$ for notational simplicity.
	\begin{propositions}[An infeasible test] \label{prop1} Suppose that $-1/2<d_1<1/2$ and Assumption~\ref{assum1} is satisfied. Then, 
		\begin{equation*}%\label{prop01}
			T^{-2} \langle { \mathcal K}_T\asd{\mathbf{Y}_{0}}\hd,\hd \rangle  \overset{d}{\to} \cd \int W(r)^2dr \quad   \text{if $d_1=0$},
		\end{equation*}
		where $ \cd  = \langle \Lambda_{v_1} \hd,\hd \rangle$. On the other hand,
		\begin{align*}
			T^{-2} \langle { \mathcal K}_T\asd{\mathbf{Y}_{0}}\hd,\hd \rangle   \overset{p}{\to}  0 \quad &\text{if $d_1 \in (-1/2,0)$},\\ %\label{prop02}\\
			T^{-2} \langle { \mathcal K}_T\asd{\mathbf{Y}_{0}}\hd,\hd \rangle  \overset{p}{\to} \infty \quad  &\text{if $d_1\in (0,1/2)$}.%\label{prop03}
		\end{align*}
	\end{propositions}
	From Proposition~\ref{prop1}, we know that the statistic $\langle { \mathcal K}_T\asd{\mathbf{Y}_{0}}\hd,\hd \rangle$ has discriminating power for the null hypothesis of I(0)-ness against stationary FI alternatives. %(i.e., the case with $d_1\in(-1/2,1/2)\setminus\{0\}$). 
%\commRV{More specifically, if we let $\eta_{\alpha}$ be the $100\times \alpha$\% quantile of $\int W(r)^2 dr$ for $\alpha \in (0,1/2)$,  it implies a two-sided $2\times 100\times \alpha\%$ test which reject the null hypothesis $H_0:d_1=0$ in favor of the alternative $H_0:d_1\neq 0$ when
%\begin{equation}T^{-2} \langle {\mathcal K}_T\asd{\mathbf{Y}_{0}}\hd,\hd \rangle < \eta_{\alpha} \quad \text{or} \quad   T^{-2} \langle {\mathcal K}_T\asd{\mathbf{Y}_{0}}\hd,\hd \rangle > \eta_{1-\alpha}, \end{equation}
%where ,}
\revlab{r2_major1_1}
\commRV{More specifically, if we let $\eta_{0.025}$ (resp.\ $\eta_{0.975}$) be the $2.5\%$ (resp.\ $97.5\%$) quantile of $\int W(r)^2 dr$, this implies a two-sided $5\%$ test that rejects the null $H_0:d_1=0$ in favor of the alternative $H_1:d_1\neq 0$ if
\begin{equation}\label{eqrejection}
T^{-2} \langle {\mathcal K}_T\asd{\mathbf{Y}_{0}}\hd,\hd \rangle < \eta_{0.025} \quad \text{or} \quad   T^{-2} \langle {\mathcal K}_T\asd{\mathbf{Y}_{0}}\hd,\hd \rangle > \eta_{0.975}. \end{equation}
As in the standard two-sided test, rejection in the lower (resp.\ upper) tail implies that $H_1:d_1<0$ (resp.\ $H_1:d_1>0$) is favored over the null.}

	However, in practice, Proposition~\ref{prop1} cannot directly be used for statistical inference since $\cd$ and $\hd$, which is an element of  $\mathcal H_1$, are unknown. Given that $\hd$ is an element of the dominant subspace, one might attempt to replace $\hd$ with a reasonable estimator $\hdd$, which can be obtained by estimating  \commRVS{the dominant subspace $\mathcal H_1$} using \commRVS{existing methods} (see, e.g., \citealp{LRS20}), and then \commRVS{employ} a standard long-run covariance estimator, if available, to replace $\cd  = \langle \Lambda_{v_1} \hd,\hd \rangle$ with its feasible counterpart. In fact, our estimator is obtained in a broadly similar manner. However, it should be noted that the asymptotic properties of the estimator of $\hd$ will inevitably and crucially depend on the value of $d_1$ and possibly on the other memory parameters, $d_2, \ldots, d_K$ (it is difficult to believe that a reasonable estimator, independent of the memory properties, exists). As a consequence, it is also not straightforward at all to replace $\cd$ with its feasible counterpart while preserving discriminating power against FI alternatives for any nonzero values of $d_1$. %\commWS{(WK: any better discussion?)} 
	
	In order to have a feasible and asymptotically valid test, we require additional conditions for the time series introduced in Assumption~\ref{assum1}. We will hereafter assume the following conditions for the stationary component of $\mathbf{Y}_0$: %that the time series $\mathbf{Y}_0$ and $q$ satisfy the properties, collected in the conditions below: 
	%, we call it a variance-ratio statistic. In order to have a feasible version of the test, we require the time series of interest to satisfy some additional conditions, which are collected below:%that the time series $\mathbf{Y}_0$ and $q$ satisfy the properties, collected in the conditions below: 
	\begin{assumpS}\label{assum2} Let $d_S := \max\{d_j: d_j < 1/2\}$ and let $P_S=\sum_{d_j \leq d_S} P_j$ in Assumption~\ref{assum1}. The stationary sequence of $P_SX_t$ satisfies the following: 
		\begin{enumerate*}[(i)]
			\item\label{assum2a} $T^{-1} \sum_{t=1}^T  P_SX_{t} = O_p(T^{-1/2+d_S})$; 
			\item \label{assum2b}
			for any $\ell \geq  0$, $T^{-1} \sum_{t=\ell+t_0}^T  P_SX_{t}\otimes P_SX_{t-\ell}  \overset{p}{\to}  C_{S,\ell} \coloneqq E[P_SX_{t}\otimes P_SX_{t-\ell}]$ and % \item \label{assum2c}  \commWS{$\|C_{\ell}\| \leq O(\ell^{-\gamma})$ for some $\gamma > 0$} and 
			\item \label{assum2d} $C_{S,0}$ is positive definite on $\mathcal H_1$.
		\end{enumerate*}
		%\bt{WK: $q^{-2d_j\asd{\mathbf{Z}}}\langle {\Lambda}_T\asd{\mathbf{Z}}h,h \rangle \to_p c_{j,h}$ implies the dominant eigenvector of $\Lambda_{T}$ is included in $\mathcal H_1$ wpa.\ 1? }
		%, where $\widehat{\Lambda}(\mathbf{Z}_T) \coloneqq  T^{-1}\sum_{t=1}^T Z_t  \otimes  Z_t +  T^{-1}\sum_{t=1}^T \sum_{s=1}^{q} w(s,q) \left(Z_{t} \otimes Z_{t-s}+ Z_{t-s} \otimes Z_{t}\right)$.
		%\item \label{assum1b} $\gamma_{j,k}(s) \sim C s^{2d-1}\sigma_{j,k}^{1/2}$ if $\phi_j, \phi_k \in \ran P_D$ and $\gamma_{j,k}(s) \leq C s^{2d-1}\sigma_{j,k}^{1/2}$ . Moreover, $\sum_{r,s,u=-\infty}^\infty |\kappa_{j,k,\ell,m} (r,s,u)|\leq C \sigma_{j,k,\ell,m}^{1/2}$ and the spectral density $f$ of $x_{t,j}$ satisfies $f(\lambda)=C|\lambda|^{-2d}$ if $d<0$, and $\sup_{r} \sum_{s,u=-N}^N |\kappa_{j,k,\ell,m} (r,s,u)| \leq C N^{2d}\sigma_{j,k,\ell,m}^{1/2}$ if $d\geq 0$.
	\end{assumpS}
	Some comments on the above  are in order. 
	Note that the time series $\{P_SX_{t}\}_{t\geq 1}$ is a stationary process, and in this case, assuming condition~\ref{assum2b}  does not \commRVS{appear restrictive}. For example, \cite{Salish2019} establish~\ref{assum2a} and~\ref{assum2b}  for FIHTS (with long-range dependence) under nonrestrictive conditions. Assumption~\ref{assum2} along with Assumption~\ref{assum1} will be maintained throughout this section.  	
	%If we have a reasonable estimator $\cdd$ satisfying $|\cdd-\cd|\to_p 0$ and consider a ratio statistic $T^{-2} {\lambda}_{\max}({\mathcal K}_T\asd{\mathbf{Y}_0})/\cdd$, then the limiting distribution when $d_1\asd{\mathbf{Y}_0}=0$ becomes a functional of the standard Brownian motion, which does not depend on $\cd$, and hence the critical values for the ratio statistic can be tabulated from standard methods. Of course, for the test based on this ratio statistic to be consistent, it needs to preserve the discriminating powers against FI alternatives with both $d>0$ and $d<0$ as in \eqref{prop02} and \eqref{prop03}. 
	
	We next show that a feasible version of the test can be obtained from ${\mathcal K}_T\asd{\mathbf{Y}_0}$ and its dominant eigenvector  $\hdd$, corresponding to the maximal eigenvalue. Specifically, consider the test statistic  % and let 
	%\begin{align}
	%\langle {\Lambda}_T\asd{\mathbf{Y}_{0}}\hdd\asd{\mathbf{Y}_0},\hdd\asd{\mathbf{Y}_0} \rangle 
	%\end{align}
	%we consider the test statistic, defined as follows: 
	\begin{equation*}
		{V}_0 =T^{-2} \frac{ \langle { \mathcal K}_T\asd{\mathbf{Y}_{0}}\hdd,\hdd\rangle}{\langle {\Lambda}_T\asd{\mathbf{Y}_{0}}\hdd,\hdd\rangle}. %
		%= T^{-2}\frac{\langle { \mathcal K}_T\asd{\mathbf{Y}_{0}}\hdd\asd{\mathbf{Y}_{0}},\hdd\asd{\mathbf{Y}_{0}} \rangle}{\langle {\Lambda}_T\asd{\mathbf{Y}_{0}}\hdd\asd{\mathbf{Y}_{0}},\hdd\asd{\mathbf{Y}_{0}}\rangle},  
	\end{equation*}
	The above statistic is used to examine the null of I($0$)-ness against FI alternatives. $	{V}_0 $ may be viewed as a ratio of two different variances (associated with ${\mathcal K}_T\asd{\mathbf{Y}_{0}}$ and ${\Lambda}_T\asd{\mathbf{Y}_{0}}$), and thus it may be called a variance-ratio statistic. The dominant eigenvector $\hdd$ can easily be computed from the standard functional principal component analysis (FPCA), and thus the test statistic is \commRVS{easy to compute}. 
	
	We will show that the test based on $V_0$ is consistent against stationary FI alternatives under the following additional assumptions: below,  we let $\mathbf{X}_1 = \{P_1 X_t\}_{t=1}^T$, where $X_t$ and $P_1$ are detailed in Assumption~\ref{assum1} (note that the subscript 1 is associated with $P_1$ and is used differently from that of $\mathbf{Y}_1$).
	\begin{assumptionV}\label{assumv1} $d_1 = d_S = \max\{d_j: d_j < 1/2\}$, and furthermore, the following hold:
		\begin{enumerate*}[(i)]
			\item\label{assumv1a}  $q^{-2d_1}\langle {\Lambda}_T\asd{\mathbf{X}_1}h,h \rangle \overset{p}{\to}   c_{j,h} > 0$ for any $h \in \mathcal H_1$ with $\|h\|=1$; 
			\item\label{assumv1b}  $q\to \infty$ and $q = o(T^{\upsilon})$ for $\upsilon \in (0, 1/2)$; if $d_1<0$, it is further assumed that $\upsilon < |(d_1-d_2)/2d_1|$.
		\end{enumerate*}

		%\bt{WK: $q^{-2d_j\asd{\mathbf{Z}}}\langle {\Lambda}_T\asd{\mathbf{Z}}h,h \rangle \to_p c_{j,h}$ implies the dominant eigenvector of $\Lambda_{T}$ is included in $\mathcal H_1$ wpa.\ 1? }
		%, where $\widehat{\Lambda}(\mathbf{Z}_T) \coloneqq  T^{-1}\sum_{t=1}^T Z_t  \otimes  Z_t +  T^{-1}\sum_{t=1}^T \sum_{s=1}^{q} w(s,q) \left(Z_{t} \otimes Z_{t-s}+ Z_{t-s} \otimes Z_{t}\right)$.
		%\item \label{assum1b} $\gamma_{j,k}(s) \sim C s^{2d-1}\sigma_{j,k}^{1/2}$ if $\phi_j, \phi_k \in \ran P_D$ and $\gamma_{j,k}(s) \leq C s^{2d-1}\sigma_{j,k}^{1/2}$ . Moreover, $\sum_{r,s,u=-\infty}^\infty |\kappa_{j,k,\ell,m} (r,s,u)|\leq C \sigma_{j,k,\ell,m}^{1/2}$ and the spectral density $f$ of $x_{t,j}$ satisfies $f(\lambda)=C|\lambda|^{-2d}$ if $d<0$, and $\sup_{r} \sum_{s,u=-N}^N |\kappa_{j,k,\ell,m} (r,s,u)| \leq C N^{2d}\sigma_{j,k,\ell,m}^{1/2}$ if $d\geq 0$.
	\end{assumptionV}
	Assumption~\ref{assumv1}\ref{assumv1a} is a high-level assumption employed to conveniently obtain a feasible version of the test. Under Assumptions~\ref{assum1} and~\ref{assum2}, primitive sufficient  conditions for this can be found in\commRVS{, e.g.,} \cite{ABADIR200956}, as detailed in Remark~\ref{remprimitive}. Requirements for the lag-truncation parameter $q$ are detailed by Assumption~\ref{assumv1}\ref{assumv1b}.	A choice of $q$ for the case when $d_1\geq 0$ is common in practical studies, \commRVS{necessitating} sample long-run covariance estimation, and thus does not place any practical restriction. The requirement for the case $d_1<0$ is introduced for mathematical convenience, related to anti-persistent components. Combined with Assumption~\ref{assumv1}\ref{assumv1a}, this assumption facilitates extending our asymptotic results developed for persistent FIHTS (with $d_1 >0$) to the case with anti-persistent FIHTS (with $d_1<0$); see our proof of Proposition~\ref{prop2}. This specifically requires that $q$ \commRVS{increase} at a sufficiently slow rate depending on the values of $d_1$ and $d_2$. A theoretical choice satisfying this condition regardless of the values of $d_1$ and $d_2$ is $q=O(\log T)$. We conjecture that this assumption is not essential for our theoretical results and can be relaxed by developing appropriate asymptotic results for anti-persistent functional time series. %However, since (i) investigating this issue requires a comprehensive study of its own, and (ii) it would be practically not common to find examples of FIHTS whose the dominant component is anti-persistent (which naturally implies that all sub-dominant components are also anti-persistent) even if we consider this possibility as a potential alternative hypothesis for completeness of the test, we do not pursue this further in the present paper.
	However, investigating this issue requires a comprehensive study of its own, and thus we do not  \commRVS{pursue it further} in the present paper. %Condition \ref{assumbw2} is not used in this section but used later in Section \ref{sec_frac02} concerning nonstationary FIHTS. Its role is overall similar to \ref{assumbw1}
	
	\begin{remarks} \label{remprimitive} \normalfont If $d_1<1/2$ (and thus $d_1=d_S$), for any $h \in \mathcal{H}_1$ with $\|h\| = 1$, $\langle {\Lambda}_T \asd {\mathbf{X}_1} h, h \rangle$ is the sample long-run covariance estimator of a fourth-order stationary I($d_1$) sequence  (Assumption~\ref{assum1}). From Theorem 2.2 of \cite{ABADIR200956}, we know that, under certain \commRVS{nonrestrictive} conditions on the spectral density and autocovariances of such a time series (detailed in equation (1.1) and Assumption M in their paper), $\langle q^{-2d_1} {\Lambda}_T \asd {\mathbf{X}_1} h, h \rangle$ converges to a positive constant depending on $d_1$ and $h$, under a requirement on $q$ implied by Assumption ~\ref{assumv1}\ref{assumv1b} (see Section~\ref{sec_app_abadir} of the Appendix for more details). In this case, Assumption~\ref{assumv1}\ref{assumv1a} is clearly satisfied. 
		%In this case,    and thus  provided by of course, replacement of the high-level condition given by Assumption \ref{assum2}\ref{assum2b} with such lower-level conditions does not affect the results to be developed. 
	\end{remarks}
	
	We next develop a feasible test of the null of I(0)-ness against stationary FI alternatives.
	\begin{propositions}[Feasible test of I(0)-ness] \label{prop2} 
		Suppose that $-1/2<d_1<1/2$ and Assumptions~\ref{assum1}, \ref{assum2} and~\ref{assumv1} are satisfied. Then,   
		\begin{align*} %\label{eqconver1}
			{V}_0  \quad \overset{p \text{ or } d}{\longrightarrow} \quad   
			\begin{cases}
				0 \quad &\text{if $d_1 \in (-1/2,0)$},\\
				\int W(r)^2 dr \quad &\text{if $d_1=0$},\\
				\infty \quad &\text{if $d_1 \in (0,1/2)$}.
			\end{cases} 
		\end{align*}
		%\begin{equation}\label{prop01a}
		%\widehat{V}_0  \to_d    \int W(r)dr^2.
		%\end{equation} On the other hand,under $H_0: d \neq 0$, 
		%\begin{eqnarray}
		%\widehat{V}_0 & \to_p 0 \quad &\text{if $d<0$},\label{prop02a}\\
		%\widehat{V}_0 & \to_p \infty \quad &\text{if $d>0$}.\label{prop03a}
		%\end{eqnarray}
	\end{propositions} 
	Proposition~\ref{prop2} shows that the $V_0$-test has power against stationary FI alternatives with $d_1 \in (-1/2,1/2)\setminus\{0\}$. Given this result, it \commRVS{is reasonable} to conjecture that this test also has power against nonstationary FI alternatives concerning more persistent FIHTS, and this is investigated in the following:
	
	\begin{propositions} \label{prop2a} 
		Suppose that $d_1 \geq 1/2$ and Assumptions~\ref{assum1} and~\ref{assum2} hold. If $d_1 = 1/2$, we further assume that, for any $h\in \mathcal H_1$ and $t=1,\ldots, T$,
		\begin{equation} \label{eqwnonstat}
			\langle P_1X_{t},h\rangle  = O_p(\sqrt{\log T}),
		\end{equation} 
		where $P_1X_{t}$ is defined in %~\eqref{eqfi01} of 
Assumption~\ref{assum1}. Then, $V_0 \overset{p}{\to}  \infty$. 
		
		%\begin{align}
		%{V}_0  \quad \overset{p \text{ or } d}{\longrightarrow} \quad   \begin{cases}
			%  0 \quad &\text{if $d_1 \in (-1/2,0)$},\\
			%   \int W(r)^2 dr \quad &\text{if $d_1=0$},\\
			%  \infty \quad &\text{if $d_1 \in (0,\infty)$}.
			%\end{cases} 
			%\end{align} 

			%\begin{equation}\label{prop01a}
			%\widehat{V}_0  \to_d    \int W(r)dr^2.
			%\end{equation} On the other hand,under $H_0: d \neq 0$, 
			%\begin{eqnarray}
			%\widehat{V}_0 & \to_p 0 \quad &\text{if $d<0$},\label{prop02a}\\
			%\widehat{V}_0 & \to_p \infty \quad &\text{if $d>0$}.\label{prop03a}
			%\end{eqnarray}
		\end{propositions}
		In fact, based on some key asymptotic results used in the proof of Proposition~\ref{prop2}, it is not difficult to show that the test statistic $V_0$ diverges to infinity under nonstationary FI alternatives with $d_1>1/2$. To conveniently address the particular case of weak nonstationarity ($d_1=1/2$), Proposition~\ref{prop2a} imposes condition~\eqref{eqwnonstat}. {This condition does not seem restrictive and is expected to hold generally given that $\langle P_1X_t,h\rangle$ can be understood as the partial sum process $\sum_{s=1}^t x_s$, where $\{x_s\}_{s\geq 1}$ is an I$(-1/2)$ time series, and $\sum_{s=1}^T x_s = O_p(\sqrt{\log T})$ holds under standard regularity conditions for such a time series; see, e.g., \citet[Lemma~2.1]{liu1998asymptotics} and \citet[Ch.~12]{Tanaka2017}.} %; some primitive sufficient conditions for an I($1/2$) process to satisfy condition \eqref{eqwnonstat} can be found in the literature (see, e.g., Lemma 6 of \citealp{nielsen2010nonparametric} and the proof of Theorem 6 of \citealp{CHO2015217}).

		Propositions~\ref{prop2} and~\ref{prop2a} give us a consistent test for examining if $d_1 \in (-1/2,0)$, $\{0\}$, or $(0,\infty)$. Rejection in the upper (resp.\ lower) tail of the $V_0$-test means that the time series is fractionally-integrated with persistence (resp.\ anti-persistence). 
		
\begin{remarks}\label{remr2major1}\revlab{r2_major1_2}\normalfont\commRV{
As discussed for the infeasible version of the test in Section \ref{sec_frac01} (see~\eqref{eqrejection}), Propositions \ref{prop2} and \ref{prop2a} provide, for $\alpha \in (0,1/2)$, a two-sided $200 \times \alpha\%$ test that rejects the null $H_0:d_1=0$ in favor of the alternative $H_1:d_1\neq 0$ if $V_0 < \eta_{\alpha}$ or $V_0>  \eta_{1-\alpha}$, where $\eta_\alpha$ is the $100 \times \alpha\%$ quantile of $\int W(r)^2dr$. Rejection in the lower (resp.\ upper) tail provides evidence of anti-persistence (resp.\ long-range dependence or nonstationarity). }
\end{remarks}
\begin{remarks} \revlab{r2_major4}\normalfont
\commRV{Our test statistic uses (asymptotic properties of) the sample covariance ${\mathcal K}_T\asd{\mathbf{Y}_{0}}$ of the cumulative series and the sample long-run covariance  ${\Lambda}_T\asd{\mathbf{Y}_{0}}$, resembling to some degree the KPSS test of \cite{kwiatkowski1992testing}. The original KPSS test examines the hypothesis of stationarity with a deterministic trend (trend-stationarity) against the I(1) alternative, using a similar ratio statistic (though simplified given their univariate setting). \cite{kokoszka2016kpss} studied examining trend-stationarity of functional time series against the I(1) alternative, developing a direct extension of the KPSS test of \cite{kwiatkowski1992testing} for functional time series. Although our test bears some resemblance to these existing tests, the purpose of our test and the setting in which it is employed differ substantially from these existing tests; consequently, the theoretical development in this paper is distinct from existing work. This remark is motivated by an observation made by an anonymous referee, to whom we are indebted.} 
\end{remarks}

\begin{remarks}\label{remrevision1}\revlab{r2major3}\normalfont
\commRV{Consider the case where $d_1\neq 0$ near $0$. The results  in Proposition \ref{prop2} rely on the fact that $(T/q)^{-2d_1}V_0$ converges to a positive random variable (see Remark \ref{remappadd1} of the Supplementary Material). % \eqref{eqpf01revision} and \eqref{eqaddrevision}). (see its proof in Section \ref{app_section_proof} of the Supplementary Material) 
From this, we see how the choice of $q$ influences the power of the test: if $d_1 > 0$ (resp.\ $d_1<0$), a smaller $q$ makes the test statistic diverge to infinity (resp. decay to zero). This aligns with the well-known property that KPSS-type tests gain power against nonstationarity with smaller bandwidth parameters. However, it is also well known that  this power gain accompanies the cost of over-rejecting the null when the process is serially correlated. Simulation results for our test also confirm this trade-off. More detailed discussion and simulation results are presented in Section \ref{app_additional_simula} of the Supplementary Material. Recent work proposes self-normalization and power enhancement to mitigate this trade-off in simpler univariate settings where I(0)-ness is tested against I(1)-ness (\citealp{PengZhou2025}). Extending these ideas to our infinite dimensional framework with fractional integration is beyond the scope of this paper due to the lack of requisite theory, but represents a promising direction for future research. This remark is inspired by a comment from an anonymous referee, to whom we are grateful.
}
\end{remarks}

\begin{remarks} \label{remrevision2}\normalfont
\commRV{It can also be deduced from our proof of Proposition \ref{prop2} that both $(T/q)^{-2d_1}V_0$ and $(T/q)^{2d_1}V_0^{-1}$ converge to nondegenerate limits (see Remark \ref{remappadd1} and Section \ref{app_additional_simulb} of the Supplementary Material). These results suggest how the value of $d_1$ affects the power properties of the test. Moreover, we consider a sequence of local-to-zero hypotheses on $d_1$ such that $V_0$ neither diverges to infinity nor decays to zero as $T \to \infty$:   
\begin{equation}
H_{1,T}: d_1 =  \log_{(T/q)^2}\mathcal{C} =  \frac{\log {\mathcal{C}}}{2\log (T/q)},
\end{equation}
where $\mathcal C \neq 1$ represents the local deviation from the null ($\mathcal C=1$). Under this sequence of local-to-zero hypotheses, the power properties depend crucially on the local deviation $\mathcal C$, while remaining similar across different values of $T$ as long as $T$ is sufficiently large. This is evidenced by our simulation experiments; a more detailed discussion is given in Section \ref{app_additional_simulb} of the Supplementary Material.}
%Under this sequence of $d_1$, $V_0$ does not diverge or converge Under this sequence of $d_1$, $(T/q)^{-2d_1}V_0$he power of the test is expected to become similar across different $T$ ($T$ is large enough) and it depend on the parameter $C$, which determines a local deviation from the null. We examine the local power properties of our test by considering the power properties of our tests under a few different values $\mathcal C$. (This is based on sequential asymptotics $d_1 \to 0$ following $T,q\to \infty$.  (MENTION NEW SIMULATION RESULTS).      We conjecture that the joint asymptotics requires a completely different theoretical framework, and to our knowledge, no requisite theory is available. However this would be an interesting future topic.)}
\end{remarks}

		\subsection{Testing I(1)-ness against FI alternatives}\label{sec_frac02}
		
		Suppose that the $V_0$-test proposed in Section~\ref{sec_frac01} rejects the null hypothesis of I$(0)$-ness in the upper tail and thus $d_1 > 0$ is concluded. Practitioners may then be interested in testing the null of I$(1)$-ness against FI alternatives as the next step. Considering the properties of an I$(1)$ time series, it may be appropriate to apply the test considered in Section~\ref{sec_frac01} to the first-differenced observations $\mathbf{Y}_{1}=\{Y_{1,t}\}_{t= 2}^T$ (see \eqref{eqinitial}). % the We then may test the null of I(1)-ness by applying a similar test to the first differenced observations $\mathbf{Y}_{1}=\{Y_{1,t}\}_{t\geq 2}$.
		Specifically, define 
		\begin{equation} \label{eqv1}
			{V}_1 %T^{-2} \frac{{\lambda}_{\max}(\widehat{ \mathcal K}(\{Y_t\}_{t=1}^T))}{\langle \widehat{\Lambda}(\{Y_t\}_{t=1}^T))\hdd,\hdd \rangle} 
			= T^{-2}\frac{\langle { \mathcal K}_T\asd{\mathbf{Y}_1}\hdd,\hdd \rangle}{\langle {\Lambda}_T\asd{\mathbf{Y}_1}\hdd,\hdd\rangle},
		\end{equation}
		where $\hdd$ is the dominant eigenvector which we used in Section~\ref{sec_frac01}. We will demonstrate that the test based on the above statistic can be used to examine I$(1)$-ness against FI alternatives, not only as a test applied after the $V_0$-test in Section~\ref{sec_frac01}, but also as a standalone test for examining I($1$)-ness against I($d$)-ness for any other possible $d$ in the admissible range $(-1/2, 2)$ (see Remark~\ref{rem1} to appear).
		
		We investigate the asymptotic properties of $V_1$. To this end, for the time series defined in Assumption~\ref{assum1}, we let $\Delta  \mathbf{X}_1 = \{P_1 \Delta X_t\}_{t=2}^T$ and employ the following conditions:
		\begin{assumptionV}\label{assumv2}  $d_1 > 1/2$ and the following hold: 
			\begin{enumerate*}[(i)]
				\item \label{assumv2a}	$q^{-2(d_1 - 1)}\langle {\Lambda}_T\asd{\Delta \mathbf{X}_1}h,h \rangle \overset{p}{\to}  c_{j,h} > 0$ for any $h \in \mathcal H_1$ with $\|h\|=1$;
				\item \label{assumv2aa} if $d_1 = 3/2$, for any $h\in \mathcal H_1$, $\langle P_1\Delta X_{t},h\rangle  = O_p(\sqrt{\log T})$; 
				\item \label{assumv2b}$q\to \infty$ and $q = o(T^{\upsilon})$ for $\upsilon \in (0, 1/2)$; if $d_1 \in (1/2, 1)$, $\upsilon < |(d_1-d_2)/2(d_1-1)|$. 
			\end{enumerate*}
		\end{assumptionV}
		
		%	\begin{enumerate}[{${\mathbf{N}}$}(i)]
			%		\item \label{assum3b}	If $d_1 = 1/2$, $\langle P_1X_{t},h\rangle  = O_p(\sqrt{\ln T})$  and $T^{-1}\sum_{t=1}^T \langle P_1X_t,h\rangle^2 \to_p \infty$ for every $h\in \mathcal H_1$.
			%	\item \label{assum3a}	
			%\bt{WK: $q^{-2d_j\asd{\mathbf{Z}}}\langle {\Lambda}_T\asd{\mathbf{Z}}h,h \rangle \to_p c_{j,h}$ implies the dominant eigenvector of $\Lambda_{T}$ is included in $\mathcal H_1$ wpa.\ 1? }
			%, where $\widehat{\Lambda}(\mathbf{Z}_T) \coloneqq  T^{-1}\sum_{t=1}^T Z_t  \otimes  Z_t +  T^{-1}\sum_{t=1}^T \sum_{s=1}^{q} w(s,q) \left(Z_{t} \otimes Z_{t-s}+ Z_{t-s} \otimes Z_{t}\right)$.
			%	\end{enumerate}

		%\item \label{assum1b} $\gamma_{j,k}(s) \sim C s^{2d-1}\sigma_{j,k}^{1/2}$ if $\phi_j, \phi_k \in \ran P_D$ and $\gamma_{j,k}(s) \leq C s^{2d-1}\sigma_{j,k}^{1/2}$ . Moreover, $\sum_{r,s,u=-\infty}^\infty |\kappa_{j,k,\ell,m} (r,s,u)|\leq C \sigma_{j,k,\ell,m}^{1/2}$ and the spectral density $f$ of $x_{t,j}$ satisfies $f(\lambda)=C|\lambda|^{-2d}$ if $d<0$, and $\sup_{r} \sum_{s,u=-N}^N |\kappa_{j,k,\ell,m} (r,s,u)| \leq C N^{2d}\sigma_{j,k,\ell,m}^{1/2}$ if $d\geq 0$.
		
		%		\begin{assumpbw}\label{assumbw2} Assumption \ref{assumbw} holds; if $d_1 \in (1/2, 1)$, it is further assumed that $\upsilon < -(d_1-d_2)/2(d_1-1)$.
			%	\end{assumpbw}
		The above conditions are adaptations of Assumption~\ref{assumv1} and~\eqref{eqwnonstat} for nonstationary FIHTS with $d_1  > 1/2$. We show that the $V_1$-test is consistent against nonstationary FI alternatives. 
		\begin{propositions} \label{prop2add} 
			Suppose that $d_1 > 1/2$ and Assumptions~\ref{assum1}, \ref{assum2} and~\ref{assumv2} are satisfied. Then, the following hold:
			\begin{align*} %\label{eqconver2}
				{V}_1  \quad \overset{p \text{ or } d}{\longrightarrow} \quad   
				\begin{cases}
					0 \quad &\text{if $d_1 \in (1/2,1)$},\\
					\int W(r)^2 dr \quad &\text{if $d_1=1$},\\
					\infty \quad &\text{if $d_1>1$}.
				\end{cases} 
			\end{align*}
		\end{propositions}
		
		That is, the $V_1$-test can distinguish the null of I($1$)-ness from nonstationary FI alternatives. Moreover, we can determine which of the following intervals \commRVS{contains} the memory parameter $d_1$: $(1/2,1)$, $\{1\}$, and $(1, 2)$.
		
		%Even if the $V_1$-test, described in Proposition \ref{prop2add}, can be used to examine the I($1$)-ness, there is one unsatisfactory property as a test used after rejection of the $V_0$-test at the upper tail and also as a test for examining I(1)-ness against I($d$) for any possible fractions in $(0,2)$		
		Even if the $V_1$-test, described in Proposition~\ref{prop2add}, can be used to examine I($1$)-ness, there is an unsatisfactory property as a test used after rejection of the $V_0$-test in the upper tail: according to Propositions~\ref{prop2} and~\ref{prop2a},  rejection of the $V_0$-test means that the memory parameter of $\mathbf{Y}_0$ is contained in the interval $(0,\infty)$, but rejection of the $V_1$-test in the lower tail in Proposition~\ref{prop2add} means that the memory of $\mathbf{Y}_0$ is included in the interval $(1/2,1)$, not $(0,1)$ containing all fractions between 0 and 1. 
		\commRVS{Thus,} it would be useful if we extend the $V_1$-test given in Proposition~\ref{prop2add} so that the test has \commRVS{discriminating} power when $\mathbf{Y}_0$ is \commRVS{a} FIHTS in Assumption~\ref{assum1} with $d_1 \in [0, 1/2]$. We hereafter investigate this issue. Given that $\mathbf{Y}_0 \sim$ I($d_1$) implies that $\Delta X_t \sim$ I($d_1-1$), this obviously requires a detailed investigation of statistical properties of an I($d_1-1$) process. A complicated case occurs particularly when $d_1 = 1/2$, and, to deal with this case conveniently, we assume the following condition for the subsequent discussion: %\commHS{$\ln$ or $\log$ choose one and be consistent}
		\begin{assumptionV}\label{assumv3}  If $d_1=1/2$, the following hold: 
			\begin{enumerate*}[(i)]
				\item \label{assumv3a}	$\langle P_1X_{t},h\rangle  = O_p(\sqrt{\log T})$;
				\item \label{assumv3b} $q\to \infty$ and $q = o(T^{\upsilon})$ for $\upsilon \in (0, 1/2)$ and there exists a divergent sequence $d_T =o(T/\log T)$ such that $d_T\langle {\Lambda}_T\asd{\Delta \mathbf{X}_1}h,h \rangle$ is bounded away from zero with probability approaching one for every $h \in \mathcal H_1$.
			\end{enumerate*}
		\end{assumptionV}
		
		Assumption~\ref{assumv3}\ref{assumv3a} is a repetition of the requirement for the case with $d_1=1/2$ given in Proposition~\ref{prop2a}. Assumption~\ref{assumv3}\ref{assumv3b} seems to be a fairly high-level condition, but it is not restrictive. For example, \citet[Theorem 6]{CHO2015217} show that Gaussian I($-1/2$) processes generally satisfy this condition with $d_T = q/ \log q$, which is of much smaller asymptotic order than $T / \log T$ under $q = o(T^{1/2})$. We thus believe that some deviations from the conditions employed by \cite{CHO2015217} can also be accommodated under Assumption~\ref{assumv3}.  The desired extension of the $V_1$-test in Proposition~\ref{prop2add} is given as a direct consequence of our next result:
		\begin{propositions} \label{prop4}
			Suppose that $0 \leq d_1 \leq 1/2$ and Assumptions \ref{assum1}, \ref{assum2} and \ref{assumv3} hold. Then $V_1\to_p 0$.
			% Suppose that $\mathbf{Y}_0$ satisfies Conditions \ref{assum1a}-\ref{assum1c} for $d_1 \asd{\mathbf{Y}_0} \in (-1/2, 1/2]$, and further satisfy Conditions \ref{assum2a}-\ref{assum2b}; if $d_1 \asd{\mathbf{Y}_0} = 1/2$, then for any $h \in \mathcal H_1$, we further assume that (i) $(T\ln T)^{-1}\sum_{t=1}^T \langle P_jX_t,h\rangle \to_p 0$ and (ii) $T^{-1}\sum_{t=1}^T \langle P_jX_t,h\rangle^2 \to_p \infty$.  Then, under Condition \ref{assumbw}, $V_1 \to_p 0$. 
			
			% $(T\ln T)^{-1}\sum_{t=1}^T \langle P_jX_t,h\rangle \to_p 0$, where $P_1X_{t}$ is defined by Condition \ref{assum1a} for $\mathbf{Y}_0$. Then, under Condition \ref{assumbw}, $V_1 \to_p 0$. 
		\end{propositions}
		%provides some primitive conditions for an I($1/2$) time series to satisfy (i) (resp.\ (ii)). Particularly, the latter requires that the I($1/2$) time series is generated by iid Gaussian innovations. Without Gaussianity, it may be difficult to verify Assumption \ref{assum3a} or find primitive sufficient conditions. 
		
		\begin{remarks} \label{rem1} \normalfont
			Proposition~\ref{prop4} can be extended to the case with $-1/2< d_1 \leq 1/2$ without requiring further assumptions. In fact, our proof of Proposition~\ref{prop4} given in the Appendix is provided to accommodate this more general case. This means that the $V_1$-test is a standalone test, which can be used to examine I$(1)$-ness against I($d$) for any other \commRVS{value} of $d$ in the admissible range $(-1/2,2)$ (see Assumption~\ref{assum1}\ref{assum1d}).
		\end{remarks}
		
		\begin{remarks}[$V_2$-test and its further extension] \label{rem2}\normalfont
			Due to simplicity and empirical relevance, we focus in this paper on the case with $d_1 \in (-1/2,2)$. However, the $V_1$-test can be extended to another test, say the $V_2$-test, using the second-differenced time series of $Y_{2,t}$ (see \eqref{eqinitial}), for examining I($2$)-ness against FI alternatives. The extension from the $V_1$-test to the $V_2$-test is quite obvious and similar to that from the $V_0$-test to the $V_1$-test, which is described by Propositions~\ref{prop2add} and~\ref{prop4}. In a similar manner,  the $V_{k}$-test for an integer $k\geq 3$ can also be considered.
		\end{remarks}
		
		\subsection{Extensions}\label{sec:4.3}
		
		\subsubsection{Sequential testing procedure}  \label{sec_seq}
		In this section, we consider a combined procedure of the two tests developed in Sections~\ref{sec_frac01} and~\ref{sec_frac02} as a way to assess the statistical plausibility of the null hypothesis of integer integration. Let $\eta_\alpha$ be the \commRVS{$100\times \alpha$}\% quantile of $\int W(r)^2 dr$ for $\alpha \in (0, 1/2)$. Consider the following test function:
		\begin{equation}\label{eqdseq}
			D_{\seq} = \mathds{1}\{\eta_{\alpha} < V_0 < \eta_{1-\alpha}\}  +  \mathds{1}\{V_0 > \eta_{1-\alpha}\} \mathds{1}\{\eta_{\alpha} < V_1 < \eta_{1-\alpha}\}, 
		\end{equation}
		where $\mathds{1}\{\cdot\}$ denotes the binary indicator function and $D_{\seq}=1$ is understood as acceptance of the null hypothesis of integer integration. If $d_1=0$, we find that $\text{pr}(\eta_{\alpha} < V_0 < \eta_{1-\alpha}) \to 1-2\alpha$ while $\text{pr}(V_0 > \eta_{1-\alpha}) \to 0$. On the other hand, if $d_1=1$, $\text{pr}(V_0 > \eta_{1-\alpha}) \to 1$ and $\text{pr}(\eta_{\alpha} < V_1 < \eta_{1-\alpha}) \to 1-2\alpha$. \commRVS{Thus,} under either of these cases, $\text{pr}\{D_{\seq} = 1\}  \to 1-2\alpha.$ If $d_1$ is a fraction in $(-1/2,2)$, then \commRVS{one} of the following holds:
		\begin{equation*}
			\text{pr}(V_0 < \eta_{\alpha}) \to 1, \quad  \text{pr}( V_0 > \eta_{1-\alpha} \text{ and } V_1 < \eta_{\alpha} ) \to 1  \quad \text{or} \quad \text{pr}( V_0 > \eta_{1-\alpha} \text{ and } V_1 > \eta_{1-\alpha}) \to 1.
		\end{equation*}
		Therefore, $\text{pr}\{D_{\seq} = 0\}  \to 1$. \commRVS{Thus}, \eqref{eqdseq} constitutes a consistent decision rule to distinguish the null hypothesis of integer integration \commRVS{from} fractional integration with a significance level \commRVS{of} $2\alpha$. These results are summarized below:
		\begin{corollarys}   \label{cor0} 
			Suppose that all the assumptions for the $V_0$- and $V_1$-tests in Sections~\ref{sec_frac01} and~\ref{sec_frac02} are satisfied for $d_1\in (-1/2, 2)$. Then, under the null of integer integration of order $0$ or $1$,
			\begin{equation*}
				\text{\normalfont{pr}}\{D_{\seq} = 1\}  \to 1-2\alpha.
			\end{equation*}
			Under the alternative of fractional integration of order $d \in (-1/2,2)\setminus\{0,1\}$,
			\begin{equation*}
				\text{\normalfont{pr}}\{D_{\seq} = 0\}  \to 1.
			\end{equation*}	
		\end{corollarys}
		
		\begin{remarks} \normalfont
			For simplicity, we only considered FIHTS with the memory parameter $d_1\in (-1/2,2)$. However, we may extend the testing procedure developed in Corollary~\ref{cor0} to accommodate integrated time series of \commRVS{higher order} by sequentially adding the $V_k$-test for $k=2,3,\ldots,M$ for a finite integer $M$, obtained from extending the $V_1$-test (see Remark~\ref{rem2}). As illustrated by Corollary~\ref{cor0}, sequential applications of these tests do not distort the asymptotic size and are consistent. For example, if we consider the case with $d_1 \in (-1/2,3)$, we can additionally apply the $V_2$-test described in Remark~\ref{rem2}, and in this case, the relevant test function will become
			\begin{align*}
				D_{\seq} = \mathds{1}\{\eta_{\alpha} < V_0 < \eta_{1-\alpha}\}  &+  \mathds{1}\{V_0 > \eta_{1-\alpha}\} \mathds{1}\{\eta_{\alpha} < V_1 < \eta_{1-\alpha}\} \\ 
				&+ \mathds{1}\{V_0 > \eta_{1-\alpha}\}  \mathds{1}\{V_1 > \eta_{1-\alpha}\}\mathds{1}\{\eta_{\alpha} < V_2 < \eta_{1-\alpha}\}.
				%	D_{\seq, \delta} =& 1\{\eta_{\alpha} < V_0 < \eta_{1-\alpha}\}  +  1\{V_0 > \eta_{1-\alpha}\} 1\{\eta_{\alpha} < V_1 < \eta_{1-\alpha}\} \\ 				&+ \cdots + \Pi_{j=0}^{\delta-1} \left(1\{V_j > \eta_{1-\alpha}\}\right) 1\{\eta_{\alpha} < V_\delta < \eta_{1-\alpha}\}.
			\end{align*}
		\end{remarks}
		As shown above, the combination of the $V_0$- and $V_1$-tests \commRVS{not only enables us to examine integer integration, but also allows} us to determine which of the following subsets of $(-1/2,2)$ contains the true memory parameter:
		%	We assume that the time series $\mathbf{Y} = \{Y_t - Y_0\}_{t\geq 1}$ is $I(\widetilde{d})$ for some $\widetilde{d} \in (-1/2,2)$; even if we here restrict the admissible values of the memory parameter $\widetilde{d}$, the specified range is sufficient to accommodate most empirical applications.  We consider 
		\begin{equation*}
			\mathcal I_1 = (-1/2,0), \quad \mathcal I_2 = \{0\}, \quad  \mathcal I_3 = (0,1), \quad \mathcal I_4 = \{1\}, \quad \mathcal I_5 = (1,2). 
		\end{equation*} 
		That is, by first applying the ${V}_0$-test, we may determine if $d_1 \in \mathcal I_1$ (when the test is rejected in the lower tail)   or $d_1 \in \mathcal I_2$ (when the test is not rejected) or $d_1 \in \mathcal I_3 \cup \mathcal I_4 \cup \mathcal I_5$ (when the test is rejected in the upper tail). If the latter is concluded from the ${V}_0$-test, then we may apply the ${V}_1$-test to determine if $d_1 \in \mathcal I_3$ (when the test is rejected in the lower tail)   or $d_1 \in \mathcal I_4$ (when the test is not rejected) or $d_1 \in \mathcal I_5$ (when the test is rejected in the upper tail). This can \commRVS{provide useful information for} implementing some statistical methods on the memory parameter \commRVS{that require} appropriate bounds of $d_1$ as an input, such as the local Whittle estimator of \cite{li2020local}, \cite{LRS}, and \cite{seoshang22} in the functional time series setup.
		
		If the time series of interest is anticipated to be similar to an I($1$) process as in many empirical examples, a reversed procedure may be more practically relevant. Specifically, we may first apply the $V_1$-test to determine if $d_1 \in \mathcal I_5$ (when the test is rejected in the upper tail)   or $d_1 \in \mathcal I_4$ (when the test is not rejected) or $d_1 \in \mathcal I_1 \cup \mathcal I_2 \cup \mathcal I_3$ (when the test is rejected in the lower tail). If the latter is concluded from the $V_1$-test, we may apply the $V_0$-test to determine if $d_1 \in \mathcal I_3$ (when the test is rejected in the upper tail)   or $d_1 \in \mathcal I_2$ (when the test is  not rejected) or $d_1 \in \mathcal I_1$ (when the test is rejected in the lower tail). This may be understood as a reversed version of the aforementioned testing procedure. %Unlike the earlier proposed testing procedure that can easily be extended to examine if $d\in \mathcal D_{5}'$, $ \mathcal D_{6}' $, or $\mathcal D_{7}'$, this reversed procedure requires a more careful approach for such an extension. For example, suppose that $d=0$, but we compute $Y_{2,t} = \Delta^2(Y_t-Y_0)$ to implement the $V_2$-test. Then obviously, the time series becomes over-differenced and its memory parameter is below $-1/2$. This case has not been studied in the previous sections due to theoretical complications. Thus this procedure needs to be used under a prior information on the memory parameter of the time series. However, our simulation results show that this reversed procedure works well even in this case.  

		\subsubsection{Deterministic terms}\label{sec_det}
		
		In the previous sections, we have used the initialized variable $Y_t-Y_0$, which is assumed to have zero mean. We in this section consider the case with an unknown intercept, where observations are given by $\widetilde{\mathbf{Y}}_0 = \mu + \mathbf{Y}_0 =  \{\mu + {Y}_{0,t}\}_{t=1}^T$, and consider a testing procedure that is robust to the existence of an unknown intercept.  
	%	We in this section consider the sequence $\widetilde{\mathbf{Y}}_0 = \mu + \mathbf{Y}_0 =  \{\mu + {Y}_{0,t}\}_{t=1}^T$. %and its realizations $\widetilde{\mathbf{Y}}_{0,T}=\{\widetilde{Y}_{0,t}\}_{t=1}^T$, where 
		%	\begin{equation}
			%	\widetilde{Y}_{0,t} = \mu + Y_{0,t}
			%	\end{equation}
		We hereafter define 
		\begin{equation*}
			\widetilde{\mathcal K}_T\asd{\mathbf{Z}} \coloneqq  \sum_{t=t_0}^T \left(\sum_{s=t_0}^t (Z_s-\bar{Z}_T)\right)  \otimes  \left(\sum_{s=t_0}^t  (Z_s-\bar{Z}_T)\right),
			%&\widehat{\Lambda}_v = \frac{1}{T} \sum_{t=1}^T \nu_t  \otimes  \nu_t +  \frac{1}{T} \sum_{t=1}^T \sum_{s=0}^{h} \left(1 - \frac{s}{q}\right) \left(\nu_t  \otimes  \nu_{t-s} + \nu_{t-s}  \otimes  \nu_{t}\right). 
		\end{equation*}
		where $\bar{Z}_T$ is the sample mean of $Z_t$ over $t=t_0,\ldots,T$. %%(T-t_0)^{-1} \sum_{t=t_0}^{T} Z_t$. 
		We also let  
		\begin{equation*}%\label{eqlrv2}
			%{\Lambda}_T\asd{\mathbf{Z}} &\coloneqq  T^{-1}\sum_{t=t_0}^T (Z_t-\bar{Z}_T)  \otimes  (Z_t-\bar{Z}_T) \\ & +  T^{-1}\sum_{t=s+t_0}^T \sum_{s=1}^{q} w(s,q) \left((Z_t-\bar{Z}_T) \otimes (Z_{t-s}-\bar{Z}_T)+ (Z_{t-s}-\bar{Z}_T) \otimes (Z_t-\bar{Z}_T)\right),
			\widetilde{\Lambda}_T\asd{\mathbf{Z}} \coloneqq  T^{-1} \sum_{s=-q}^q w(s,q) \sum_{t_0 \leq t, t-s \leq T} (Z_t-\bar{Z}_T)  \otimes  (Z_{t-s}-\bar{Z}_T),
		\end{equation*}
		where $w(s,q)=1 - {|s|}/(q+1)$. We then consider the test statistic given as follows:
		\begin{equation*}
			\widetilde{V}_0 =T^{-2} \frac{ \langle { \widetilde{\mathcal K}}_T\asd{\widetilde{\mathbf{Y}}_0} \hddd,\hddd \rangle}{\langle \widetilde{\Lambda}_T\asd{\widetilde{\mathbf{Y}}_0}\hddd,\hddd\rangle}, %
			%= T^{-2}\frac{\langle { \mathcal K}_T\asd{\mathbf{Y}_{0}}\hdd\asd{\mathbf{Y}_{0}},\hdd\asd{\mathbf{Y}_{0}} \rangle}{\langle {\Lambda}_T\asd{\mathbf{Y}_{0}}\hdd\asd{\mathbf{Y}_{0}},\hdd\asd{\mathbf{Y}_{0}}\rangle},  
		\end{equation*}
		where $\widetilde{h}_T$ is the dominant eigenvector of $\widetilde{\mathcal K}_{T}\asd{\mathbf{Y}_{0}}$. 
		
		We can similarly define $\widetilde{V}_1$ as in \eqref{eqv1}. However, since we use the first-differenced sequence for this test, the test statistic does not change from $V_1$ in \eqref{eqv1}, except that $\hdd$ is replaced by $\hddd$. As will be discussed in our proof of Proposition~\ref{prop2add2}, under Assumption~\ref{assum1}, $\hddd$ satisfies all the necessary properties required for $\hdd$, which are used to establish the asymptotic properties of the test. Hence, the two tests have the same limiting behaviors under the null and alternative hypotheses. We thus focus only on the limiting behavior of $\widetilde{V}_0$ in this section. 
		
		\begin{propositions} \label{prop2add2} 
			Suppose that  $-1/2<d_1<1/2$, Assumptions~\ref{assum1} and~\ref{assum2} hold, and Assumption~\ref{assumv1} is satisfied when $\Lambda_T(\mathbf{X}_1)$ %in  Assumptions \ref{assum1}\ref{assumv1a} 
			is replaced by $\widetilde{\Lambda}_T(\mathbf{X}_1)$. Then,
			\begin{align*}
				\widetilde{V}_0  \quad \overset{p \text{ or } d}{\longrightarrow} \quad   \begin{cases}
					0 \quad &\text{if $d_1 \in (-1/2,0)$},\\
					\int \widetilde{W}(r)^2 dr \quad &\text{if $d_1=0$},\\
					\infty \quad &\text{if $d_1 \in (0,1/2)$},
				\end{cases} 
			\end{align*}
			where $\widetilde{W}$ is the standard Brownian bridge on $[0,1]$, given by $\widetilde{W}(r) = W(r)-rW(1)$. If $d_1\geq 1/2$, Assumptions~\ref{assum1} and~\ref{assum2} hold, and \eqref{eqwnonstat} is satisfied when $d_1=1/2$, then $\widetilde{V}_0 \overset{p}{\to}  \infty$.
		\end{propositions}
		
		\begin{remarks} \label{remprimitive2} 
			\normalfont We discuss primitive sufficient conditions for Assumption~\ref{assumv1}\ref{assumv1a} to be satisfied when $\Lambda_T(\mathbf{X}_1)$ is replaced by $\widetilde{\Lambda}_T(\mathbf{X}_1)$ in Section~\ref{sec_app_abadir} of the Supplementary Material.
			%In this case,    and thus  provided by of course, replacement of the high-level condition given by Assumption \ref{assum2}\ref{assum2b} with such lower-level conditions does not affect the results to be developed. 
		\end{remarks}
		
		\section{Simulation studies}\label{sec_sim}
		
		\subsection{Simulation data generating process}\label{sec_sim1}
		We provide simulation results for the proposed test. The simulation data generating process (DGP) considered in this section is similar to the curve-valued processes considered by \cite{seoshang22}. In our subsequent experiments, $d_1$ is controlled to be a specified value in $[-0.45,0.45]$, and depending on this value, we generate FIHTS so that it exhibits various degrees of memory reduction on $\mathcal H$.  
		
		Let $\{e_k\}_{k\geq 1}$ be an orthonormal basis of $\mathcal H$, where $\{e_k\}_{k=1}^{r_1}$ is an orthonormal set, spanning the dominant subspace $\mathcal H_1 = \ran P_1$ of dimension $r_1>0$. We next define $\mathcal H_2 = \ran P_2$ as the $(r_2-r_1)$-dimensional subspace spanned by $\{e_k\}_{k=r_1+1}^{r_2}$, and successively define $\mathcal H_j = \ran P_j$ as the span of %$\spn\{e_{r_{j-1}+1},\ldots,e_{r_j}\}$ 
		$\{e_{k}\}_{k=r_{j-1}+1}^{r_j}$ for $j=1,\ldots,K$ and $r_1< r_2 <\ldots < r_{K}$. 	
		
		For each $j$, we let $\Delta^{d_j} P_j X_t$ in~\eqref{eqfi01} --~\eqref{eqfi03} be generated as follows: 
		\begin{equation} \label{eqsim01}
			\Delta^{d_j} P_j X_t = \sum_{\ell=r_{j-1}+1}^{r_j}{a}_{\ell,t} \sigma_{\ell}  e_{\ell}, \quad a_{\ell,t} \sim \text{ARMA}(1,1),  
		\end{equation}
		where  the AR coefficient $\phi_{\ell}$ and MA coefficient $\theta_{\ell}$, characterizing the $\text{ARMA}(1,1)$ process $\{{a}_{\ell,t}\}$, are determined as follows in each simulation run:
		\begin{equation*}
			\phi_{\ell} \sim U[-\mathsf{b},\mathsf{b}] \quad \text{and} \quad \theta_{\ell} \sim U[-\mathsf{b},\mathsf{b}],
		\end{equation*}
		for some $\mathsf{b} > 0$ without dependence on any other variables. In the subsequent simulation experiments, we consider the cases with $\mathsf{b}=0.15$ and $\mathsf{b}=0.6$, and obviously, the former case is closer to white noise than the latter. Using~\eqref{eqsim01} and an appropriate anti-differencing operation, we construct $P_jX_t$ and compute $Y_{0,t}$ as $ Y_{0,t} = \sum_{j=1}^{K} P_jX_t$ for large enough $K>0$. $\sigma_{\ell}$ in \eqref{eqsim01} is a sequence decaying to zero as $\ell$ gets larger, introduced to ensure the required summability condition for the resulting time series to be $\mathcal H$-valued (\commRVS{see, e.g.,} \citealp{Nielsen2023}); we specifically let $\sigma_{\ell} =\ell^{-2}$ so that the variances of ${a}_{\ell,t}$ for $\ell \geq 1$ become summable. 
		
		For each simulation run where $d_1$ is specified, we let $d_2,\ldots,d_K$ and the dimensions of $\ran P_j$, $p_j=r_j-r_{j-1}$ (for $j\geq 1$ with $r_0=0$), be randomly determined. By doing so, we can evaluate the average performance of the proposed test to some degree for various values of those parameters. 			
		%	Specifically, in our simulation concerning the stationary FIHTS (i.e., the case where $d_1 \in (-1/2,1/2)$), we let $p_j \sim U\{1,2\}$,  $d_2 \sim U(\max\{d_1-0.2,-0.5\},\max\{d_1-0.1,-0.5 +\eta \})$, and  $d_3,\ldots,d_K$ are ordered sequence of iid copies of $U(\max\{d_1-0.5,-0.5\},\max\{d_1-0.2,-0.5 +\eta \})$, where $\eta$ is set to a small constant so that $d_1> -0.5+\eta$; in any of he subsequent simulation experiments, we restrict $d_1\geq -0.485$ and hence we set $\eta = 0.01$ so that $d_1 > d_2$ in any case. Moreover, note that since $d_2$ has a crucial role in our asymptotic analysis, we purposely generate $d_2$ so that it tends to be relatively closer to $d_1$ than the others. 
		Specifically, in our simulation concerning the stationary FIHTS (i.e., the case where $d_1 \in (-1/2,1/2)$), we let $p_j \sim [U(1,4)]$, where $[a]$ denotes the integer part of $a$. The memory parameters $d_2,\ldots,d_K$ are determined as the ordered (from the largest to \commRVS{the} smallest) realizations $\{\bar{d}_j\}_{j=1}^{K}$ from uniform random variables given by  
		\begin{equation} \label{eqdj1}
			\bar{d}_j \sim \begin{cases}
				U((d_1-0.2) \vee -0.5, (d_1-0.1)   \vee -0.49 ) \quad &\text{if $j=2$},\\
				U((d_1-0.5)\vee-0.5, (d_1-0.2) \vee -0.49) \quad &\text{if $j\geq 3$},
			\end{cases}
		\end{equation}
		where $a\vee b = \max\{a,b\}$ for $a,b \in \mathbb{R}$.  Since we consider the case where $d_1 \in [-0.45,0.45]$, $d_2$ (i.e., the largest $\bar{d}_j$) obtained as above satisfies that $0.04< |d_1-d_2| < 0.2$, and hence the simulation DGP always allows a moderate memory reduction, which may be regarded as a realistic assumption in many empirical applications.    
		%and  $d_3,\ldots,d_K$ are ordered sequence of iid copies of $$U((d_1-0.5)\vee-0.5, (d_1-0.2) \vee -0.49),$$  
		%; in any of he subsequent simulation experiments, we restrict $d_1\geq -0.45$ and hence we set $\eta = 0.01$ so that $d_1 > d_2$ in any case. Moreover, note that since $d_2$ has a crucial role in our asymptotic analysis, we purposely generate $d_2$ so that it tends to be relatively closer to $d_1$ than the others unless $d_1$ is not close to the lower boundary of admissible values, $-0.5$. 	
		
		%	In the cases where nonstationary FIHTS are considered for a specified $d_1 > 1/2$, we let $p_1 \sim U\{1,2\}$, $p_2 \sim U\{1,2\}$, and $d_2 \sim U(\max\{d_1-0.2,0.5\},\max\{d_1-0.1,0.5+\eta\})$. Then we let $P_1X_t$ and $P_2X_t$ be consist of the nonstationary part of the time series. Then the stationary part is set to the FIHTS with memory 0.25, which is generated exactly as in the case concerning stationary FIHTS with $d_1=0.25$. 
		
		In the cases where nonstationary FIHTS are considered, we let $d_1\geq 0.55$, and let $p_1(=r_1)$, $p_2 (=r_2-r_1)$ and $d_2$ be determined as follows:  
		\begin{equation}\label{eqdj2}
			p_1 \sim [U(1,4)], \quad p_2 \sim [U(1,4)], \quad d_2 \sim U((d_1-0.2)\vee 0.5, (d_1-0.1) \vee 0.51).
		\end{equation}
		We then let $P_1X_t$ and $P_2X_t$ constitute the nonstationary part $X_t^N$ of the time series, with $\Delta^{d_1} P_1X_t$ and $\Delta^{d_2} P_2X_t$ being generated as in \eqref{eqsim01}. The time series used in the analysis is obtained by adding a stationary component $X_t^S$ to this nonstationary time series; \commRVS{specifically,} we let the stationary part $X_t^S$ be generated exactly as in the above case concerning stationary FIHTS with $d_1=0.25$. 
		
		We let the basis system $\{e_k\}_{k\geq 1}$ vary to some degree across different simulation runs \citep[see e.g.,][]{ARS18}. Specifically, we let $\{e_k\}_{k\geq 1}$ be the Fourier basis functions of which the first five basis functions are randomly permuted, so that the dominant parts of the time series take values in different subspaces across different simulation runs. This is also to avoid \commRVS{evaluating performance of the test with} specific shapes of the dominant components of the generated functional observations. In all the simulation experiments, $q$ is set to the greatest integer that does not exceed $T^{0.2}$.
		
		\subsection{Simulation results} \label{sec_sim2}
		
		We first examine the sizes and correct rejection rates of the proposed $V_0$- and $V_1$-tests in various scenarios regarding the integration order. The correct rejection rate is slightly different from the power of the test. Specifically, it measures the relative frequency that the null is rejected at lower (resp.\ upper) tail when $d_1<0$ (resp.\ $d_1>0$) for the $V_0$-test. For the $V_1$-test, it measures  the relative frequency that the test is rejected at lower (resp.\ upper) tail when $d_1<1$ (resp.\ $d_1>1$). Considering that our testing results may also be used to conjecture a reasonable range for $d_1$, the correct rejection rate represents a performance measure as the relative frequency of the correct decision being made by the proposed test.	

	Table~\ref{tabemp1} shows the simulated sizes and correct rejection rates of the $V_0$- and $V_1$-tests in various scenarios. It may be noticeable that the $V_0$-test (resp.\ $V_1$-test) performs better in the case where $d_1>0$ (resp.\ $d_1 >1$) compared to the case where $d_1<0$ (resp.\ $d_1<1$). This may be partly due to the restriction on admissible values of $d_j$ in the DGP in each case. As described in~\eqref{eqdj1} and~\eqref{eqdj2}, in the simulation DGP concerning the stationary (resp.\ nonstationary) case, the second most persistent I($d_2$) component is set to a stationary (resp.\ nonstationary) process, and thus $d_2$ needs to satisfy $d_2>-1/2$ (resp.\  $d_2>1/2$). If $d_1$ is lower than the specified integer value under the null in each case, the gap between $d_1$ and $d_2$ tends to be smaller, and overall $d_1,\ldots,d_K$ tend to become more homogeneous (and hence less distinguishable from the data) compared to the case where $d_1$ is greater than the specified integer value.
		\begin{table}[!htb]	
			\caption{Size and correct rejection rates} \label{tabemp1}	
			\begin{subtable}[b]{1\linewidth}
				\centering
				\caption{$V_0$-test for $H_0: d_1=0$}\label{tabemp1a}	
			\commRV{	\begin{tabular*}{1\linewidth}{@{\extracolsep{\fill}}l|l|ccccccc@{}}
					\thickhline
					&  \backslashbox{$T$}{$d_1$}	& -0.45 & -0.3 & -0.15  & 0 & 0.15 & 0.3 & 0.45 \\ \hline
					$\mathsf{b}=0.15$  	& $125$	& 0.693 &0.372& 0.102& 0.041& 0.240& 0.616& 0.931\\	
                    & $250$				& 0.832&0.470&0.128&0.031&0.264&0.659&0.946 \\ 
					& $500$\,\, 		& 0.942		& 0.654&0.195&0.038&0.322&0.762&0.968\\ 
					& $750$\,\, 		&0.979		& 0.757&0.248&0.036&0.360&0.804&0.984 \\
					& $1000$\,\, 		&0.994		& 0.812&0.282&0.041&0.392&0.834&0.991 \\ \hline 
					$\mathsf{b}=0.60$ & $125$	& 0.616& 0.357 &0.124& 0.077& 0.258& 0.632& 0.934 \\	
                    & $250$ 	&  0.781&0.471&0.155&0.062&0.281&0.674&0.943 \\ 
					& $500$\,\, 		&  0.910		& 0.643&0.220&0.064&0.342&0.762&0.968\\ 
					& $750$\,\, 		& 0.952		& 0.736&0.269&0.062&0.378&0.802&0.981\\ 
					& $1000$\,\, 		& 0.978		& 0.791&0.306&0.064&0.398&0.841&0.990\\ 
					\thickhline
				\end{tabular*}} \quad \vspace{0.75em}
			\end{subtable}
			\,\,
			\begin{subtable}[b]{1\linewidth}
				\centering
				\caption{$V_1$-test for $H_0: d_1=1$}
				\commRV{	\begin{tabular*}{1\linewidth}{@{\extracolsep{\fill}}l|l|ccccccc@{}}
					\thickhline
					&  \backslashbox{$T$}{$d_1$} & 0.55 & 0.7 & 0.85 & 1 & 1.15 & 1.3 & 1.45 \\ \hline
					$\mathsf{b}=0.15$ & $125$ &  0.653& 0.352& 0.082& 0.032& 0.248& 0.624 &0.938 \\ 
& $250$ &0.800&0.460&0.117&0.032&0.280&0.676&0.945 \\ 
					& $500$\,\, & 0.935&0.658&0.168&0.035&0.358&0.790&0.971\\ 
					& $750$\,\, &0.975&0.757&0.206&0.028&0.404&0.834&0.981\\ 
					& $1000$\,\, &0.990&0.809&0.238&0.033&0.432&0.862&0.987 \\ \hline
					$\mathsf{b}=0.60$ & $125$ & 0.614 &0.354& 0.128 &0.073& 0.260& 0.625& 0.938\\
& $250$ &0.752&0.465&0.148&0.062&0.294&0.687&0.939 \\ 
					& $500$\,\, &0.901&0.634&0.206&0.064&0.373&0.789&0.970\\ 
					& $750$\,\, & 0.955&0.730&0.235&0.058&0.411&0.830&0.981\\ 
					& $1000$\,\, &0.974&0.788&0.264&0.054&0.431&0.857&0.985\\ 
					\thickhline
				\end{tabular*}}
			\end{subtable}
			\begin{minipage}{1\linewidth}
				\footnotesize{Notes: In the case with either $d_1=0$ or $1$, the size of the test is reported; the nominal size is $5\%$. Except for these cases, the correct rejection rates are shown. The number of Monte Carlo replications in each case is~2000.}
			\end{minipage}
		\end{table}
		
		Overall, the $V_0$- and $V_1$-tests perform better as the sample size $T$ gets larger, and both tests have good size control. The correct rejection rates also increase as $d_1$ moves away from the hypothesized integer value for both tests, which may well be expected from the theoretical results presented earlier. The parameter $\mathsf{b}$, which measures the tendency of departure of the underlying ARMA process $\{a_{\ell,t}\}$ from white noise, seems to affect the performance of the tests, but the difference is moderate for the considered values of $\mathsf{b}$. 		
		
				We examine the performance of the sequential testing procedure discussed in Section~\ref{sec_seq}. We study how accurately the testing procedure identifies integer integration (the cases where $d = 0$ or $1$) in the simulation DGP. We also assess its performance in the case of fractional integration by computing the relative frequency of correctly identifying the interval that includes the fractional order. The simulation results are presented in Table~\ref{tabemp2}, along with additional details of the Monte Carlo experiments. Overall, the testing procedure performs reasonably well, even with a moderate sample size of $T=250$, and its performance improves as the sample size increases.
				
				\vspace{-.09in}
		
				\begin{table}[!htb]	
			\caption{Relative frequencies of correct determination of integer/fractional integration} \label{tabemp2}
			\begin{subtable}[b]{1\linewidth}
				\centering
				\caption{Case with integer integration}
				\commRV{	\begin{tabular*}{1\linewidth}{@{}c|c@{\extracolsep{\fill}}|ccccc@{}}
					\thickhline
					&  \phantom{CC} Case \phantom{CC} & $T=125$ & $T=250$ & $T=500$ & $T=750$ & $T=1000$ \\  \hline 
					$\mathsf{b}=0.15$ & $d_1 \in \{0,1\}$ &0.962 &0.970& 0.960& 0.959 &0.957 \\
					& $d_1=0$ &0.968& 0.969& 0.959& 0.959& 0.952\\
					& $d_1=1$ &0.908& 0.950& 0.955 &0.957 &0.962 \\ \hline
					$\mathsf{b}=0.60$ & $d_1 \in \{0,1\}$ &0.933& 0.947& 0.941& 0.936& 0.934 \\
					& $d_1=0$ & 0.941 &0.948 &0.935& 0.934& 0.931 \\
					& $d_1=1$ &0.883 &0.924& 0.941& 0.937& 0.938 \\ 
					\thickhline
				\end{tabular*}	}
			\end{subtable}
			\begin{subtable}[b]{1\linewidth}
				\centering
				\caption{Case with fractional integration}
				\commRV{\begin{tabular*}{1\linewidth}{@{}c|c@{\extracolsep{\fill}}|ccccc@{}}
					\thickhline
					& \phantom{CC} Case \phantom{CC}& $T=125$& $T=250$ & $T=500$ & $T=750$  & $T=1000$ \\  \hline 
					$\mathsf{b}=0.15$ & $d_1<0$ & 0.387& 0.520& 0.669& 0.756& 0.794 \\
					& $d_1\in(0,1)$ &0.476& 0.552& 0.714& 0.770& 0.794   \\
					& $d_1>1$ &0.689 &0.715& 0.757& 0.800& 0.827 \\ \hline
					$\mathsf{b}=0.60$ & $d_1<0$ &0.356& 0.533& 0.657& 0.718& 0.754 \\
					& $d_1\in(0,1)$ &0.477& 0.563& 0.702& 0.749& 0.776  \\
					& $d_1>1$ &0.697& 0.724& 0.767 &0.788& 0.806 \\ 
					\thickhline
				\end{tabular*} }
			\end{subtable}	
			\begin{minipage}{1\linewidth}
				\footnotesize{Notes: In the top panel, $d_1$ is set to $0$ or $1$ in the simulation DGP, and the case $d_1 \in \{0,1\}$ reports the relative frequency of the cases where $d_1 = 0$ or $1$ is concluded by the testing procedure for each sample size; the total number of Monte Carlo replications is set to 2000. Within the same simulation results, the case $d_1=0$ (resp.\ $d_1=1$) reports the frequency of correctly determining $d_1$ by the testing procedure, relative to the frequency of the cases where $d_1=0$ (resp.\ $d_1=1$) in the simulation DGP. The bottom panel reports the frequency of correctly identifying the subset $\mathcal I_j$ for each sample size, relative to the frequency of the cases where $d_1 \in \mathcal I_j$ in the simulation DGP. In each simulation run concerning this case, $d_1$ is randomly chosen; specifically, $d_1$ is determined by one of the following four uniform random variables, each of which can occur with a probability $1/4$: $U(-0.485,-0.15)$, $U(0.15,0.5)$,  $U(0.5,0.85)$, or $U(1.15, 1.5)$.}
		\end{minipage}
	\end{table}

	We additionally examine the finite sample performance of the $\widetilde{V}_0$-test, which is discussed in Section~\ref{sec_det}.  In this simulation experiment, we added a nonzero mean $\mu_{\ell}$ to each of the ARMA \commRVS{processes} $a_{\ell,t}$ in~\eqref{eqsim01}, and $\mu_{\ell}$ is drawn from $N(0,1)$ in each simulation, without dependence \commRVS{on} any other variables. The simulation results are reported in Table~\ref{tabemp1add}. The results are qualitatively similar to those of Table~\ref{tabemp1}.
	\begin{table}[!htb]
		\caption{Size and correct rejection rates for the $\widetilde{V}_0$-test for $H_0: d_1=0$} \label{tabemp1add}
		\centering
	\commRV{	\begin{tabular*}{1\linewidth}{@{\extracolsep{\fill}}l|l|ccccccc@{}}
			\thickhline
			&  \backslashbox{$T$}{$d_1$} 	& -0.45 & -0.3 & -0.15  & 0 & 0.15 & 0.3 & 0.45 \\ \hline
			$\mathsf{b}=0.15$ & $T=125$&  0.670& 0.326& 0.094& 0.039 &0.162 &0.431& 0.672 \\ 
           & $T=250$& 0.845& 0.454 &0.136& 0.041& 0.185& 0.484& 0.750 \\ 
			& $T=500$\,\, & 0.970& 0.711& 0.232& 0.042& 0.260& 0.645& 0.884\\ 
			& $T=750$\,\, &0.994& 0.826& 0.292& 0.036& 0.317& 0.715& 0.935 \\
			& $T=1000$\,\, &0.994& 0.826& 0.292& 0.036 &0.317& 0.715& 0.935 \\ \hline 
			$\mathsf{b}=0.60$ & $T=125$&   0.594& 0.342& 0.144& 0.093& 0.199& 0.456& 0.698 \\ 
           & $T=250$ &   0.780& 0.476& 0.192 &0.069& 0.211& 0.498& 0.772 \\ 
			& $T=500$\,\, & 0.932 &0.674& 0.279& 0.078& 0.290& 0.649& 0.892\\ 
			& $T=750$\,\, &  0.977& 0.772& 0.318 &0.082 &0.334 &0.725& 0.944\\ 
			& $T=1000$\,\, & 0.977& 0.772& 0.318& 0.082& 0.334& 0.725 &0.944\\ 
			\thickhline
		\end{tabular*} }\quad \vspace{0.75em}
		\begin{minipage}{1\linewidth}
			\footnotesize{Notes: The nominal size is $5\%$. The number of Monte Carlo replications in each case is 2000.}
		\end{minipage}
	\end{table}

	\section{Empirical applications}\label{sec:6}
	
	\subsection{Canadian yield curves for zero-coupon bonds}\label{sec:6.1}
	
	We apply our proposed test to the end-of-day Canadian yield curves for the period spanning 04 July 2022 to 11 September 2024; the data used in this section is publicly available at \url{https://www.bankofcanada.ca/rates/interest-rates/bond-yield-curves/}. \revlab{r1_major3}\commRV{In the considered dataset, yields are observed daily (excluding non-trading days) on a fine grid of maturities: specifically, zero-coupon bond yields at 120 regularly spaced maturities from 0.25 to 30 years. We treat these as functional data. In total, the number of all  available daily observations over the entire time span is 545.}  In Figure~\ref{fig:1}, we present a rainbow plot of the Canadian yield curve data by maturities. A similar dataset is used by \cite{seoshang22}, although their analysis employs a long-span, monthly frequency time series, whereas the analysis here considers a daily frequency.  
	\begin{figures}[!htb]
		\centering
		\includegraphics[width=8.4cm]{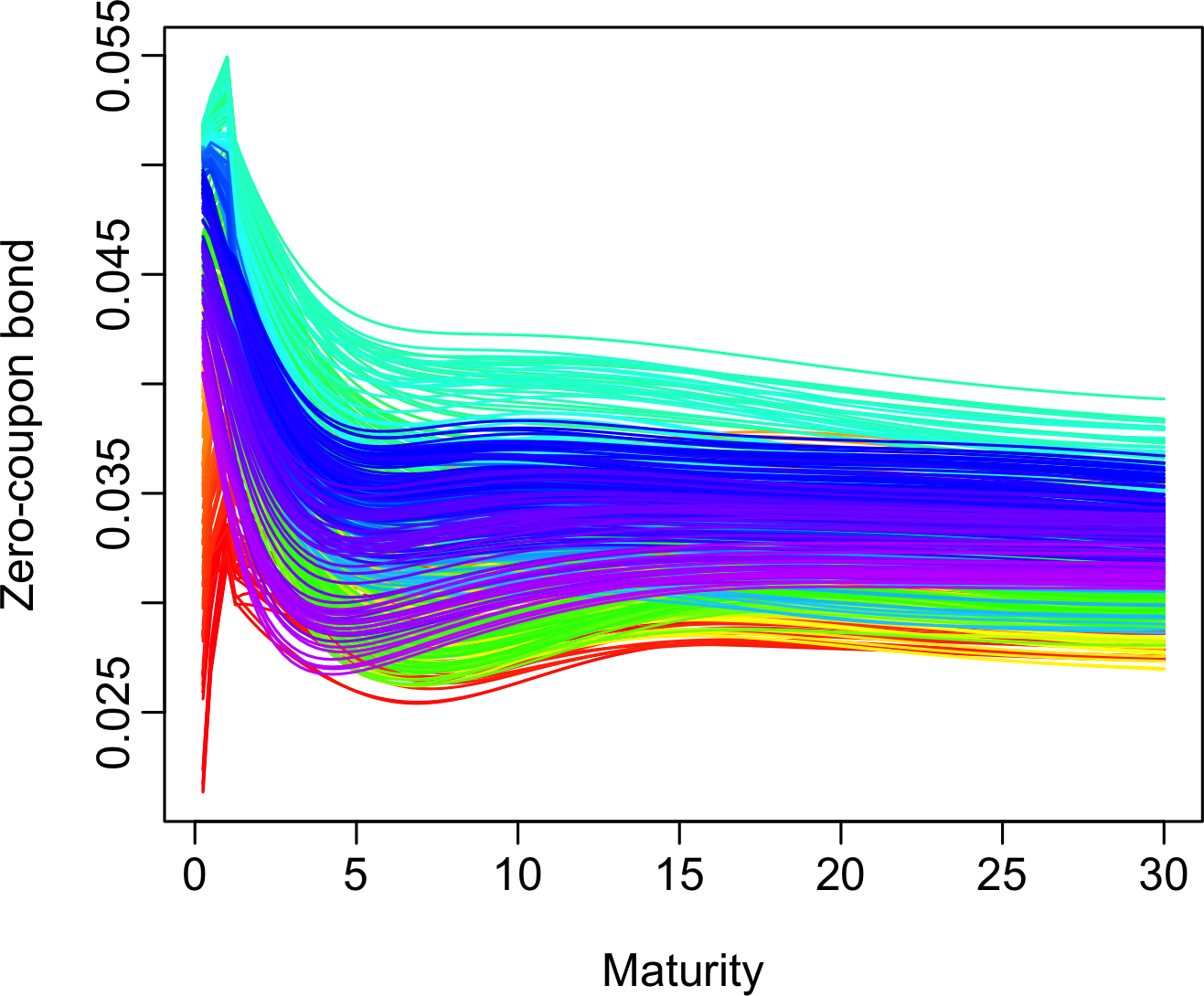}
		\caption{Visualization of the Canadian yield curves %by term-specific maturities from 0.25 to 30 years.
		}\label{fig:1}
	\end{figures}
	
	In some recent articles, a time series of yield curves is often mentioned as an example of \commRVS{a} nonstationary functional time series (\commRVS{see, e.g.,} \citealp{LRS, Nielsen2023, seoshang22}), and empirical evidence suggests that such a time series behaves similarly to an I(1) process. We examine this in detail using our methodology. 
	
		\begin{table}[!htb]
		\caption{Testing results for Canadian yield curves}\label{tab:yield}
		\centering
		\begin{tabular*}{1\linewidth}{@{}l@{\extracolsep{\fill}}ccc@{}}
			\toprule
			&$V_0$-test    				&  $V_1$-test   &  $V_2$-test 	\\ \midrule
			Statistic 	 & 58.06  					& 1.82 		& 0.004 		\\ 
			$p$-value  & $< 0.001$ 				&  $0.08$  	&  $<0.001$  	\\
			Test result & Rejection in the upper tail  & Accept 		& Rejection in the lower tail \\
			\bottomrule
		\end{tabular*}
		\begin{minipage}{1\linewidth}
			\footnotesize{Notes: The $p$-values in the second row are derived from 200,000 Monte Carlo replications of the approximate limiting distribution. The third row presents the test results at the 5\% significance level, where the null hypothesis is rejected if the test statistic falls below the 2.5\textsuperscript{th} percentile (\commRVS{$q_{0.025}$}) or exceeds the 97.5\textsuperscript{th} percentile ($q_{0.975}$) of the limiting distribution, approximately 0.045 and 2.126, respectively. \commRVV{The reported $p$-value is computed as $2 \min \{\text{pr} (V_j > \hat{V_j}), \text{pr} (V_j <\hat{V_j})\},$ since each of the considered tests is two-sided, where $V_j$ denotes the limiting random variable and $\hat{V}_j$ the test statistic. The $p$-value is computed using the sample quantiles from 200,000 Monte Carlo realizations of the limiting distribution, generated by the standard simulation method.}}
		\end{minipage}
	\end{table}	
	
	Specifically, we apply the sequential testing procedure based on the $V_0$- and $V_1$-tests to the yield curve data. The test results are reported in the first two columns of Table~\ref{tab:yield}. As reported in the table, the $V_0$-test is rejected in the upper tail with a $p$-value nearly equal to zero, meaning that I$(0)$ stationarity is strongly rejected against a \commRVS{higher-order} integration. On the other hand, the null hypothesis of I$(1)$-ness is not rejected at \commRVS{the} 5\% significance level, from which we conclude that the time series of interest is I($1$). Given that our testing procedure can easily be extended to examine \commRVS{higher-order} integration (see Remark~\ref{rem2}), one may want to confirm this test result by implementing the $V_2$-test and checking if the null of I(2)-ness is rejected in the lower tail. The test result is reported in the third column of Table~\ref{tab:yield}, and the null of I(2)-ness is strongly rejected against a \commRVS{lower-order} integration.  

\vspace{-.1in}
	
If the considered tests are implemented at the 10\% significance level, then the $V_1$-test rejects the null hypothesis of I$(1)$ integration as well, and thus the testing procedure concludes that this time series is fractionally-integrated of order $d \in (1,2)$. Overall, our test results confirm the nonstationarity of the Canadian yield curve time series with an integration order of $d=1$ or $d \in (1,2)$.
	
	\subsection{French sub-national mortality rates}\label{sec:6.2}
	
	We next apply our proposed tests to age- and sex-specific French sub-regional mortality data from 1901 to 2021, with the data sourced from the French Human Mortality Database at \url{https://frdata.org/en/french-human-mortality-database/} \citep[see][for data description]{Bonnet20}. Depending on the territorial units for statistics, there exist NUT1, NUT2 and NUT3 ranging from the largest to the smallest units. For illustration, we consider the middle level, i.e., NUT2, which has 22 regions. The number of samples is 121 for all regions except one (Alsace), which allows records with many missing or overly imputed entries from 1901 to 1920; for that region, we use data from 1921 to 2021. As is common in the literature on proportional rate data, we use the logit transformation of the mortality rate in the following analysis \citep[\commRVS{see, e.g.,}][]{cairns2011mortality}. The logit transformation ensures the transformed data are adequately defined. Mortality rates for each year are observed for ages ranging from 0 to 110 (and older) for both genders over time. We treat the logit transformation of the mortality rates at various ages as functional observations, as in \cite{hyndman2007robust} and \cite{Shang2017}, before applying our tests. 
	
	In Figure~\ref{fig:2a}, we display a rainbow plot of the logit-transformed age-specific female mortality for the region Île de France, along with its functional autocorrelation (ACF) plot in Figure~\ref{fig:2b}. From the rainbow plot, we observe a decreasing trend in age-specific mortality rates over the years. By examining the functional autocorrelation plot of \cite{MPR+21}, we notice a strong persistence exhibited in the data set.
	\begin{figure}[!htb]
		\centering
		\subfloat[Rainbow plot]
		{\includegraphics[width=7.5cm]{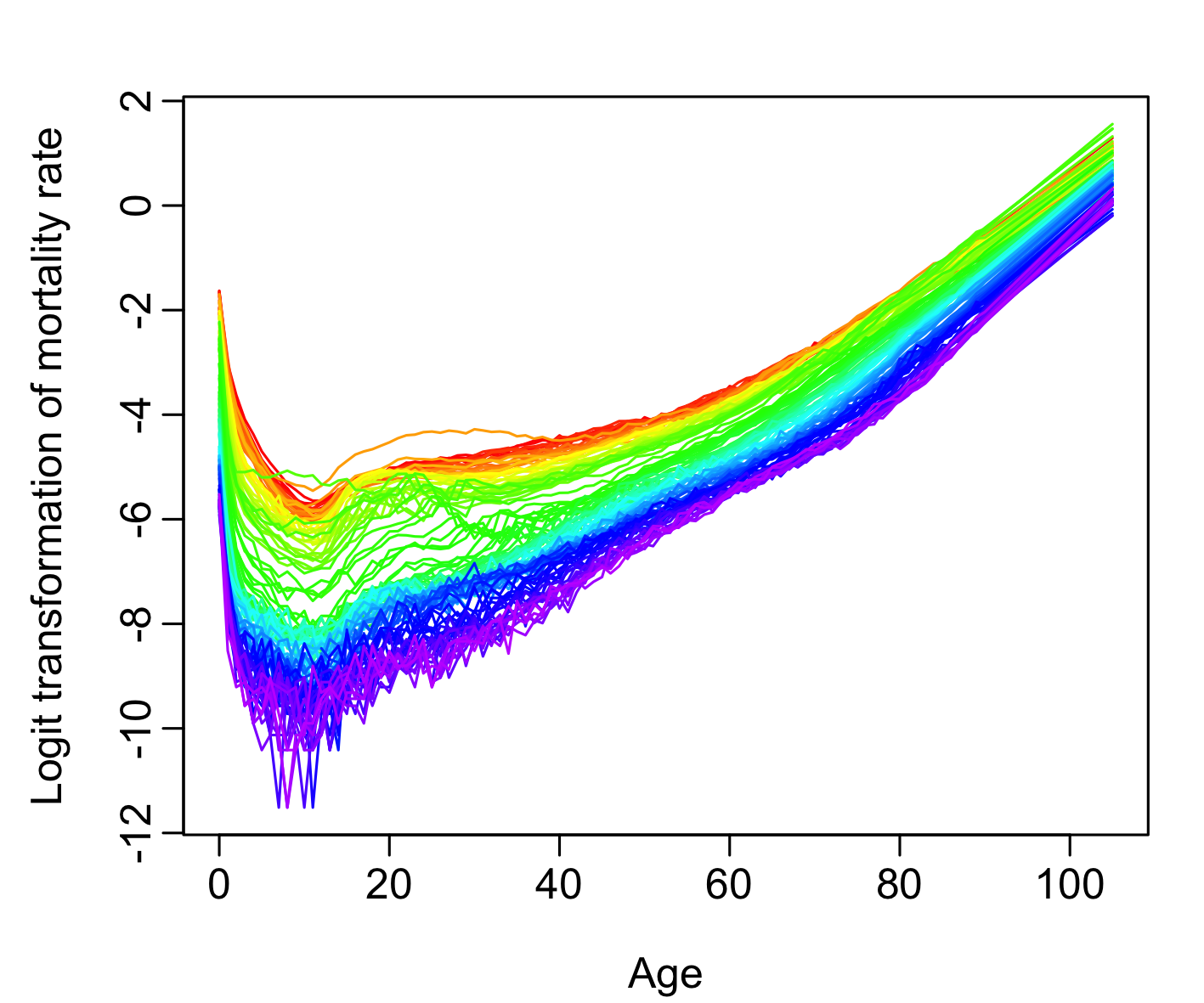}\label{fig:2a}}
		\qquad
		\subfloat[Functional ACF plot]
		{\includegraphics[width=7.5cm]{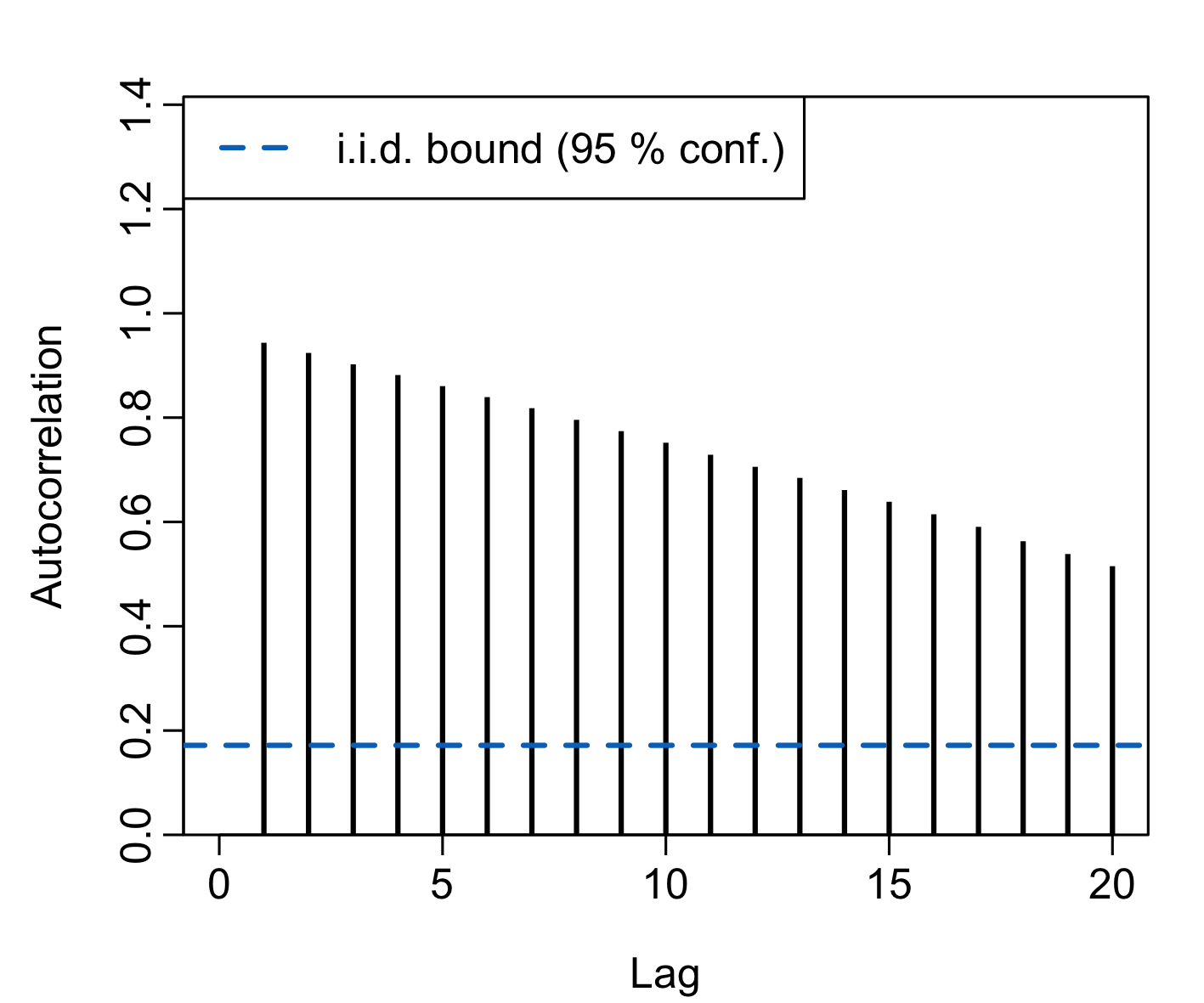}\label{fig:2b}}
		\caption{Graphical displays of the logit-transformed age-specific female mortality rates in region Île de France.}\label{fig:2}
	\end{figure}
	
	Age-specific mortality rates are often considered to be I($d$) functional time series of integer integration order $d$ (mostly $d=1$). As in Section~\ref{sec:6.1}, we assess this using the $V_0$-, $V_1$- and $V_2$-tests, covering 0, 1 and 2 as possible integer integration orders. 
	%\begin{table}[!htb]
	%\caption{Testing results for 13 subnational age- and gender-specific logit-mortality rates}\label{tab:regions}
	%\centering
	%\begin{tabular}{@{}llcc@{}}
	%\toprule
	%\textbf{No.} & \textbf{Region}  &\textbf{$d_1$ for female data \,\,} &\textbf{$d_1$ for male data} \\
	%\midrule
	%1  & Île de France 				& 	$\in (1,2)$ & $1$ \\
	%2  & Centre-Val de Loire			&  	$1$ 		& $1$ \\
	%3  & Bourgogne-Franche-Comté 	&  	$1$ 		& $1$  \\
	%4  & Normandie					&  	$1$ 		& $1$ \\
	%5  & Hauts-de-France			&  	$1$ 		& $\in (0,1)$ \\
	%6  & Grand Est					& 	$1$ 		& $1$ \\
	%7  & Pays de la Loire				&	$1$ 		& $1$ \\
	%8  & Bretagne					&	$1$ 		& $1$ \\
	%9  & Nouvelle-Aquitaine			&	$1$ 		& $1$ \\
	%10 & Occitanie 					&	$1$ 		& $1$ \\
	%11 & Auvergne-Rhône-Alpes		&	$1$ 		& $1$ \\
	%12 & Provence-Alpes-Côte d’Azur	& 	$1$ 		& $1$ \\
	%13 & Corse 					&	$1$ 		& $\in (0,1)$ \\
	%\bottomrule
	%\end{tabular}
	%\end{table}
	
	\begin{table}[!htb]
		\centering
		\caption{Testing results for 22 sub-national  age- and gender-specific logit-mortality rates}\label{tab:regions}
		\begin{tabular*}{1\linewidth}{@{}ll@{\extracolsep{\fill}}cc@{}}
			\toprule
			{No.} & {Region} &{$d_1$ for female data \,\,} &\commRV{$d_1$ for male data} \\
			\midrule
			1  & Île de France 				&$\in (1,2)$	& \commRV{$1$} \\
			2  & Centre-Val de Loire 			&$1$			& \commRV{$1$} \\
			3  & Bourgogne  				&$1$			& \commRV{$1$} \\
			4  & Franche-Comté  			&$1$			& \commRV{$1$} \\
			5  & Basse-Normandie 			&$1$			& \commRV{$1$} \\
			6  & Haute-Normandie 			&$1$			& \commRV{$1$} \\
			7  & Nord-Pas-de-Calais			&$\in (1,2)$	& \commRV{$1$} \\
			8  & Picardie 					&$1$			&\commRV{$1$} \\
			9  & Alsace 					&$1$			& \commRV{$1$} \\
			10 & Champagne-Ardenne 		&$1$			& \commRV{$1$} \\
			11 & Lorraine  					&$1$			& \commRV{$1$} \\
			12 & Pays de la Loire 			&$1$			& \commRV{$1$}  \\
			13 & Bretagne 					&$1$			& \commRV{$1$}  \\
			14 & Aquitaine 					&$1$			& \commRV{$1$}  \\
			15 & Limousin 					&$1$			& \commRV{$1$}  \\
			16 & Poitou-Charentes			&$1$			&\commRV{$1$} \\
			17 & Languedoc-Roussillon 		&$1$			& \commRV{$1$} \\
			18 & Midi-Pyrénées 				&$1$			& \commRV{$1$} \\
			19 & Auvergne					&$1$			&\commRV{$1$} \\
			20 & Rhône-Alpes 				&$1$			& \commRV{$1$} \\
			21 & Provence-Alpes-Côte d’Azur	&$1$			& \commRV{$\in (1,2)$} \\
			22 & Corse 					&$1$			& \commRV{$1$} \\
			\bottomrule
		\end{tabular*}
		\begin{minipage}{1\linewidth}
			\footnotesize{Notes : \commRV{For the 1st and 7th regions of female data and the 21th region of male data, we conducted the $V_2$-test and found that the tests are rejected in the lower tail, and concluded that $d_1\in (1,2)$.}}
		\end{minipage}
	\end{table}
	
		For each region, we apply our testing procedure to identify the order of integration of age-specific mortality rates for each gender, and the test results are reported in Table~\ref{tab:regions}. For most regions, our testing procedure concludes that the considered time series is I$(1)$, which may be understood as evidence supporting the popularity of using the random walk with drift as the forecasting technique in \citeauthor{LC92}'s \citeyearpar{LC92} model. %For some regions, the test results suggest fractional integration orders exceeding one (Île de France and Nord-Pas-de-Calais for female data) or falling below one (five regions for male data). However, there is no case where I($0$)-ness is supported in the considered dataset. This finding is consistent with the existing empirical evidence of nonstationarity or long-range dependence in a time series of age-specific mortality rates.
\revlab{r1_major4_2}\commRV{For some regions, the test results suggest fractional integration orders exceeding one (Île de France and Nord-Pas-de-Calais for female data and Provence-Alpes-Côte d’Azur for male data). However, there is no case where I($0$)-ness is supported in the dataset. This finding is consistent with the existing empirical evidence of nonstationarity or persistence in a time series of age-specific mortality rates.}

\revlab{r1_major4_1}\commRV{One caveat is that the sample size in this empirical example is not very large; we used 121 samples for each of 21 regions except Alsace which has a smaller sample size due to issues in data entries. Even though our simulation results show that the proposed tests perform reasonably even with a small sample size, similar to this case, practitioners must be aware of possible inaccuracy resulting from the limited sample size. }

\begin{remarks}\normalfont \label{remaddr1}
%We, in the empirical study, consider the logit-mortality rates, since the raw mortality rates must take nonnegative values and are also bounded above by one, and in this case, there are mathematical difficulties in modeling these time series as nonstationary (fractionally) integrated linear processes; see, e.g., \cite{Beare2017} and \cite{SEO2019} concerning density-valued time series with integer integration. There are some recent articles where other transformations are employed (see, e.g., \citealp{Shang2017}), and practitioners may prefer using untransformed mortality rates, possibly for better interpretability in practice. We thus consider other cases (probit-mortality, log-mortality, and untransformed mortality) in Section \ref{sec_add_num} of the Supplementary Material; the findings in that section are consistent with the existing empirical evidence of nonstationarity or long-range dependence in time series of age-specific mortality rates.
\commRV{In the empirical study, we use logit-mortality rates (e.g., \cite{CBD+11}), since raw mortality rates are nonnegative and bounded above by one, which poses mathematical challenges for modeling them as nonstationary (fractionally) integrated linear processes (see, e.g., \citealp{Beare2017,SEO2019}). Alternative transformations have been considered in recent studies (e.g., \citealp{Shang2017}), and practitioners may prefer untransformed rates for better interpretability. Accordingly, we also examine log-mortality \citep{LC92}, untransformed mortality \citep{LNC11}, and probit-mortality in Section \ref{sec_add_num1} of the Supplementary Material. As expected, the test results are not invariant to the employed transformation, but the results are overall consistent with existing empirical evidence of nonstationarity or long-range dependence in age-specific mortality rates.}

\end{remarks}

	\section{Concluding remarks}\label{sec:7}
	
	In this paper, statistical tests for examining if a functional time series of interest is integer-integrated or fractionally-integrated are proposed, and these are extended to a sequential testing procedure, which can be used to identify the integration order or an interval including it in a unified framework. Given recent interest in curve-valued time series exhibiting strong persistence, we believe that the proposed methods are useful for practitioners. To illustrate our methodology and its practical relevance, we apply the proposed tests to the Canadian yield curve data and the French sub-national age-specific mortality data. The results overall support the widespread assumption of I(1) for such time series in the literature, although evidence of fractional integration is also observed in certain cases.
	
	There are a few ways in which the methodology presented here can be useful in practice, and we briefly mention two. First, our hypothesis testing procedure can assist in identifying the admissible range of fractional integration orders of functional time series, which, in turn, may help in estimating the fractional order, a crucial component for statistical inference on such time series;  \commRVS{see, e.g.,} \cite{li2020local,LRS} concerning the local Whittle estimator of the fractional order for functional time series.  Second, to illustrate, we considered yield curve and mortality datasets; however, the developed testing procedure can also be applied to various economic/statistical functional time series, which are often assumed to be integer-integrated, and can provide statistical evidence either in support of or against this assumption in favor of a fractional integration order. Some existing examples in the recent literature include earning densities (\citealp{chang2016}) and their transformations (\citealp{seo2020functional}), age-specific employment rates (\citealp{Nielsen2023}), and high-frequency financial time series, such as cumulative intraday returns (\citealp{gabrys2010tests}) and intraday records of a volatility index (\citealp{shang2019intraday}).

	  %The R code for the simulation experiments and data analyses is available online at \url{https://github.com/wonkiseo86/Testing_Integer_Integration}.
	%\appendix
\appendix

\section*{Supplementary Material}

\section{Preliminaries}\label{sec_lem}
To reduce notational burden, for any linear operator $A_T$ and a real number $B_T$ depending on $T$, we write $A_T = O_p(B_T)$ (resp.\ $A_T = o_p(B_T)$) if $\|A_T\|_{\mathcal L_{\mathcal H}}$ is $O_p(B_T)$ (resp.\ $o_p(B_T))$, whenever it is convenient, as in \citet[Supplementary Material]{seo2020functional}. Moreover, we hereafter let 
\begin{align}\notag
\widehat{\mathcal K}_0 = {\mathcal K}_T\asd{\mathbf{Y}_0}, \quad  \widehat{\mathcal K}_1 ={\mathcal K}_T\asd{\mathbf{Y}_1}, \quad \widehat{\Lambda}_0={\Lambda}_T\asd{\mathbf{Y}_0}, \quad \widehat{\Lambda}_1={\Lambda}_T\asd{\mathbf{Y}_1}. 
\end{align}

We collect some preliminary results, which will be repeatedly used in the subsequent discussion, as follows:
\begin{lemmas} \label{lem1} Suppose that Assumption~\ref{assum1} holds. Then the following hold:
\begin{enumerate}[(i)]
	%		\item \label{lem1a} $\|{\mathcal K}_T\asd{\mathbf{Y_0}} \|_{\mathcal L_{\mathcal H}}$ and $\|P_1{\mathcal K}_T\asd{\mathbf{Y_0}}  P_1\|_{\mathcal L_{\mathcal H}}$ are $O_p(T^{2+2d_1}).$ 
\item \label{lem1a} $\|\widehat{\mathcal K}_0 \|_{\mathcal L_{\mathcal H}}$ and $\|P_1\widehat{\mathcal K}_0 P_1\|_{\mathcal L_{\mathcal H}}$ are $O_p(T^{2+2d_1}).$ 
\item \label{lem1b} % The dominant eigenvector $\hdd$ of ${\mathcal K}_T\asd{\mathbf{Y_0}}$ converges to a random vector $\htd$ taking values in $\mathcal H_1$ in the sense that  $\|\hdd- \sgn(\langle \hdd,\htd \rangle) \hdd\| \overset{p}{\to}  0$. More specifically, $\|\hdd - \sgn(\langle \hdd,\htd \rangle)\htd\| = O_p(T^{-(d_1-d_2)})$.
 The dominant eigenvector $\hdd$ of $\widehat{\mathcal K}_0$ converges to a random vector $\htd$ taking values in $\mathcal H_1$ in the sense that  $\|\hdd- \sgn(\langle \hdd,\htd \rangle) \hdd\| \overset{p}{\to}  0$. More specifically, $\|\hdd - \sgn(\langle \hdd,\htd \rangle)\htd\| = O_p(T^{-(d_1-d_2)})$.
\end{enumerate}  
\end{lemmas}

\begin{proofs}[Proof of Lemma~\ref{lem1}]
We first show (i). Since $d_1 >-1/2$, under Assumption~\ref{assum1}, $\sum_{s=1}^t X_s$ is a Type-II fractionally-integrated process of integration order $\widetilde{d}_1 = 1+d_1 >1/2$. Then from Theorem 3.1 of \cite{LRS} and similar arguments used in its proof, we deduce that $T^{-2\widetilde{d}_1}\langle \widehat{\mathcal K}_0v_1,v_2  \rangle$ converges to a well defined limit. \commRVS{Particularly, if} $v_1,v_2 \in \ran P_1$, it converges to an almost surely nonzero (random) constant, but if either of $v_1$ or $v_2$ is contained in $[\ran P_1]^\perp$, it decays to zero.  From these results, % (and also similar arguments used in the proof of Theorem 3.1 of \citealp{LRS}), 
we find that $\|\widehat{\mathcal K}_0\|_{\mathcal L_{\mathcal H}} = O_p(T^{2\widetilde{d}_1})$ and $\|P_1\widehat{\mathcal K}_0P_1\|_{\mathcal L_{\mathcal H}} = O_p(T^{2\widetilde{d}_1})$, from which the \commRVS{desired} results immediately follow.  
		
Given that $\sum_{s=1}^t P_j X_s$ is a Type-II fractionally-integrated process of integration order $\widetilde{d}_j = d_j+1>1/2$, we deduce from similar arguments used in Proposition 2.1 and Theorem 3.1 of \cite{LRS} that $ \sum_{s=1}^t P_j X_s = O_p(T^{\widetilde{d}_j-1/2})$ uniformly in $t=1,\ldots,T$, and thus find that $\|P_j \widehat{\mathcal K}_0 P_k\|_{\mathcal L_{\mathcal H}}= O_p(T^{\widetilde{d}_j+\widetilde{d}_k})$ for all $j, k \in \{1,\ldots,K\}$. Combining this result with~\ref{lem1a} and the fact that $d_1>d_2>\ldots>d_K$, we find that 
\begin{equation*}
\|T^{-2\widetilde{d}_1}(\widehat{\mathcal K}_0 - P_1\widehat{\mathcal K}_0P_1)\|_{\mathcal L_{\mathcal H}} = O_p(T^{-(d_1-d_2)}).  
\end{equation*}
From nearly identical arguments used in the proof of Proposition~3.2 of \cite{CKP16}, we further find that (a) the dominant eigenvector $\hdd$ of $\widehat{\mathcal K}_0$ converges to the dominant eigenvector $\htd$ of $P_1\widehat{\mathcal K}_0 P_1$, which is random (and also dependent on $T$) but takes values in $\mathcal H_1 = \ran P_1$ with probability one and (b) $\|\hdd - \sgn(\langle \hdd,\htd \rangle)\htd \| = O_p(T^{-(d_1-d_2)})$. 
\end{proofs}

\section{Proofs}\label{app_section_proof}

	%and also use $\hdd$ (resp.\ $\hd$) to denote $\hdd\asd{\mathbf{Y}_0}$ or $\hdd\asd{\mathbf{Y}_1}$ (resp.\ $\hd\asd{\mathbf{Y}_0}$ or $\hd\asd{\mathbf{Y}_1}$), introduced in Lemma \ref{lem1}, depending on the context, whenever there is no risk of confusion.
\begin{proofs}[Proof of Proposition~\ref{prop1}]
From Lemma~\ref{lem1}, we know that $T^{-2} \widehat{\mathcal K}_{0} = T^{-2}P_1  \widehat{\mathcal K}_{0}  P_1 + o_p(1)$.  This implies that 
\begin{equation}\notag %\label{pfadd01}
T^{-2} \langle \widehat{\mathcal K}_{0}\hd,\hd \rangle =T^{-2} \langle P_1 \widehat{\mathcal K}_{0} P_1\hd,\hd\rangle  + o_p(1).
\end{equation}
Since $d_1=0$, we find that
\begin{align}\notag 
T^{-2}P_1 \widehat{\mathcal K}_{0}  P_1 = \frac{1}{T} \sum_{t=1}^T \left(\frac{1}{\sqrt{T}}\sum_{s=1}^t  P_1v_{1,s}\right)  \otimes  \left(\frac{1}{\sqrt{T}}\sum_{s=1}^t  P_1v_{1,s}\right). 
\end{align}
When $d_1=0$ and the summability condition imposed on $\{\psi_{1,j}\}_{j\geq 0}$ is given, $\{ P_1v_{1,t}\}_{t\geq 1}$ becomes the so-called $L^4$-$m$-approximable sequence (Proposition 2.1 of \citealp{hormann2010}). We then deduce from Theorem 1.1 of \cite{berkes2013weak}  and Theorem 2.1 of \cite{horvath2014test} that $\sup_{0\leq r \leq 1}\|T^{-1/2}\sum_{t=1}^{\lfloor Tr \rfloor} P_1v_{1,t} - \mathcal W(r)\| = o_p(1)$, where $\mathcal W(r)$ is a $\mathcal H$-valued Brownian motion whose covariance operator is given by $P_1\Lambda_{v_1}P_1$. Moreover, using similar arguments used in the proof of Lemma 1 of \cite{NSS23}, we may also assume that  $\|T^{-2}{P}_1 \widehat{\mathcal K}_{0} P_1 - \int \mathcal W(r)\otimes \mathcal W(r) dr\|_{\mathcal L_{\mathcal H}} \overset{p}{\to}  0$. Given that $P_1\Lambda_{v_1}P_1$ is the covariance operator of  $\mathcal W$ and $\hd \in \mathcal H_1$, we  find that 
\begin{equation}\label{eqprop1a}
T^{-2} \langle \widehat{\mathcal K}_{0}\hd,\hd \rangle \overset{d}{\to}  \int \langle \mathcal W(r),\hd \rangle^2 \overset{d}{=}  \langle \Lambda_{v_1} \hd,\hd \rangle \int W(r)^2 dr.
\end{equation} 
From~\eqref{eqprop1a}, the desired result is established. 
		
Now suppose that $d_1<0$. Then, we know from Lemma~\ref{lem1}\ref{lem1a} that $T^{-2}\widehat{\mathcal K}_0 = o_p(1)$, from which we find that $T^{-2} \langle \widehat{\mathcal K}_{0}\hd,\hd \rangle \overset{p}{\to}  0$. On the other hand, in the case where $d_1>0$, we may deduce from Theorem 3.1 of \cite{LRS} and Lemma~\ref{lem1}\ref{lem1a} %\ref{lem1a} and \ref{lem1b}
that  $T^{-2-2d_1} \langle \widehat{\mathcal K}_{0}\hd,\hd \rangle$ converges to a well defined  nondegenerate limit, from which, $T^{-2} \langle \widehat{\mathcal K}_{0}\hd,\hd \rangle \overset{p}{\to}  \infty$ is deduced as desired. 
\end{proofs}
	
\begin{proofs}[Proof of Proposition~\ref{prop2}]
		%Note that $\rho_j$ is the long-run variance of $\langle P_D^Mv_t, h_j\rangle$, where $h_j$ is the eigenvector corresponding to the $j$-th largest eigenvalue of the limiting operator of $T^{-2d}\widehat{\mathcal K}$. 
Noting that $\widehat{\Lambda}_0$ is self-adjoint nonnegative definite, we deduce from Assumptions~\ref{assum1}, \ref{assum2} and~\ref{assumv1} that $q^{-2d_1}\|\widehat{\Lambda}_0\|_{\mathcal L_{\mathcal H}} = O_p(1)$. 	Since $\|\hdd - \sgn(\langle \hdd,\hd \rangle)  \hd\| \overset{p}{\to}  0$ (see Lemma~\ref{lem1}), we find that  $\langle\widehat{\Lambda}_{0}\hdd,\hdd\rangle  = \langle P_1\widehat{\Lambda}_{0}P_1\hd,\hd\rangle + o_p(1)$ when $d_1=0$. Moreover, under the summability condition on $\{\psi_{1,j}\}_{j\geq1}$  given by Assumption~\ref{assum1}, $\{P_1X_t\}_{t\geq 1}$ is the so-called $L^4$-$m$-approximable sequence and $ P_1\widehat{\Lambda}_0P_1$ is its sample long-run covariance. From Theorem~2 of \cite{horvath2013estimation}, it is deduced that  $\|P_1\widehat{\Lambda}_0P_1-\Lambda_{v_1}\|_{\mathcal L_{\mathcal H}} \overset{p}{\to}  0$. We thus find that  
\begin{equation}\label{addproof1}
\langle \widehat{\Lambda}_0\hdd,\hdd \rangle \overset{p}{\to}  \langle \Lambda_{v_1}\hd,\hd \rangle.
\end{equation} 
On the other hand, we note that 
\begin{equation}\label{addproof2}
\langle T^{-2} \widehat{\mathcal K}_{0} \hdd,\hdd \rangle=  \langle T^{-2}P_1 \widehat{\mathcal K}_{0}  P_1 \hd,\hd \rangle + o_p(1)  \overset{d}{\to}  \int \langle \mathcal W(r),\hd \rangle^2 \overset{d}{=}  \langle \Lambda_{v_1}\hd,\hd \rangle \int W(r)^2 dr.
\end{equation}
Thus the desired result when $d_1=0$ follows from~\eqref{addproof1} and~\eqref{addproof2}.
		
We then consider the case where $d_1\neq 0$. $V_0$ can be written as
\begin{equation} \label{eqpf01revision}
V_0 = ((T/q)^{-2d_1}q^{-2d_1}\langle\widehat{\Lambda}_{0}\hdd,\hdd\rangle)^{-1} \langle T^{-2-2d_1} \widehat{\mathcal K}_0 \hdd, \hdd\rangle, 
\end{equation}
where $T^{-2-2d_1} \langle \widehat{\mathcal K}_0 \hdd, \hdd\rangle$ can be shown to converge to a well defined nondegenerate limit under the employed assumptions,  \revlab{r2major3_1}\commRV{using Lemma \ref{lem1}\ref{lem1a} and similar arguments used in Theorem 3.1 of \cite{LRS}; more specifically, 
\begin{equation}\label{eqrevision0}
T^{-2-2d_1} \langle \widehat{\mathcal K}_0 \hdd, \hdd\rangle  \overset{d}{\to}  \int \langle \mathcal W_{d_1}(r), \hd \rangle^2, 
\end{equation}
where $\mathcal W_{d_1}$ is an $\mathcal H$-valued Type-II fractional Brownian motion, defined as $\mathcal W_{d_1} = \frac{1}{\Gamma(d_1+1)} \int_{0}^r(r-s)^{d_1} d\mathcal W(s)$, with $\mathcal W$ defined earlier in the proof of Proposition~\ref{prop1}; see, e.g., \cite{LRS} and \citet[Section 5.2]{hassler2018time}, noting the difference in the meaning of the parameter $d_1$ in their papers). %Due to the properties of $\mathcal W_{d_1}$ and $\hd$, the limit in \eqref{eqrevision0} is almost surely positive.
    To establish the desired properties of the test, we thus study the limiting behavior of $(T/q)^{-2d_1}q^{-2d_1}\langle\widehat{\Lambda}_{0}\hdd,\hdd\rangle$ when $d_1>0$ or $d_1<0$.} Using the results given in Lemma~\ref{lem1}, we deduce that $q^{-2d_1}\langle\widehat{\Lambda}_{0}\hdd,\hdd\rangle = q^{-2d_1}\langle P_1\widehat{\Lambda}_{0}P_1\hd,\hd\rangle  + O_p(q^{-2d_1}T^{-(d_1-d_2)})$. From this result and Assumptions~\ref{assum1}, \ref{assum2} and~\ref{assumv1}, it is deduced that $q^{-2d_1}\langle\widehat{\Lambda}_{0}\hdd,\hdd\rangle = q^{-2d_1}\langle P_1\widehat{\Lambda}_{0}P_1\hd,\hd\rangle + o_p(1) \overset{p}{\to}  \commRVS{\tilde{c}_{1,h}}$ holds under either of (a) $d_1>0$ or (b) $d_1<0$ (since $O_p(q^{-2d_1}T^{-(d_1-d_2)})=o(1)$ under Assumption~\ref{assumv1}\ref{assumv1b}), where  $\commRVS{\tilde{c}_{1,h}}$ is a positive random constant whose randomness results from \commRVS{the fact} that $\hd$ is random. %; \commRV{if $\widehat{d}$ is fixed at $h \in \mathcal{H}_1$, its definition coincides with $c_{1,h}$ in Assumption~\ref{assumv1}.}  
Combining this result with the fact that $T/q \to \infty$, \revlab{r2major3_2}\commRV{we find that 
\begin{equation}\label{eqaddrevision}
(T/q)^{-2d_1}q^{-2d_1}\langle\widehat{\Lambda}_{0}\hdd,\hdd\rangle = (T/q)^{-2d_1} (\tilde{c}_{1,h}+o_p(1));
\end{equation} this converges in probability to zero if $d_1>0$ (and thus $V_0 \overset{p}{\to}  \infty$) while it diverges to infinity if $d_1<0$ (and thus $V_0 \overset{p}{\to}  0$)} %$(T/q)^{-2d_1}q^{-2d_1}\langle\widehat{\Lambda}_{0}\hdd,\hdd\rangle \overset{p}{\to}  \infty$ if $d_1<0$  .
\end{proofs}
	
\begin{proofs}[Proof of Proposition~\ref{prop2a}]
Note that the test statistic can be written as
\begin{equation} \label{eqpf01a0}
(T^{-2d_1}\langle\widehat{\Lambda}_{0}\hdd,\hdd\rangle)^{-1} \langle T^{-2-2d_1} \widehat{\mathcal K}_0 \hdd, \hdd\rangle. 
\end{equation}
As discussed in our proof of Proposition~\ref{prop2}, $T^{-2-2d_1} \langle \widehat{\mathcal K}_0 \hdd, \hdd\rangle$ converges to a well defined nondegenerate limit. From Assumptions~\ref{assum1}, \ref{assum2}, similar arguments used in the proof of Theorem 3.1 of \cite{LRS} and the condition~\eqref{eqwnonstat}, it can be deduced that $T^{-2d_1}\sum_{t=1}^T X_{t}  \otimes  X_{t} = T^{-2d_1}\sum_{t=1}^T P_1X_{t}  \otimes  P_1X_{t} + o_p(1) = O_p(1)$ if $d_1>1/2$  while $(T\log T)^{-1}\sum_{t=1}^T X_{t}  \otimes  X_{t} = (T\log T)^{-1}\sum_{t=1}^T P_1X_{t}  \otimes  P_1X_{t} + o_p(1) =O_p(1)$ if $d_1=1/2$. From the construction of $\widehat{\Lambda}_0$, we may also deduce that $\widehat{\Lambda}_0 = O_p(qT^{2d_1-1})$ if $d_1>1/2$ and $\widehat{\Lambda}_0 = O_p(q \log T )$ if $d_1=1/2$; that is, $\widehat{\Lambda}_0 = O_p(q \max\{T^{2d_1-1},\log T\})$ for any $d_1 \geq 1/2$.  This in turn implies that $T^{-2d_1}\langle\widehat{\Lambda}_{0}\hdd,\hdd\rangle = O_p(\max\{qT^{-1}, qT^{-2d_1}\log T\})$, which converges to zero given that $d_1\geq 1/2$ and $T/q \to \infty$ (Assumption~\ref{assumv1}\ref{assumv1b}) hold. From these findings, we know that the test statistic in \eqref{eqpf01a0} diverges to infinity. 
\end{proofs}
	
\begin{proofs}[Proof of Proposition~\ref{prop2add}] 
In the case where $d_1=1$, we find from only a slight modification of the arguments used in our proof of Proposition~\ref{prop2} that 
\begin{equation*}%\label{eqprop1aadd}
T^{-2} \langle \widehat{\mathcal K}_{1}\hdd,\hdd \rangle \overset{d}{\to}  \int \langle \mathcal W(r),\hd \rangle^2 \overset{d}{=}  \langle \Lambda_{v_1} \hd,\hd \rangle \int W(r)^2 dr.
\end{equation*}
Since $\|\hdd - \sgn(\langle \hdd,\hd \rangle)  \hd\| \overset{p}{\to}  0$, we note that $\langle \widehat{\Lambda}_1\hdd,\hdd\rangle =  \langle P_1\widehat{\Lambda}_1P_1\hd,\hd\rangle +o_p(1)$ (see Lemma~\ref{lem1}). Note that $P_1\widehat{\Lambda}_1P_1$ is the sample long-run covariance of $\{P_1\Delta X_t\}_{t\geq 1}$, which is $L^4$-$m$-approximable under the conditions employed in Assumption~\ref{assum1}, and thus  $\|P_1\widehat{\Lambda}_1P_1-\Lambda_{v_1}\|_{\mathcal L_{\mathcal H}} \overset{p}{\to}  0$ (\citealp{horvath2013estimation}, Theorem~2). From these results, the desired result when $d_1=1$ is obtained. 
		
We next consider the case where  $d_1 \in (1/2,3/2) \setminus \{1\}$. We write the test statistic as follows: 
\begin{equation*} %\label{eqpf01adds}
V_1=((T/q)^{-2(d_1-1)}q^{-2(d_1-1)}\langle\widehat{\Lambda}_{1}\hdd,\hdd\rangle)^{-1} \langle T^{-2-2(d_1-1)} \widehat{\mathcal K}_1 \hdd, \hdd\rangle, 
\end{equation*}
where $\langle T^{-2-2(d_1-1)} \widehat{\mathcal K}_1 \hdd, \hdd\rangle$ converges to a well defined nondegenerate limit as in the previous proofs. We observe that $\|\hdd - \sgn(\langle \hdd,\hd \rangle)  \hd\| = O_p(T^{-(d_1-d_2)})$, and thus $q^{-2(d_1-1)}\langle\widehat{\Lambda}_{1}\hdd,\hdd\rangle = q^{-2(d_1-1)}\langle P_1{\Lambda}_{1}P_1\hd,\hd\rangle + O_p(q^{-2(d_1-1)} T^{-(d_1-d_2)})$ (see Lemma~\ref{lem1}). From similar arguments used in our proof of Proposition~\ref{prop1}, we find that $(T/q)^{-2(d_1-1)}q^{-2(d_1-1)}\langle\widehat{\Lambda}_{1}\hdd,\hdd\rangle \overset{p}{\to}  0$ if $d_1 \in (1,3/2)$\commRVS{, while} $(T/q)^{-2(d_1-1)}q^{-2(d_1-1)}\langle\widehat{\Lambda}_{1}\hdd,\hdd\rangle \overset{p}{\to}  \infty$ if $d_1\in(1/2,1)$, from which the desired results follow. 
		
In the case where $d_1 \geq 3/2$, it can be shown that $V_1 \overset{p}{\to} \infty$ using nearly identical arguments used in our proof of Proposition~\ref{prop2a}. 
\end{proofs}
\begin{proofs}[Proof of Proposition~\ref{prop4}] 
First consider the case with $d_1\in (-1/2,1/2)$. Then the test statistic can be written as
\begin{equation} \label{eqpf01a}
(T\langle\widehat{\Lambda}_{1}\hdd,\hdd\rangle)^{-1} \langle T^{-1} \widehat{\mathcal K}_1 \hdd, \hdd\rangle.
\end{equation}
Let $y_t = \langle X_{t}, \hdd \rangle$. We find that 
\begin{equation}\label{eqpf01b}
\langle T^{-1} \widehat{\mathcal K}_1 \hdd, \hdd\rangle = \frac{1}{T} \sum_{t=2}^T (y_t-y_1)^2 =  \frac{1}{T} \sum_{t=2}^T y_t^2 + y_1^2 + o_p(1) =  \langle \widehat{C}_0\hdd,\hdd \rangle  + y_1^2 + o_p(1),
\end{equation}
where $\widehat{C}_0 = T^{-1}\sum_{t=2}^T X_t \otimes X_t$. On the other hand, from similar algebra used in the proof of Theorem~4 of \citet[][pp.\ 234--5]{CHO2015217}, we find that 
\begin{align*}
\langle\widehat{\Lambda}_{1}\hdd,\hdd\rangle =&  O_p(qT^{-1}) +  \left(\frac{1}{T}\sum_{t=2}^{T-1}(2-2w(1,q))y_t^2\right) \\
& + \sum_{r=2}^{q} \left(\frac{1}{T}\sum_{t=r+1}^{T-1}2(2w(r-1,q)-w(r-2,q)-w(r,q))y_ty_{t-r+1}\right) \\
& + \left(\frac{1}{T}\sum_{t=2}^{T-1}2(2w(q,q)-w(q-1,q))y_ty_{t-q}\right) - \frac{1}{T}\sum_{t=q+2}^{T}2w(q,q)y_ty_{t-q-1}.
\end{align*}
We observe, as in \cite{CHO2015217}, that $w(r-1,q)-w(r-2,q)-w(r,q)=0$ for every $r=1,\ldots,q$ and also find that  $2w(q,q)-w(q-1,q)=0$.  From a little bit of algebra, we find that
\begin{align}
q\langle\widehat{\Lambda}_{1}\hdd,\hdd\rangle  &= O_p(q^2T^{-1})+ \frac{q}{T}\sum_{t=2}^{T-1}(2/(q+1))y_t^2 - \frac{q}{T}\sum_{t=q+2}^{T}(2/(q+1))y_ty_{t-q-1}  \notag\\ 
&= O_p(q^2T^{-1}) + \frac{2}{T}\sum_{t=2}^{T-1}y_t^2- \frac{2}{T}\sum_{t=q+2}^{T}y_ty_{t-q-1}  =   o_p(1)+ {2}(\langle \widehat{C}_0 -  \widehat{C}_{q+1},\hdd,\hdd \rangle.  \label{eq001} 
			%\\ & =o_p(1) + 2(\langle {C}_0\hdd,\hdd \rangle- \langle {C}_{q+1}\hdd,\hdd \rangle).
\end{align}
Under Assumption~\ref{assum2}, we find that $\langle \widehat{C}_0 -  \widehat{C}_{q+1},\hdd,\hdd \rangle = \langle {C}_0 -  {C}_{q+1},\hdd,\hdd \rangle + o_p(1) = \langle {C}_0,\hd,\hd \rangle  + o_p(1)$ for large enough $T$ and $q$, where $\hd$ is a unit-norm vector taking values in $\mathcal H_1$ (see Lemma~\ref{lem1}).  Since $C_0$ is positive definite on a finite dimensional subspace $\mathcal H_1$, we may write $P_1C_0P_1 = \sum_{j=1}^{\dim(\mathcal H_1)} a_j f_j \otimes f_j$ with $a_{\dim(\mathcal H_1)}>0$. We then observe that 
\begin{align}\label{eqpfadd01}
\langle {C}_0,\hd,\hd \rangle \geq \min_{h\in \mathcal H_1,\|h\|=1}\langle {C}_0,h,h \rangle \geq a_{\dim(\mathcal H_1)}.
\end{align}
Combining this result with~\eqref{eqpf01a}, \eqref{eq001} and the fact that $T/q \to \infty$, we find that 
\begin{equation}\label{eqpf01c}
T \langle\widehat{\Lambda}_{1}\hdd,\hdd \rangle \overset{p}{\to}  \infty.
\end{equation}
From~\eqref{eqpf01a}, \eqref{eqpf01b} and \eqref{eqpf01c}, we find the test statistic converges to zero if $-1/2<d_1<1/2$.
		
To establish the desired result for the case with $d_1=1/2$,  we write the inverse of the test statistic as follows:
\begin{equation} \label{eqpf01add}
V_1^{-1}=	d_T \langle\widehat{\Lambda}_{1}\hdd,\hdd\rangle ( d_T\langle 	T^{-2}\widehat{\mathcal K}_1 \hdd, \hdd\rangle)^{-1}.
\end{equation}
		%	From Assumption \ref{assum3b} and Lemma \ref{lem1}, we first observe that $\langle\widehat{\Lambda}_{1}\hdd,\hdd\rangle = \langle P_1\widehat{\Lambda}_{1}P_1 \hd,\hd\rangle + O_p(T^{-(d_1-d_2)})$. 	
		%	$(T\ln T)^{-1}\sum_{t=1}^T \langle P_jX_t,h_j\rangle \langle P_kX_t,h_k\rangle = O_p(1)$ holds for every $h_j \in \ran P_j$ and $h_k \in \ran P_k$, where $j, k \in \{1,\ldots,K\}$. This means that $d_T\langle\widehat{\Lambda}_{1}\hdd,\hdd\rangle = d_T\langle P_1\widehat{\Lambda}_{1}P_1\hd,\hd\rangle+ O_p(d_T T^{-(d_1-d_2)})$, which is bounded away from zero with probablity approaching one.	
We first observe that  
\begin{align}
\langle (T\log T)^{-1} \widehat{\mathcal K}_1 \hdd, \hdd\rangle &= \frac{1}{T \log T} \sum_{t=2}^T ({y}_t-{y}_1)^2  \notag  \\
& =\frac{1}{T \log T} \sum_{t=2}^T {y}_t^2- \frac{2}{T \log T} {y}_1\sum_{t=2}^T {y}_t + \frac{1}{\log T} {y}_1^2. \label{eqpfadd001} 
\end{align}
Note that under Assumption~\ref{assumv3} we have $(T\log T)^{-1}\sum_{t=1}^T y_t =  O_p(1 / \sqrt{\log{T}})$, and thus, for large enough $q$ and $T$, 
\begin{equation}%\label{eqpfadd002} 
\frac{1}{T \log T} {y}_1\sum_{t=2}^T {y}_t + \frac{1}{\log T} y_1^2 = O_p(1 / \sqrt{\log{T}}),
\end{equation} and 
\begin{align}\label{eqpfadd003} 
\frac{1}{T \log T} \sum_{t=2}^T {y}_t^2 &= \frac{1}{T\log T} \sum_{t=1}^T \langle X_t,\hdd\rangle^2  = \frac{1}{\log T}\langle P_1\widehat{C}_0P_1\hd,\hd \rangle + o_p(1) =  O_p(1) . %\\ &  = \frac{1}{\log T}\langle P_1\widehat{C}_0P_1\hdd,\hdd \rangle + o_p(1)\\ & \geq \frac{1}{\log T} \min_{\|h\|=1, h \in \mathcal H_1}\langle P_1\widehat{C}_0P_1h,h \rangle + o_p(1). 
\end{align}
Combining \eqref{eqpfadd001}-\eqref{eqpfadd003}, we find that $d_T T^{-2}\widehat{\mathcal K}_1= O_p(d_TT^{-1}\log T)$. Since $\widehat{\Lambda}_1$ is self-adjoint nonnegative definite, we find from Assumptions~\ref{assumv3} that $d_T \|\widehat{\Lambda}_{1}\|_{\mathcal L_{\mathcal H}} = O_p(1)$ and thus  $d_T\langle\widehat{\Lambda}_{1}\hdd,\hdd\rangle = d_T \langle P_1\widehat{\Lambda}_{1}P_1 \hd,\hd\rangle + o_p(1)$, which is bounded away from zero for any $h\in \mathcal H_1$ with probability approaching one. Since $d_T T^{-1}\log T \to 0$, we thus find that the inverse statistic considered in \eqref{eqpf01add} diverges to infinity, implying that $V_1 \overset{p}{\to}  0$.  
		%	Thus the first  term asymptotically dominates the other terms. Thus the first  term asymptotically dominates the other terms, and it diverges to infinity under the employed assumptions. This implies that \eqref{eqpf01add} converges to zero, meaning that ${V}_1 \to_p 0$. %Furthermore, note that $\langle P_1\widehat{C}_0P_1h,h\rangle = T^{-1}\sum_{t=1}^T \langle P_1X_t,h\rangle^2$, which is almost surely greater than or equal to the minimal eigenvalue of $P_1\widehat{C}_0P_1$, viewed as a linear operator acting on $\mathcal H_1$. Since $P_1\widehat{C}_0P_1$ is the sample covariance operator of the I($1/2$) time series of $P_1X_t$, we know from the properties given by Condition \ref{assum1a} that the minimal eigenvalue of $P_1\widehat{C}_0P_1$ (viewed as an operator acting on $\mathcal H_1$) diverges in probability to infinity.  Observe that the time series of $P_1X_t$ is I($1/2$) described by Condition \ref{assum1a}. This implies that \eqref{eqpf01add} converges to zero, meaning that ${V}_1 \to_p 0$.
\end{proofs}

\begin{proofs}[Proof of Proposition~\ref{prop2add2}] 
Let $\widetilde{\Lambda}_0=\widetilde{\Lambda}(\mathbf{Y}_0)$ and $\widetilde{\mathcal K}_0=\widetilde{\mathcal K}(\mathbf{Y}_0)$ for notational simplicity. With slight modifications of our proofs of Lemma~\ref{lem1}, the following can be shown: (a) $\|\widetilde{\mathcal K}_0 \|_{\mathcal L_{\mathcal H}}=O_p(T^{2+2d_1})$, (b) $\|P_1\widetilde{\mathcal K}_0 P_1\|_{\mathcal L_{\mathcal H}}=O_p(T^{2+2d_1})$, (c) $\hddd$ converges to a random vector $\htd$ taking values in $\mathcal H_1$ and (d) $\|\hddd - \sgn(\langle  \hddd,\htd\rangle)\htd\| = O_p(T^{-(d_1-d_2)})$. 
		
First consider the case where $d_1 \in (-1/2, 1/2)$. We know from Assumptions~\ref{assum1}  and~\ref{assum2} that, for $d_1$ in the considered range,  $T^{-1}\sum_{s=1}^T P_1X_s = O_p(T^{{d}_1-1/2})$, and hence $T^{-1}\sum_{s=1}^T X_s = O_p(T^{{d}_1-1/2})$. Using this result, we find that
\begin{align*}%\label{eqlrv2add}
\widetilde{\Lambda}_0 % &\coloneqq  T^{-1} \sum_{s=-q}^q w(s,q) \sum_{t_0 \leq t, t-s \leq T} (X_t-\bar{X}_T)  \otimes  (X_{t-s}-\bar{X}_T), 	\\&
&=T^{-1} \sum_{s=-q}^q w(s,q) \sum_{t:1 \leq t, t-s \leq T} \{X_t\otimes  X_{t-s} - \bar{X}_T \otimes X_{t-s} - X_{t}   \otimes \bar{X}_T + \bar{X}_T \otimes  \bar{X}_T  \},\\
&= T^{-1} \sum_{s=-q}^q w(s,q) \sum_{t:1 \leq t, t-s \leq T} X_t\otimes  X_{t-s} + O_p(q T^{2d_1-1}).
\end{align*}
Thus,
\begin{equation}\label{eqfnal}
q^{-2d_1}\widetilde{\Lambda}_0  = q^{-2d_1}\widehat{\Lambda}_0  +  O_p((T/q)^{2d_1-1}).
\end{equation}
Since $T/q \to \infty$, $O_p((T/q)^{2d_1-1}) = o_p(1)$  for any $d_1 \in (-1/2,1/2)$, we have $q^{-2d_1}\widetilde{\Lambda}_0  = q^{-2d_1}\widehat{\Lambda}_0 + o_p(1)$. 
		% $q^{-2d_1} \|\widetilde{\Lambda}_{0} -P_1 \widehat{\Lambda}_{0}P_1\|_{\op} \to_p 0$. 
		%From this result, we also find from similar arguments used in our proof of Lemma \ref{lem1} that  $\|\hdd-\hd\| \to_p 0$ for a possibly random vector $\hd$ taking values in $\mathcal H_1$. 
From similar arguments and the existing asymptotic results used in our proof of Proposition~\ref{prop1}, we deduce that 
\begin{equation*}
\langle T^{-2} \widetilde{\mathcal K}_{0} \hddd,\hddd \rangle=  \langle T^{-2}P_1 \widetilde{\mathcal K}_{0}  P_1 \hddd,\hddd \rangle + o_p(1)  \overset{d}{\to}  % \int \langle \widetilde{\mathcal W}(r),\htd \rangle^2 	=
 \langle \Lambda_{v_1}\htd,\htd \rangle \int \widetilde{W}(r)^2 dr.
\end{equation*}
Then the desired result when $d_1=0$ immediately follows from \commRVS{the fact} that $\langle \widetilde{\Lambda}_0\hddd,\hddd \rangle =\langle \widehat{\Lambda}_0\hddd,\hddd \rangle + o_p(1)$, as shown above, and that $ \langle \widehat{\Lambda}_0\hddd,\hddd \rangle  \overset{p}{\to}  \langle \Lambda_{v_1}\htd,\htd \rangle$ as we earlier established in our proof of Proposition~\ref{prop2}. 
		
In the case where  $d_1 \in (-1/2, 1/2)\setminus \{0\}$, we write the test statistic as follows:
\begin{equation*} %\label{eqpf01addadd}
((T/q)^{-2d_1}q^{-2d_1}\langle\widetilde{\Lambda}_{0}\hddd,\hddd\rangle)^{-1} \langle T^{-2-2d_1} \widetilde{\mathcal K}_0 \hddd, \hddd\rangle. 
\end{equation*}
With some obvious modification of our proofs of Proposition~\ref{prop2}, we find that $\langle T^{-2-2d_1} \widetilde{\mathcal K}_0 \hddd, \hddd\rangle$ converges to a well defined nondegenerate limit. Moreover, $(T/q)^{-2d_1}q^{-2d_1}\langle\widetilde{\Lambda}_{0}\hddd,\hddd\rangle = (T/q)^{-2d_1}q^{-2d_1}\langle\widehat{\Lambda}_{0}\hddd,\hddd\rangle + o_p(1)$ (see \eqref{eqfnal} and note that $T/q \to \infty$), which  converges to zero (resp.\ diverges to infinity) if $d_1 \in (0,1/2)$ (resp.\ $d_1\in (-1/2,0)$), as earlier shown in our proof of Proposition~\ref{prop2}.   From these results, the desired results when $d_1 \in (-1/2, 1/2)\setminus\{0\}$ follow. % $\|\widetilde{\mathcal K}_0 \|_{\mathcal L_{\mathcal H}}$ and $\|P_1\widetilde{\mathcal K}_0  P_1\|_{\mathcal L_{\mathcal H}}$ are $O_p(T^{2+2d_1\asd{\mathbf{Y}_0}})$ and (ii) the dominant eigenvector $\hdd$ of $\widetilde{\Lambda}_0$ converges to a possibly random vector $\hd$ taking values in $\mathcal H_1$, i.e., $\|\hdd- P_1\hd\| \to_p 0$. Combining these results, it is straightforward to show that \eqref{eqconver2add} holds when $d_1\asd{\mathbf{Y}_0} < 1/2$. 
		
To show that $\widetilde{V}_0 \overset{p}{\to}  \infty$ when $d_1 \geq 1/2$, we use nearly identical arguments used in our proof of Proposition~\ref{prop2a}, noting that, under the employed assumptions, $T^{-2d_1}\sum_{t=1}^T (X_{t}-\bar{X}_T)  \otimes  (X_{t}-\bar{X}_T) = O_p(1)$ if $d_1>1/2$  while $(T\log T)^{-1}\sum_{t=1}^T (X_{t}-\bar{X}_T)  \otimes  (X_{t}-\bar{X}_T) = O_p(1)$ if $d_1=1/2$. 
\end{proofs} 
	
	%\begin{align}
	% \langle q(T\log T)^{-1} \widehat{\mathcal K}_1 \hdd, \hdd\rangle &=  q \langle (T\log T)^{-1} \widehat{\mathcal K}_1 \hd, \hd\rangle + o_p(q) \\&= q \left(\frac{1}{T \log T} \sum_{t=2}^T (\widetilde{y}_t-\widetilde{y}_1)^2 + o_p(1)\right) \\ & 
	%=\frac{q}{T \log T} \sum_{t=2}^T \widetilde{y}_t^2- \frac{2q}{T \log T} \widetilde{y}_1\sum_{t=2}^T \widetilde{y}_t + \frac{q}{\log T} \widetilde{y}_1^2.
	%\end{align}
	%For large enough $q$ and $T$, the first term asymptotically dominates the other terms, and it diverges to infinity under Assumption \ref{assum3a}.

\section{Detailed discussion on Remarks~\ref{remprimitive} and~\ref{remprimitive2}}\label{sec_app_abadir}
	
	A set of primitive sufficient conditions for Assumption~\ref{assumv1}\ref{assumv1a} (and also a similar condition used in Proposition~\ref{prop2add2}) can be deduced from \cite{ABADIR200956}. For notational simplicity, let $x_{j,t} \coloneqq \langle X_t,h_j \rangle$ for $h_j \in \mathcal H_j$ (see Assumption~\ref{assum1}) and let $f_j$ be the spectral density of $x_{j,t}$. Suppose that 
	\begin{equation}\label{eqapp01}
		f_j(\omega) = a_j|\omega|^{-2d_j} + o(|\omega|^{-2d_j}) \quad \text{as $\omega \to 0$},
	\end{equation}
	where $a_j > 0$ and $d_j \in (-1/2,1/2)$ as considered in Assumption~\ref{assum2}. If $d_j<0$, the following is further assumed: 
	\begin{equation}\label{eqapp01add}
		f_j(\omega) \leq C|\omega|^{-2d_j} \quad \text{for $\omega \in [-\pi, \pi]$}.
	\end{equation}
	Define the lag-$s$ autocovariance of $x_{j,t}$ as $\gamma_{j}(s)$ (i.e., $\gamma_{j}(s)=Cov(x_{j,t},x_{j,t+s})$) and let $\kappa(s,u,r)$ be the \commRVS{fourth-order} cumulant defined by $\kappa_j(s,u,r)=E[x_tx_{t+s}x_{t+u}x_{t+r}]-[\gamma_j(s)\gamma_j(u-r)+\gamma_j(u)\gamma_j(s-r)+\gamma_j(r)\gamma_j(s-u)]$. As in Assumption M of \citet[][p.57]{ABADIR200956}, we consider the following conditions: 
	\begin{align} \label{eqapp02}
		\gamma_j(s) \sim c s^{2d_j-1} \quad \text{if $d_j\neq 0$}, \quad \sum_{s=-\infty}^\infty |\gamma_j(s)| < \infty \quad \text{if $d_j= 0$},
	\end{align}
	and 
	\begin{align}\label{eqapp03}
		\sum_{s,u,r=-\infty}^\infty |\kappa_j(s,u,r)| \leq C \quad \text{if $d_j< 0$}, \quad \sup_{s}\sum_{u,r=-n}^n \kappa_j(s,u,r)| \leq Cn^{2d_j} \quad \text{if $d_j\geq 0$}.
	\end{align}
	Then from Theorem 2.2 of \cite{ABADIR200956}, Assumptions~\ref{assum1}, the condition $q=o(T^{1/2})$ and  \eqref{eqapp01} -- \eqref{eqapp03},  we find that  both $q^{-2d_j}\langle \widehat{\Lambda}_0 h_j,h_j \rangle$ and $q^{-2d_j}\langle \widetilde{\Lambda}_0 h_j,h_j \rangle$ converge to $a_j g(d_j)$, where 
\begin{align}
		g(d_j) = \begin{cases}
			\frac{2\Gamma(1-2d_j)\sin(\pi d_j)}{d_j (1+2d_j)} & \text{if $d_j \neq 0$},\\
			2\pi& \text{if $d_j = 0$,}
		\end{cases}
\end{align} 
where $\Gamma(\cdot)$ is the gamma function.

\section{\commRV{Additional numerical results}}\label{sec_add_num}
\subsection{\commRV{Supplementary results for Section \ref{sec:6.2}}}\label{sec_add_num1}
\commRV{We report additional testing results using the age- and gender-specific French sub-regional mortality dataset. Instead of applying the logit transformation of mortality rates, as in Section~\ref{sec:6.2} \citep[see also][]{CBD+11}, we consider alternative transformations commonly used in the literature, including the logarithmic transformation \citep{LC92} and the probit transformation \citep{DM07}, as well as the untransformed mortality rates \citep{LNC11}. To the best of our knowledge, this set of transformations encompasses most empirical studies on the dynamics of age-specific mortality rates. }

\subsubsection{Untransformed mortality rates}\label{mor_trans1}
\commRV{
We apply our test to the original mortality rates without any transformation.  A caveat is that modeling functional time series subject to a nonnegativity constraint (i.e., $Y_t(u) \geq 0$ for all $u$ in the domain) as nonstationary (fractionally) integrated linear processes entails mathematical difficulties, as noted by \cite{Beare2017} and \cite{SEO2019} in the context of density-valued time series with integer integration. Nonetheless, test results for such time series may be of practical value to some degree, since practitioners often prefer raw mortality rates for interpretability. For this reason, we also consider the untransformed mortality rates in this section. The results are reported in Table~\ref{tab:regions_supp1}.}
	\begin{table}[!htb]
		\centering
		\caption{Testing results for age- and gender-specific mortality rates -- without transformation}\label{tab:regions_supp1}
		\commRV{	\begin{tabular*}{1\linewidth}{@{}ll@{\extracolsep{\fill}}cc@{}}
			\toprule
			{No.} & {Region} &{$d_1$ for female data \,\,} &\commRV{$d_1$ for male data} \\
			\midrule
			1  & Île de France 				&$1$	& $1$ \\
			2  & Centre-Val de Loire 			&$1$			&  $1$  \\
			3  & Bourgogne  				&$1$			&  $1$  \\
			4  & Franche-Comté  			&$1$			&  $1$ \\
			5  & Basse-Normandie 			&$\in(0,1)$	&  $1$  \\
			6  & Haute-Normandie 			&$1$			&  $1$  \\
			7  & Nord-Pas-de-Calais			&$\in(0,1)$	& $\in(0,1)$ \\
			8  & Picardie 					&$\in(0,1)$	&$\in(0,1)$ \\
			9  & Alsace 					&$1$			& $\in(0,1)$ \\
			10 & Champagne-Ardenne 		&$1$			&  $1$ \\
			11 & Lorraine  					&$1$			&  $1$ \\
			12 & Pays de la Loire 			&$1$			&  $1$  \\
			13 & Bretagne 					&$1$			&  $1$  \\
			14 & Aquitaine 					&$1$			&  $1$ \\
			15 & Limousin 					&$\in(0,1)$	& $\in(0,1)$ \\
			16 & Poitou-Charentes			&$1$			& $1$ \\
			17 & Languedoc-Roussillon 		&$1$			&  $1$ \\
			18 & Midi-Pyrénées 				&$1$			&  $1$ \\
			19 & Auvergne					&$1$			& $1$ \\
			20 & Rhône-Alpes 				&$1$			&  $1$ \\
			21 & Provence-Alpes-Côte d’Azur	&$1$			& $1$ \\
			22 & Corse 					&$1$			& $\in(0,1)$\\
			\bottomrule
		\end{tabular*}}
		\begin{minipage}{1\linewidth}
		%	\footnotesize{Notes : \commRV{For the 1st and 7th regions of female data and the 21th region of male data, we conducted the $V_2$-test and found that the tests are rejected in the lower tail, and concluded that $d_1\in (1,2)$.}}
		\end{minipage}
	\end{table}\\
\commRV{
The results are similar to those for the logit-transformed mortality rates in that our tests conclude the time series are I$(1)$ for most regions. They differ, however, in that the time series for a few regions in each gender are found to be fractionally integrated of order less than one. Overall, the results support the empirical finding that age-specific mortality rates are persistent (nonstationary or long-range dependent).}

\subsubsection{Log-mortality rates}\label{mor_trans2}
	
\commRV{In Table~\ref{tab:regions_supp2}, we now apply our tests to the logarithmically transformed data, which is a common and popular choice in the literature; see, e.g., \cite{Shang2017} and the references therein. When the results are compared to those for the logit-transformed mortality rates, the test outcomes are generally similar, but more cases of fractional integration with order exceeding one are observed (Basse-Normandie, Alsace, Bretagne, and Provence-Alpes-Côte d’Azur for female data, and Île-de-France for male data). For Alsace, where the sample size is smallest due to many missing and heavily imputed entries from 1901–1920 (see Section \ref{sec:6.2}), our testing procedure concludes that the integration order is 2; the $p$-value for the null hypothesis of I(2) is approximately 14.8\%, which is computed using the sample quantiles from 200,000 Monte Carlo realizations of the limiting distribution generated by the standard simulation method.}
\begin{table}[!htb]
\centering
\caption{Testing results for age- and gender-specific mortality rates -- log transformation}\label{tab:regions_supp2}
\commRV{	\begin{tabular*}{1\linewidth}{@{}ll@{\extracolsep{\fill}}cc@{}}
			\toprule
			{No.} & {Region} &{$d_1$ for female data \,\,} &{$d_1$ for male data} \\
			\midrule
			1  & Île de France 				&$\in (1,2)$	& $\in (1,2)$ \\
			2  & Centre-Val de Loire 			&$1$			&$1$ \\
			3  & Bourgogne  				&$1$			& $1$ \\
			4  & Franche-Comté  			&$1$			& $1$\\
			5  & Basse-Normandie 			&$\in (1,2)$	& $1$ \\
			6  & Haute-Normandie 			&$1$			& $1$ \\
			7  & Nord-Pas-de-Calais			&$\in (1,2)$	& $1$ \\
			8  & Picardie 					&$1$			&$1$ \\
			9  & Alsace 					&$2$			&$1$ \\
			10 & Champagne-Ardenne 		&$1$			& $1$ \\
			11 & Lorraine  					&$1$			& $1$ \\
			12 & Pays de la Loire 			&$1$			& $1$  \\
			13 & Bretagne 					&$\in (1,2)$	& $1$  \\
			14 & Aquitaine 					&$1$			&$1$  \\
			15 & Limousin 					&$1$			& $1$  \\
			16 & Poitou-Charentes			&$1$			&$1$ \\
			17 & Languedoc-Roussillon 		&$1$			&$1$ \\
			18 & Midi-Pyrénées 				&$1$			& $1$ \\
			19 & Auvergne					&$1$			&$1$\\
			20 & Rhône-Alpes 				&$1$			&$1$ \\
			21 & Provence-Alpes-Côte d’Azur	&$\in (1,2)$	& {$\in (1,2)$} \\
			22 & Corse 					&$1$			& {$1$} \\
			\bottomrule
\end{tabular*}}
\begin{minipage}{1\linewidth}
		%	\footnotesize{Notes : \commRV{For the 1st and 7th regions of female data and the 21th region of male data, we conducted the $V_2$-test and found that the tests are rejected in the lower tail, and concluded that $d_1\in (1,2)$.}}
\end{minipage}
\end{table}

\subsubsection{Probit-mortality rates}\label{mor_trans3}

\commRV{We also applied our proposed tests following the application of the probit transformation, which, analogous to the logit transformation, maps any variable with bounded range $[0,1]$ into $\mathbb{R}$. In this context, for every region and gender, the tests indicate that the time series are integer-integrated of order one; consequently, the results are not presented in tabular form. The outcomes are broadly consistent with those obtained using logit-transformed mortality rates. Whereas a few instances of fractional integration exceeding order one are observed for logit-transformed data, all series in the probit-transformed case are classified as integer-integrated of order one. Overall, when compared to results based on log-mortality rates or untransformed mortality rates, the findings for probit-transformed data most closely resemble those for logit-transformed rates. This may reflect the similarity between these two transformations.}

\subsection{Additional simulation results and discussion}\label{app_additional_simul}
\commRV{We provide additional discussion and simulation results that supplement those presented in Section~\ref{sec_sim}.}
\subsubsection{Simulation results for different choices of $q$} \label{app_additional_simula}
\commRV{
In our proof of Proposition~\ref{prop2}, we show that $(T/q)^{-2d_1}V_0$ converges to a positive random variable for $d_1$ near $0$ (more specifically, $d_1 \in (-1/2,1/2)$; see \eqref{eqpf01revision} and \eqref{eqaddrevision}), from which we deduce that
\begin{equation}\label{eqappenadd}
V_0 = O_p((T/q)^{2d_1}) \quad \text{and} \quad V_0^{-1} = O_p((T/q)^{-2d_1});
\end{equation}
see Remark \ref{remappadd1} for more details.
From this result, we can further investigate how the choice of $q$, used to compute the sample long-run covariance ${\Lambda}_T\asd{\mathbf{Y}_{0}}$, affects the power properties of the $V_0$-test. Specifically, a smaller $q$ increases the power since $V_0$ (resp.\ $V_0^{-1}$) diverges (resp.\ decays) faster when $d_1>0$ (resp.\ $d_1<0$). This theoretical finding aligns with the well-known power behavior of KPSS-type tests (e.g., \citealp{kwiatkowski1992testing,nyblom2000tests}). However, using a smaller $q$ is not always advisable because, as shown in our proof, under the null of I(0)-ness, $q$ plays a crucial role in the consistent estimation of the population long-run covariance of $P_1X_t$ (see \eqref{addproof1} and the related discussion). In particular, for more serially correlated processes, a larger $q$ is preferred. It is well established that the power gain from a smaller $q$ in KPSS-type tests comes at the cost of over-rejection of the correct null hypothesis, a trade-off that has been extensively discussed even in the recent literature on functional time series (see, e.g., \citealp{seo2020functional}, Section S4.1).

We replicate the simulation design from Section~\ref{sec_sim} and examine the properties of the $V_0$-test under two alternative choices of $q$: the first, $q = \lfloor 0.4 \log T \rfloor$, and the second, $q = \lfloor T^{1/4} \rfloor$. The first (resp.\ second) choice corresponds to a smaller (resp.\ larger) bandwidth compared to $q = \lfloor T^{1/5} \rfloor$ used in Section~\ref{sec_sim}. 

The results in Table~\ref{tabemp1add1} are obtained analogously to those in Table~\ref{tabemp1}($\subref{tabemp1a}$), but with a smaller choice of $q$. As expected, a smaller $q$ yields higher power; however, when the stationary sequence is more strongly correlated (e.g., $\mathsf{b}=0.6$), the test tends to over-reject the null of I(0)-ness compared to the weakly correlated case ($\mathsf{b}=0.15$). This over-rejection is particularly pronounced when the sample size is small ($T=125$). Although the issue becomes less severe as $T$ increases, it remains noticeable for highly correlated series. To further illustrate this, we conducted additional experiments with $\mathsf{b}=0.75$ and varying choices of $q$. The results, reported in Table~\ref{tabemp1add3}, confirm substantial over-rejection even when $T=1000$ if $q$ is small ($q = \lfloor 0.4 \log T \rfloor$). In contrast, the larger choice ($q = \lfloor T^{1/4} \rfloor$) does not exhibit severe over-rejection, even under strong serial correlation. Overall, as $q$ increases, the degree of over-rejection diminishes, though this improvement comes at the cost of lower power when the alternative holds. These findings align closely with the theoretical properties discussed in Remark~\ref{r2major3}.

\begin{remarks}\label{remappadd1}\normalfont
%The rationale behind the local-to-zero hypotheses is based on a type of sequential approximation. Suppose that we consider a sequence of $d_1$ which shrinks to zero, but with $d_1-d_2$ bounded away from zero (as required for the convergence of $\hdd$ in Lemma \ref{lem1}\ref{lem1b}), after the asymptotic approximation via $T \to \infty$ is established; of course, it may be more useful to consider an asymptotic theory when $d_1 \to 0$ and $T \to \infty$ jointly, but, to the best of the authors' knowledge, there is no requisite theory to do this in the considered setup. 
From \eqref{eqpf01revision}--\eqref{eqaddrevision}, %, \eqref{eqrevision0} and \eqref{eqaddrevision}, 
we find that \begin{equation}\label{eqapprem01}
(T/q)^{-2d_1}V_0 \overset{d}{\to} \tilde{c}_{1,h}^{-1} \int \langle \mathcal W_{d_1}(r), \hd \rangle^2,
\end{equation} 
which is (almost surely) positive ($\tilde{c}_{1,h}$ and $\mathcal W_{d_1}$ are defined in the proof of Proposition \ref{prop2}). Regardless of the value of $d_1$ in the considered range, the limit appearing in \eqref{eqapprem01} is $O_p(1)$, from which we find that $V_0 = O_p((T/q)^{2d_1})$ and also $V_0^{-1} = O_p((T/q)^{-2d_1})$. 
%Thus,  $V_0 = O_p(1)$ if $d_1$ shrinks to zero so that $(T/q)^{-2d_1} \to \mathcal C$ for some positive constant $\mathcal C$, and the local-to-zero alternatives are considered to satisfy $(T/q)^{-2d_1} = 1$ for every $T$, $q$ and $d_1$. 
\end{remarks}}

	\begin{table}[!tb]	
		\caption{\commRV{Size and correct rejection rates for the ${V}_0$-test, with smaller $q$ ($q\sim \log T$).}} \label{tabemp1add1}
		\centering
	\commRV{	\begin{tabular*}{1\linewidth}{@{\extracolsep{\fill}}l|l|ccccccc@{}}
			\thickhline
			&  \backslashbox{$T$}{$d_1$} 	& -0.45 & -0.3 & -0.15  & 0 & 0.15 & 0.3 & 0.45 \\ \hline
			$\mathsf{b}=0.15$ & $T=125$& 0.818 &0.500 &0.153& 0.051 &0.288& 0.701& 0.963 \\ 
           & $T=250$&  0.885& 0.556 &0.160 &0.037& 0.298 &0.719& 0.963\\ 
			& $T=500$\,\, &  0.970& 0.727& 0.232& 0.038& 0.364& 0.809& 0.979\\ 
			& $T=750$\,\, &0.993& 0.814& 0.288& 0.035& 0.399& 0.843& 0.990 \\
			& $T=1000$\,\, &0.999 &0.865 &0.322 &0.039& 0.428& 0.865& 0.995 \\ \hline 
			$\mathsf{b}=0.60$ & $T=125$&   0.702 &0.471& 0.236& 0.156& 0.314& 0.689 &0.951\\ 
           & $T=250$ &  0.823& 0.526& 0.184& 0.079& 0.314 &0.729& 0.964\\ 
			& $T=500$\,\, & 0.926 &0.678 &0.246& 0.077& 0.374& 0.807& 0.979\\ 
			& $T=750$\,\, &  0.960 &0.763& 0.295 &0.076& 0.414& 0.834 &0.990\\ 
			& $T=1000$\,\, &  0.980& 0.812& 0.331 &0.075 &0.432& 0.866& 0.994\\ 
			\thickhline
		\end{tabular*} }\quad \vspace{0.75em}
		\begin{minipage}{1\linewidth}
			\footnotesize{\commRV{Notes: The nominal size is $5\%$. The number of Monte Carlo replications in each case is 2000. $q = \lfloor 0.4 \log T \rfloor$.}}
		\end{minipage}
	\end{table}

	\begin{table}[!tb]	
		\caption{\commRV{Size and correct rejection rates for the ${V}_0$-test, with larger $q$ ($q\sim T^{1/4}$).}} \label{tabemp1add2}
		\centering
	\commRV{	\begin{tabular*}{1\linewidth}{@{\extracolsep{\fill}}l|l|ccccccc@{}}
			\thickhline
			&  \backslashbox{$T$}{$d_1$} 	& -0.45 & -0.3 & -0.15  & 0 & 0.15 & 0.3 & 0.45 \\ \hline
			$\mathsf{b}=0.15$ & $T=125$& 0.573 &0.283& 0.074 &0.038 &0.204& 0.563& 0.901 \\ 
           & $T=250$& 0.832& 0.470& 0.128& 0.031& 0.264 &0.659 &0.946\\ 
			& $T=500$\,\, & 0.921 &0.600& 0.168&0.035& 0.298& 0.724& 0.957\\ 
			& $T=750$\,\, &0.947 &0.666& 0.197& 0.034& 0.317 &0.753 &0.969 \\
			& $T=1000$\,\, & 0.975& 0.736& 0.226 &0.038& 0.347& 0.785 &0.983 \\ \hline 
			$\mathsf{b}=0.60$ & $T=125$&  0.541& 0.286 &0.101& 0.059& 0.220 &0.576 &0.904\\ 
           & $T=250$ &  0.781 &0.471 &0.155& 0.062& 0.281& 0.674& 0.943\\ 
			& $T=500$\,\, & 0.889 &0.591 &0.180& 0.051& 0.316& 0.732 &0.960\\ 
			& $T=750$\,\, &  0.926 &0.661 &0.212& 0.049 &0.334& 0.760 &0.963\\ 
			& $T=1000$\,\, & 0.959 &0.723& 0.244& 0.051& 0.364& 0.797 &0.981\\ 
			\thickhline
		\end{tabular*} }\quad \vspace{0.75em}
		\begin{minipage}{1\linewidth}
			\footnotesize{\commRV{Notes: The nominal size is $5\%$. The number of Monte Carlo replications in each case is 2000. $q = \lfloor T^{1/4} \rfloor$.}}
		\end{minipage}
	\end{table}

\begin{table}[!tb]	
\caption{\commRV{Size and correct rejection rates for the ${V}_0$-test, when $\mathsf{b}=0.75$.}} \label{tabemp1add3}
\centering
\commRV{	\begin{tabular*}{1\linewidth}{@{\extracolsep{\fill}}l|l|ccccccc@{}}
\thickhline
			&  \backslashbox{$T$}{$d_1$} 	& -0.45 & -0.3 & -0.15  & 0 & 0.15 & 0.3 & 0.45 \\ \hline
			$q\sim\log T$ & $T=125$& 0.659& 0.460& 0.252 &0.221& 0.341& 0.685& 0.939 \\ 
           & $T=250$& 0.761 &0.482& 0.189 &0.110& 0.339& 0.732& 0.965\\ 
			& $T=500$\,\, &0.890& 0.623& 0.254& 0.110& 0.388& 0.803 &0.978\\ 
			& $T=750$\,\, & 0.927& 0.704& 0.288& 0.106& 0.428& 0.837& 0.988 \\
			& $T=1000$\,\, & 0.949& 0.759& 0.322& 0.109& 0.450& 0.868& 0.994\\ \hline 
			$q\sim\log T^{1/5}$ & $T=125$& 0.573& 0.338& 0.133& 0.117& 0.284& 0.645& 0.929\\ 
           & $T=250$ &  0.737 &0.454 &0.178 &0.092 &0.292& 0.679& 0.946\\ 
			& $T=500$\,\, &  0.882& 0.601& 0.234 &0.090& 0.348& 0.758& 0.969\\ 
			& $T=750$\,\, &0.927& 0.689&0.281& 0.091& 0.381 &0.801& 0.982\\ 
			& $T=1000$\,\, &0.953 &0.746& 0.314& 0.096& 0.406& 0.837& 0.991\\ \hline 
	$q\sim\log T^{1/4}$ & $T=125$& 0.511& 0.296 &0.120& 0.087 &0.235& 0.582& 0.903\\ 
           & $T=250$ & 0.737 &0.454& 0.178& 0.092& 0.292 &0.679& 0.946\\ 
			& $T=500$\,\, & 0.858& 0.560& 0.190& 0.072& 0.326& 0.732& 0.959\\ 
			& $T=750$\,\, &0.903& 0.626& 0.220& 0.068 &0.342 &0.760& 0.963\\ 
			& $T=1000$\,\, &0.936& 0.688& 0.258& 0.068 &0.372& 0.799& 0.980\\ 
			\thickhline
		\end{tabular*} }\quad \vspace{0.75em}
		\begin{minipage}{1\linewidth}
			\footnotesize{\commRV{Notes: The nominal size is $5\%$. The number of Monte Carlo replications in each case is 2000. $q$ is set to $q = \lfloor 0.4\log T \rfloor$ (case $q \sim \log T$), $q = \lfloor T^{1/5} \rfloor$ (case $q \sim T^{1/5}$), or $q = \lfloor T^{1/4} \rfloor$ (case $q \sim T^{1/4}$).}}
		\end{minipage}
	\end{table}

%%%%%% WK NOTE: TENTATIVE, NEED TO CHECK, AND SUBJECT TO CHANGE. 
\subsubsection{Power properties under local-to-zero alternative hypotheses}\label{app_additional_simulb}
\commRV{It is shown that $V_0=O_p((T/q)^{2d_1})$ and $V_0^{-1}=O_p((T/q)^{-2d_1})$ (see \eqref{eqappenadd} and Remark \ref{remappadd1}) for $d_1 \in (-1/2,1/2)$. From these results, it is possible to examine how $d_1$, when near $0$, affects the power properties. Moreover, the following may be understood as a local-to-zero sequence of hypotheses, which ensures that ${V}_0$ neither diverges to infinity nor decays to zero (see Remark \ref{remappadd2}):   
\begin{equation}\label{eqlab}
H_{1,T}: d_1 =  \log_{(T/q)^2}\mathcal{C} =  \frac{\log {\mathcal{C}}}{2 \log (T/q)},
\end{equation}
where $\mathcal C > 1$ (resp.\ $\mathcal C \in (0,1)$) if $d_1>0$ (resp.\ $d_1<0$), and $\mathcal C \neq 1$ corresponds to a local deviation from the null. Under these local-to-zero hypotheses, the power properties are expected to depend crucially on $\mathcal C$ and become similar across different values of $T$ as long as $T$ is sufficiently large. This is evidenced by our simulation experiments, where we computed rejection rates for the $V_0$-test under local alternatives for various values of $\mathcal C$. The results are reported in Table~\ref{tabemp1add5} (with $q = \lfloor 0.4 \log T \rfloor$), Table~\ref{tabemp1add4} (with $q = \lfloor T^{1/5} \rfloor$), and Table~\ref{tabemp1add6} (with $q = \lfloor T^{1/4} \rfloor$), presented in order of increasing bandwidth.  Figure \ref{figadd1} provides a visualization of the rejection rates reported in Tables \ref{tabemp1add5}--\ref{tabemp1add6} for varying values of $c = 0.5 \log \mathcal{C}$; the results for $c = 0$ correspond to the simulated size of the test.

	\begin{table}[!htb]	
		\caption{\commRV{Rejection rates for the ${V}_0$-test against local alternatives ($q \sim \log T$)}} \label{tabemp1add5}
		\centering
	\commRV{	\begin{tabular*}{1\linewidth}{@{\extracolsep{\fill}}l|l|ccccccc@{}}
			\thickhline
			&  \backslashbox{$T$}{$c$} 	& -0.9 & -0.6 & -0.3  & 0 & 0.3 & 0.6 & 0.9\\ \hline
			$\mathsf{b}=0.15$ & $T=125$& 0.208& 0.100& 0.044& 0.045& 0.112& 0.224 &0.388\\ 
           & $T=250$&0.220& 0.116& 0.046& 0.034& 0.120& 0.244& 0.418\\ 
			& $T=500$\,\, & 0.270& 0.132&0.050& 0.044& 0.102 &0.228 &0.420\\ 
			& $T=750$\,\,  &0.298 &0.140 &0.053& 0.040 &0.102& 0.227& 0.444 \\
			& $T=1000$\,\, &0.292 &0.140& 0.053& 0.045 &0.100& 0.255 &0.422 \\ \hline 
			$\mathsf{b}=0.60$ & $T=125$&  0.269& 0.192& 0.148& 0.153 &0.196& 0.282 &0.414\\ 
           & $T=250$ &0.246 &0.152 &0.082& 0.077& 0.140 &0.266& 0.432\\ 
			& $T=500$\,\, &0.292 &0.152 &0.080&0.079& 0.143 &0.262& 0.424\\ 
			& $T=750$\,\, & 0.288& 0.173 &0.084 &0.078& 0.138& 0.251 &0.452\\ 
			& $T=1000$\,\, &  0.297 &0.164 &0.096& 0.078& 0.134& 0.273 &0.445\\ 
			\thickhline
		\end{tabular*} }\quad \vspace{0.75em}
		\begin{minipage}{1\linewidth}
			\footnotesize{Notes: The nominal size is $5\%$. The number of Monte Carlo replications in each case is 2000. \commRV{$c$ is defined as $0.5 \log \mathcal{C}$, appearing in Remark \ref{remrevision2}, and in the DGP, $d_1$ is set to ${c}/{\log (T/q)}$ for each sample size $T$}.}
		\end{minipage}
	\end{table}

	\begin{table}[!htb]	
		\caption{\commRV{Rejection rates for the ${V}_0$-test against local alternatives ($q \sim T^{1/5}$)}} \label{tabemp1add4}
		\centering
	\commRV{	\begin{tabular*}{1\linewidth}{@{\extracolsep{\fill}}l|l|ccccccc@{}}
			\thickhline
			&  \backslashbox{$T$}{$c$} 	& -0.9 & -0.6 & -0.3  & 0 & 0.3 & 0.6 & 0.9\\ \hline
			$\mathsf{b}=0.15$ & $T=125$&   0.197& 0.092& 0.034& 0.034& 0.105& 0.226 &0.408 \\ 
           & $T=250$&  0.213 &0.110& 0.043 &0.033 &0.117& 0.244& 0.422\\ 
			& $T=500$\,\, &0.266& 0.123& 0.046& 0.043& 0.103& 0.232 &0.426\\ 
			& $T=750$\,\, &0.292& 0.136& 0.049& 0.039 &0.101& 0.229& 0.450\\
			& $T=1000$\,\, &0.291& 0.138& 0.049 &0.045 &0.100& 0.254& 0.428 \\ \hline 
			$\mathsf{b}=0.60$ & $T=125$&  0.203 &0.126 &0.068& 0.070 &0.138& 0.264& 0.436\\ 
           & $T=250$ &0.244 &0.146& 0.068& 0.059& 0.134& 0.260& 0.432\\ 
			& $T=500$\,\, &0.289& 0.149 &0.070& 0.066& 0.128& 0.250& 0.426\\ 
			& $T=750$\,\, &0.302& 0.172& 0.074 &0.064 &0.124& 0.244 &0.454\\ 
			& $T=1000$\,\, &0.302 &0.154& 0.086 &0.068 &0.118& 0.266 &0.444\\ 
			\thickhline
		\end{tabular*} }\quad \vspace{0.75em}
		\begin{minipage}{1\linewidth}
			\footnotesize{Notes: The nominal size is $5\%$. The number of Monte Carlo replications in each case is 2000. \commRV{$c$ is defined as $0.5 \log \mathcal{C}$, appearing in Remark \ref{remrevision2}, and in the DGP, $d_1$ is set to ${c}/{\log (T/q)}$ for each sample size $T$}.}
		\end{minipage}
	\end{table}

	\begin{table}[!htb]	
		\caption{\commRV{Rejection rates for the ${V}_0$-test against local alternatives ($q \sim T^{1/4}$)}} \label{tabemp1add6}
		\centering
	\commRV{	\begin{tabular*}{1\linewidth}{@{\extracolsep{\fill}}l|l|ccccccc@{}}
			\thickhline
			&  \backslashbox{$T$}{$c$} 	& -0.9 & -0.6 & -0.3  & 0 & 0.3 & 0.6 & 0.9\\ \hline
			$\mathsf{b}=0.15$ & $T=125$& 0.172& 0.078 &0.030& 0.029& 0.100& 0.225& 0.414 \\ 
           & $T=250$& 0.213& 0.110& 0.043& 0.033 &0.117 &0.244& 0.422\\ 
			& $T=500$\,\, & 0.258&0.117 &0.044& 0.043& 0.102 &0.236& 0.425\\ 
			& $T=750$\,\, & 0.284 &0.130 &0.046 &0.037 &0.102& 0.228 &0.446 \\
			& $T=1000$\,\, &0.286 &0.132 &0.048 &0.043 &0.100& 0.256 &0.430 \\ \hline 
			$\mathsf{b}=0.60$ & $T=125$&  0.194 &0.112 &0.058& 0.053& 0.116 &0.257 &0.436\\ 
           & $T=250$ & 0.244& 0.146& 0.068 &0.059 &0.134& 0.260& 0.432\\ 
			& $T=500$\,\, &0.274& 0.134& 0.059& 0.056& 0.118& 0.248 &0.436\\ 
			& $T=750$\,\, & 0.292& 0.152& 0.061& 0.051& 0.110& 0.244& 0.464\\ 
			& $T=1000$\,\, &0.286& 0.144 &0.061& 0.051 &0.110& 0.266& 0.442\\ 
			\thickhline
		\end{tabular*} }\quad \vspace{0.75em}
		\begin{minipage}{1\linewidth}
			\footnotesize{Notes: The nominal size is $5\%$. The number of Monte Carlo replications in each case is 2000. \commRV{$c$ is defined as $0.5 \log \mathcal{C}$, appearing in Remark \ref{remrevision2}, and in the DGP, $d_1$ is set to ${c}/{\log (T/q)}$ for each sample size $T$}.}
		\end{minipage}
	\end{table}
	
\begin{remarks}\label{remappadd2}\normalfont
The rationale behind the local-to-zero hypotheses is based on \eqref{eqappenadd} (see Remark \ref{remappadd1}) and a sequential approximation. Specifically, we consider a sequence of $d_1$ that shrinks to zero with $d_1-d_2$ bounded away from zero (as required for the convergence of $\hdd$ in Lemma \ref{lem1}\ref{lem1b}), after establishing the asymptotic approximation via $T \to \infty$. While a joint asymptotic theory in which $d_1 \to 0$ and $T \to \infty$ simultaneously may be more useful, to the best of the authors' knowledge, no requisite theory exists for the setup considered, which represents a potential avenue for future research. Given \eqref{eqappenadd}, we preserve $V_0 = O_p(1)$ and $V_0^{-1} = O_p(1)$ by considering $d_1$ such that $(T/q)^{-2d_1} \to \mathcal C$ for some positive constant $\mathcal C$. The local-to-zero hypotheses are constructed so that $(T/q)^{-2d_1} = \mathcal C$ for every $T$, $q$, and $d_1$.
\end{remarks}}

\begin{figure}[!tb]
  \centering\caption{\commRV{Rejection rates for the ${V}_0$-test against local alternatives}} \label{figadd1}
 \begin{subfigure}[t]{0.48\linewidth}
    \centering
    \includegraphics[width=\linewidth]{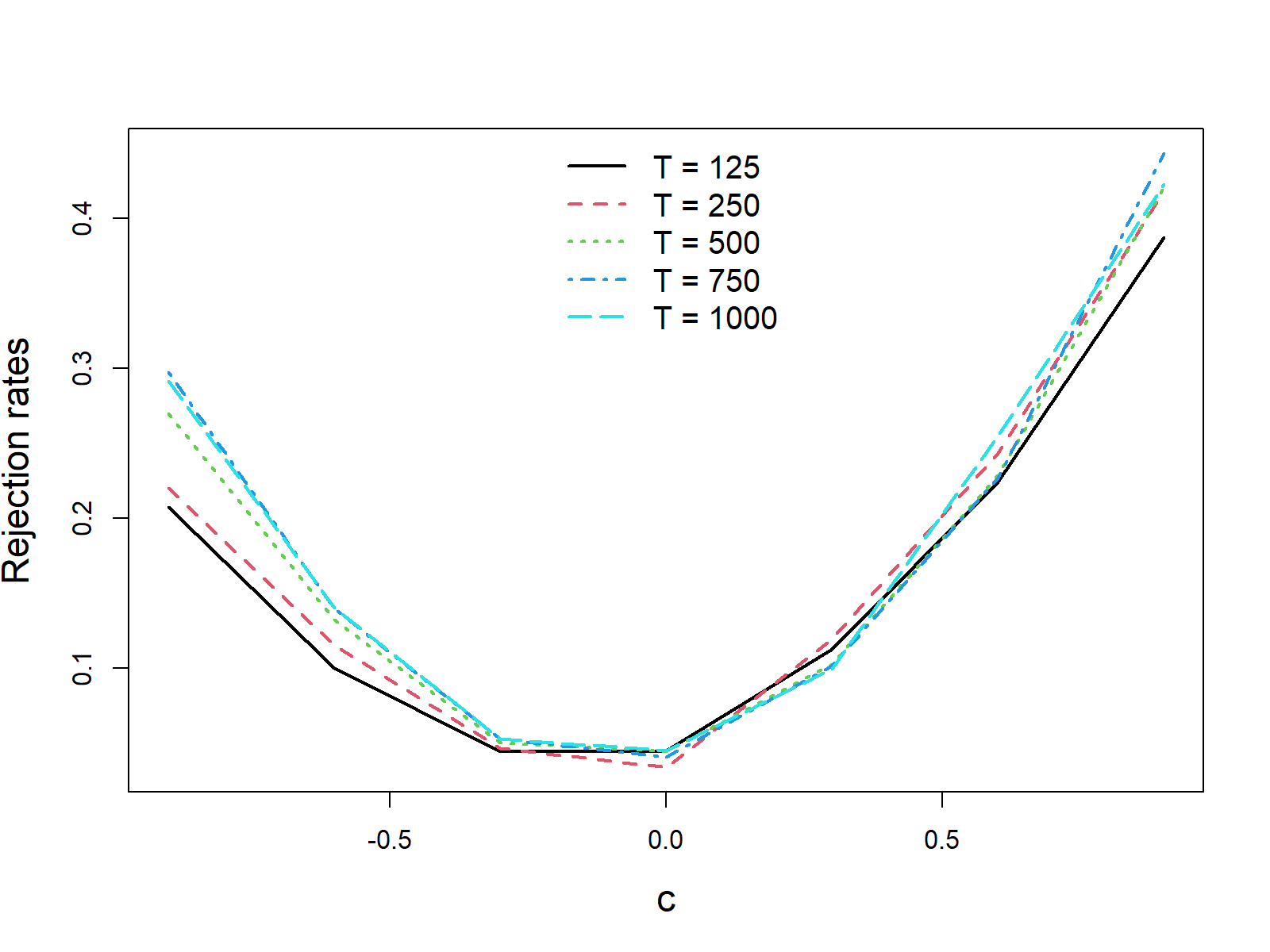}
    \caption{$\mathbf{b}=0.15$ and $q \sim \log T$}
    \label{fig:rejection_rates5}
  \end{subfigure}
  \hfill
  \begin{subfigure}[t]{0.48\linewidth}
    \centering
    \includegraphics[width=\linewidth]{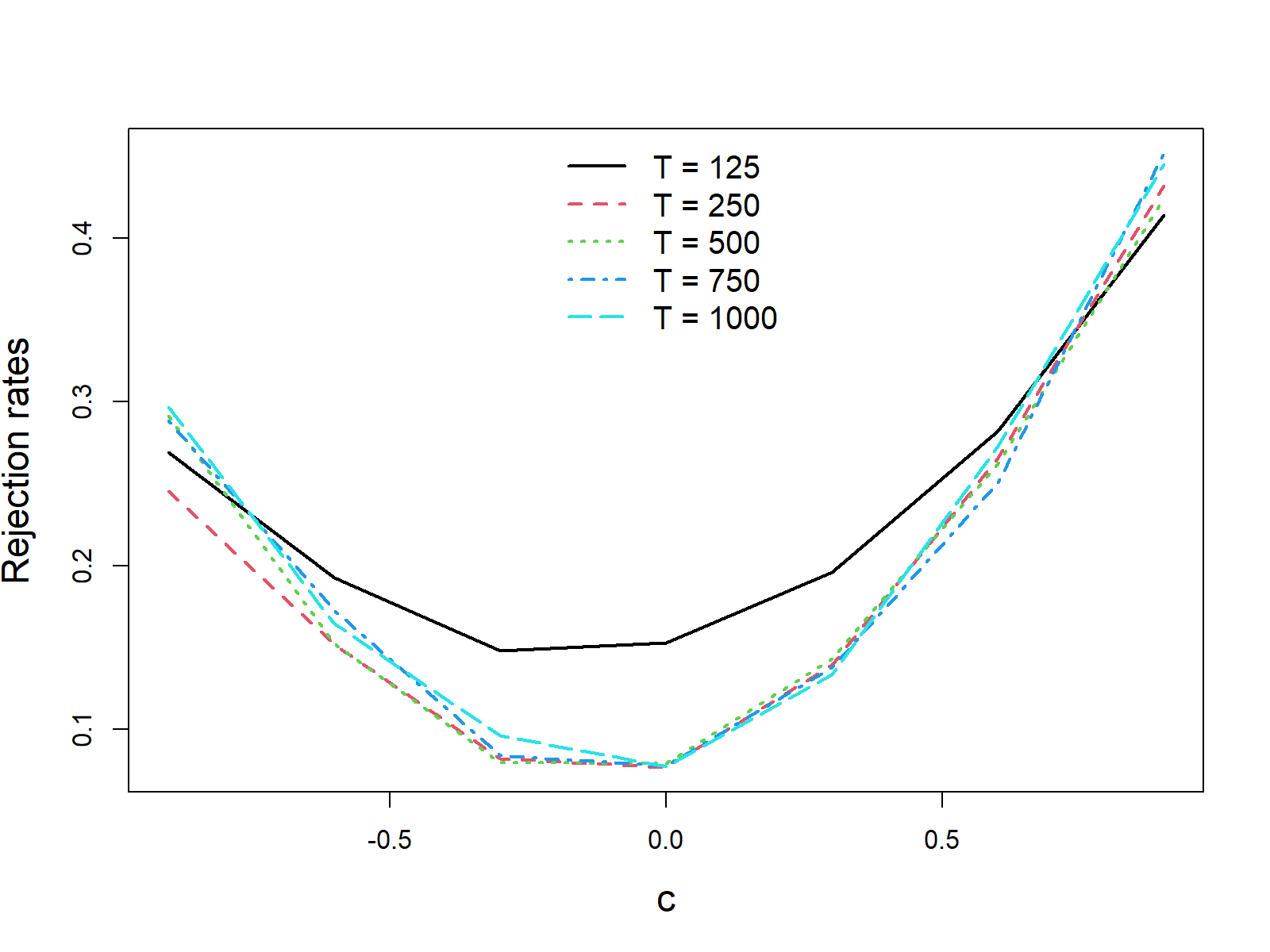}
    \caption{$\mathbf{b}=0.6$ and $q \sim \log T$}
    \label{fig:rejection_rates6}
  \end{subfigure}
 \begin{subfigure}[t]{0.48\linewidth}
    \centering
    \includegraphics[width=\linewidth]{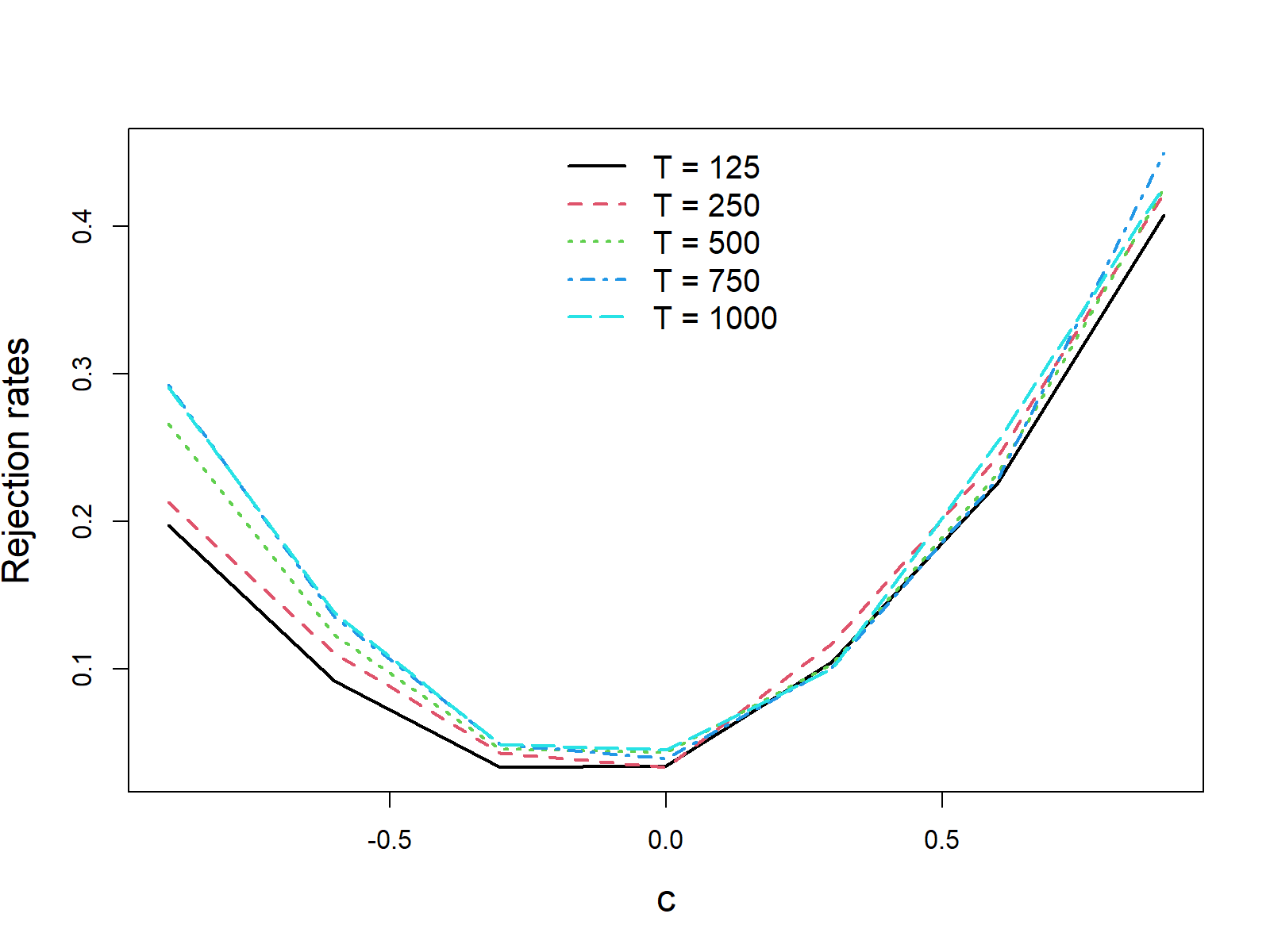}
    \caption{$\mathbf{b}=0.15$ and $q \sim T^{1/5}$}
    \label{fig:rejection_rates1}
  \end{subfigure}
  \hfill
  \begin{subfigure}[t]{0.48\linewidth}
    \centering
    \includegraphics[width=\linewidth]{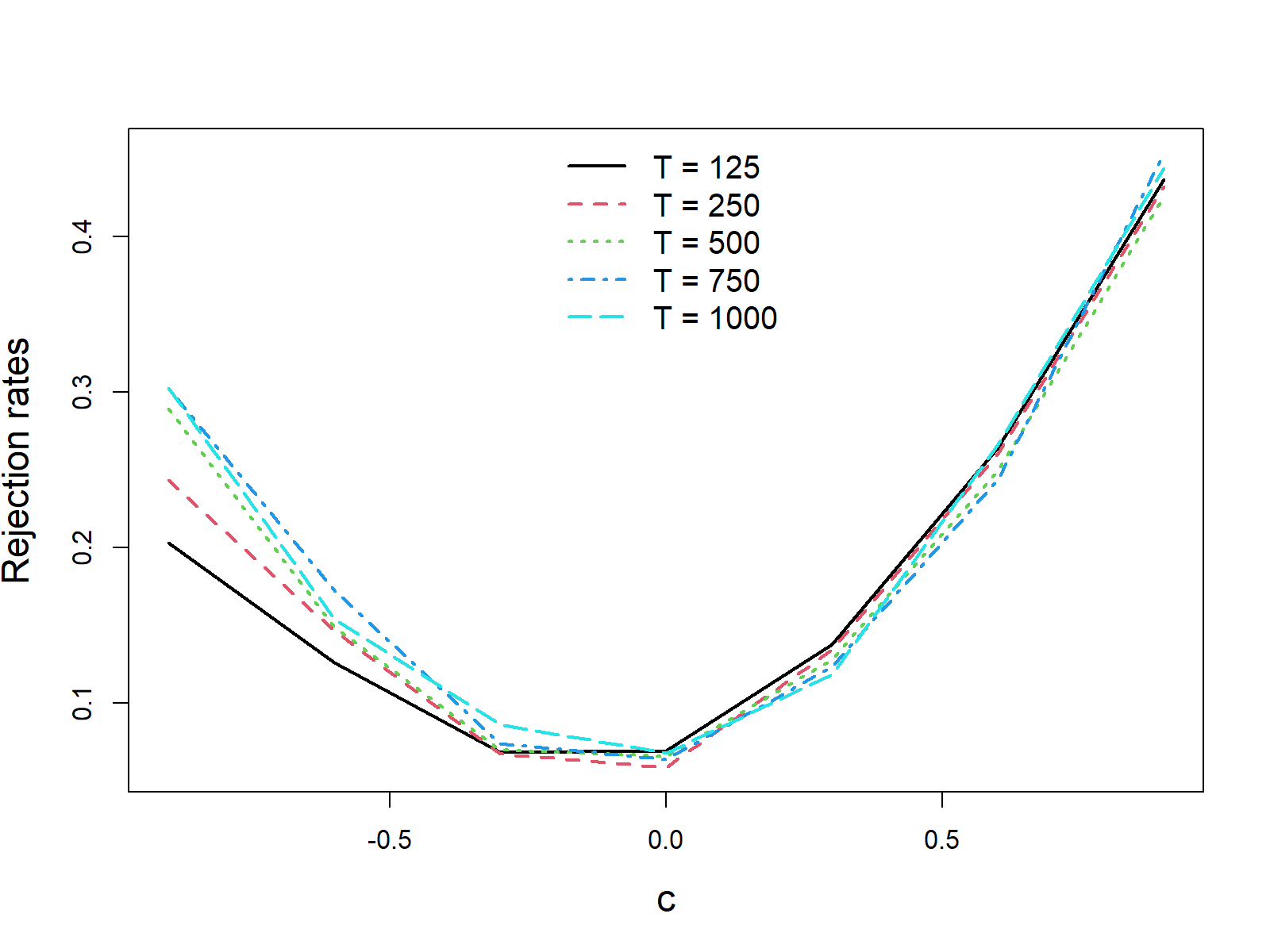}
    \caption{$\mathbf{b}=0.6$ and $q \sim T^{1/5}$}
    \label{fig:rejection_rates2}
  \end{subfigure}
 \begin{subfigure}[t]{0.48\linewidth}
    \centering
    \includegraphics[width=\linewidth]{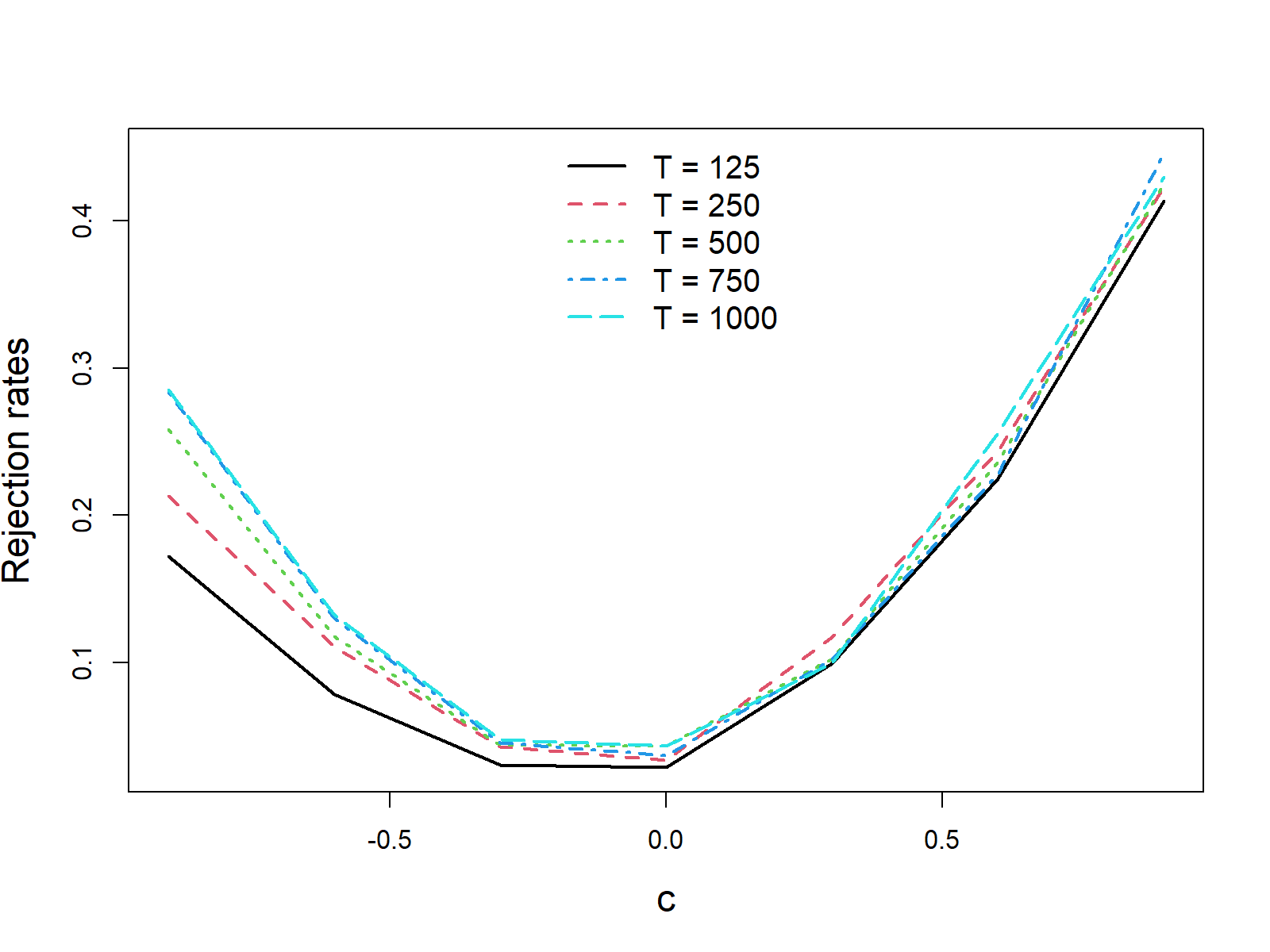}
    \caption{$\mathbf{b}=0.15$ and $q \sim T^{1/4}$}
    \label{fig:rejection_rates3}
  \end{subfigure}
  \hfill
  \begin{subfigure}[t]{0.48\linewidth}
    \centering
    \includegraphics[width=\linewidth]{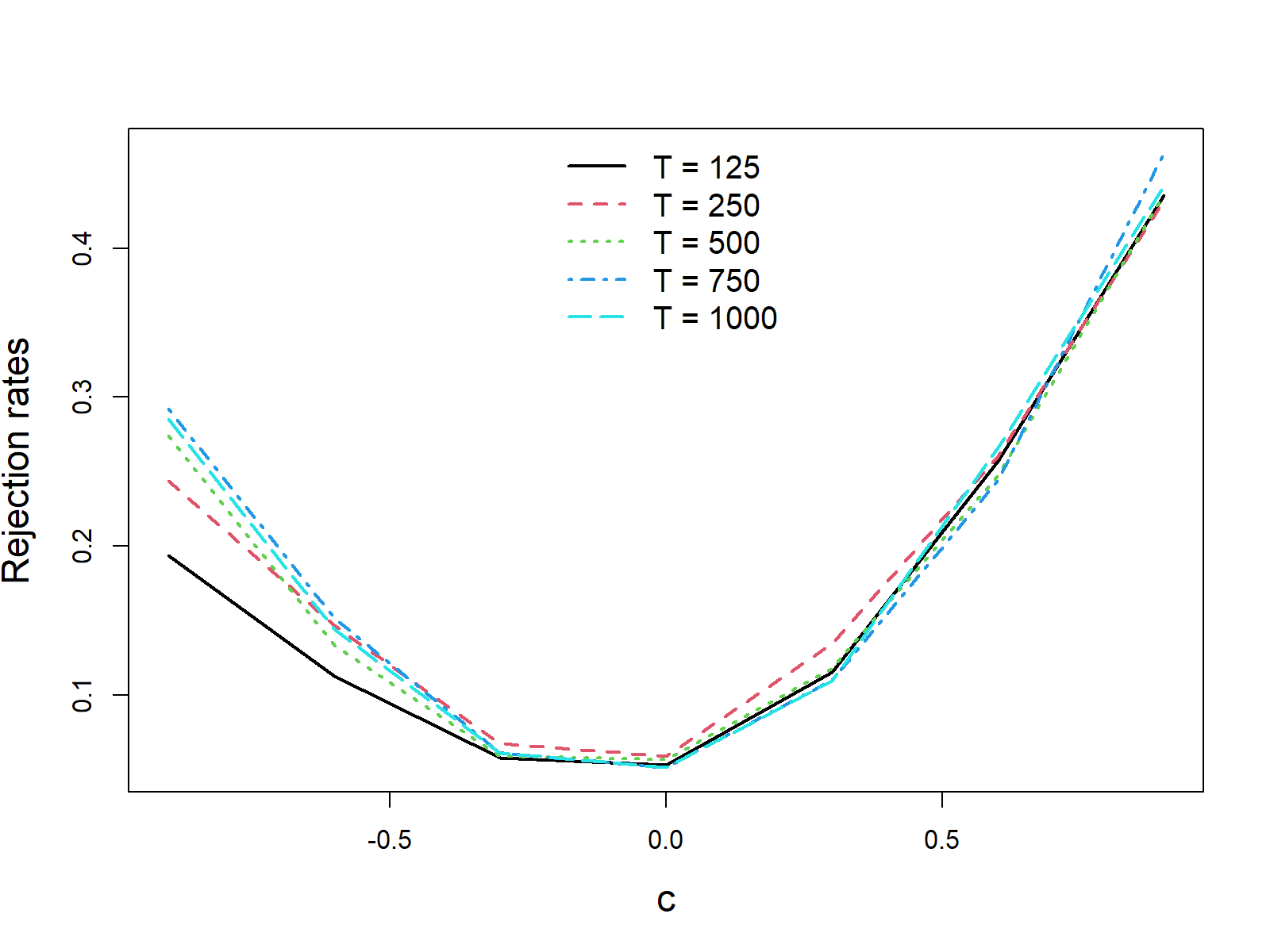}
    \caption{$\mathbf{b}=0.6$ and $q \sim T^{1/4}$}
    \label{fig:rejection_rates4}
  \end{subfigure}
\end{figure} 
\newpage
\phantomsection{}

  \bibliography{integer_integration}

\end{document}